\documentclass[fleqn,usenatbib]{mnras}
\usepackage{newtxtext,newtxmath}
\usepackage[T1]{fontenc}
\DeclareRobustCommand{\VAN}[3]{#2}
\let\VANthebibliography\thebibliography
\def\thebibliography{\DeclareRobustCommand{\VAN}[3]{##3}\VANthebibliography}

\usepackage{graphicx}
\usepackage{amsmath}	
\usepackage{orcidlink}
\usepackage{hyperref}
\usepackage{physics}
\usepackage{booktabs}

\def\grs{GRS\,1915+105}
\def\gx{GX\,339--4}

\def\ha{H$\alpha$}

\def\micron{\textmu m}
\def\p{$\pm$}

\def\lbol{$L_{\rm Bol}$}
\def\ledd{$L_{\rm Edd}$}

\def\av{$A_{\rm V}$}

\def\rxte{{\em RXTE}}
\def\jwst{{\em JWST}}
\def\spitzer{{\em Spitzer}}
\def\wise{{\em WISE}}

\def\astrosat{{\em AstroSat}}

\def\nustar{{\em NuSTAR}}

\def\mbh{$M_{\rm BH}$}
\def\Msun{M$_{\odot}$}
\def\Rsun{R$_{\odot}$}

\def\xspec{{\tt XSPEC}}
\def\gtsim{\mathrel{\hbox{\rlap{\hbox{\lower3pt\hbox{$\sim$}}}\hbox{$>$}}}}
\def\ltsim{\mathrel{\hbox{\rlap{\hbox{\lower3pt\hbox{$\sim$}}}\hbox{$<$}}}}

\def\fmirfx{$F_{\rm MIR}/F_{\rm X-ray}$}

\defcitealias{john24}{J24}
\defcitealias{gandhi25}{G25}

\title[Mid-IR Spectral-Timing of \gx]{
Infrared Synchrotron Emission in the Soft State of \gx\ and the Mid-Infrared/X-ray Luminosity Plane of Black Hole X-ray Binaries}
\author[JWST Timing Consortium]{\parbox{\textwidth}{
P.~Gandhi,\orcidlink{0000-0003-3105-2615}$^{1}$\thanks{E-mail: poshak.gandhi@soton.ac.uk (PG)}
D.~M.~Russell,$^{\orcidlink{0000-0002-3500-631X}}$$^{2}$
M.~C.~Baglio,$^{\orcidlink{0000-0003-1285-4057}}$$^{3}$
Y.~Bhargava,$^{\orcidlink{0000-0002-5967-8399}}$$^{4,\,5}$
R.~Duncan,$^{6,\,7}$
A.~G\'urpide,$^{\orcidlink{0000-0002-2256-2704}}$$^{1}$
C.~O. Heinke,$^{\orcidlink{0000-0003-3944-6109}}$$^{8}$
C.~Knigge,$^{1}$
K.~S.~Long,$^{9,\,10}$
T.~J. Maccarone,$^{11}$
G.~Mastroserio,$^{12}$
T.~D.~Russell,$^{\orcidlink{0000-0002-7930-2276}}$$^{13}$
A.~W. Shaw,$^{\orcidlink{0000-0002-8808-520X}}$$^{14}$
A.~J.~Tetarenko,$^{\orcidlink{0000-0003-3906-4354}}$$^{15}$
F.~M. Vincentelli,$^{\orcidlink{0000-0002-1481-1870}}$$^{1,\,16}$
E.~S.~Borowski,$^{\orcidlink{0000-0002-7004-9956}}$$^{17}$
D.~A.~H.~Buckley,$^{\orcidlink{0000-0002-7004-9956}}$$^{18}$
P.~Casella,$^{\orcidlink{0000-0002-0752-3301}}$$^{19}$
C.~Dashwood Brown,$^{1}$
G.~C.~Dewangan,$^{20}$
R.~I.~Hynes,\orcidlink{0000-0003-3318-0223}$^{17}$
S.~Markoff,$^{\orcidlink{0000-0001-9564-0876}}$$^{6,\,7}$
J.~A.~Tomsick,$^{\orcidlink{0000-0001-5506-9855}}$$^{21}$
K.~Alabarta,$^{\orcidlink{0000-0003-0168-9906}}$$^{2}$
F.~Carotenuto, $^{\orcidlink{0000-0002-0426-3276}}$$^{19}$
E.~Carver,$^{\orcidlink{0009-0008-1454-163X}}$$^{15}$
N.~Castro-Segura,$^{1}$
P.~Charles,$^{1}$
F.~Lewis,$^{\orcidlink{0000-0003-3352-2334}}$$^{22,\,23}$
J.~A.~Paice,$^{24}$
D.~Pawar,$^{\orcidlink{0009-0000-9571-4361}}$$^{25}$
M.~E.~Ressler,$^{\orcidlink{0000-0001-5644-8330}}$$^{26}$
S.~K.~Rout, $^{\orcidlink{0000-0001-7590-5099}}$$^{2}$
P.~Saikia,$^{\orcidlink{0000-0002-5319-6620}}$$^{2}$
T.~Shahbaz,$^{\orcidlink{0000-0003-1331-5442}}$$^{27,\,28}$
G.~R.~Sivakoff$^{\orcidlink{0000-0001-6682-916X}}$$^{8}$
\begin{center} (JWST Timing Consortium) \end{center}
\begin{center} {\small \textit{Affiliations can be found at the end}}\end{center}
}
}

\date{Received in original form 2025 Sep 26}
\pubyear{\the\year{}}

\begin{document}
\label{firstpage}
\pagerange{\pageref{firstpage}--\pageref{lastpage}}
\maketitle
\begin{abstract}
Progress in understanding the growth of accreting black holes remains hampered by a lack of sensitive coordinated multiwavelength observations. In particular, the mid-infrared (MIR) regime remains ill-explored except for jet-dominant states. Here, we present comprehensive follow-up of the black hole X-ray binary \gx\ during a bright disc-dominated state in its 2023/24 outburst as part of a multi-wavelength campaign coordinated around {\em JWST}/MIRI. The X-ray properties are fairly typical of soft accretion states with no significant X-ray variability, though with a weak high-energy Comptonised power-law tail. The source is significantly detected between 5--10\,\micron, albeit at a faint flux level requiring MIR compact jet emission to be quenched by a factor of $\sim$\,300 or more relative to previous hard-state detections. The MIRI spectrum can be described as a simple power-law with slope $\alpha$\,=\,+0.39\,\p\,0.07  ($F_\nu$\,$\propto$\,$\nu^\alpha$), but surprisingly matches neither the radio/sub-mm nor the optical broadband slopes. Significant MIR stochastic variability is detected. Synchrotron radiation from the same medium responsible for high-energy Comptonisation can self-consistently account for the observed MIRI spectral-timing behaviour, offering new constraints on the physical conditions in the soft-state accretion disc atmosphere/corona. Alternative explanations, including a circumbinary disc or emission from a warm wind, fail to cleanly explain either the spectral properties or the variability. Multiwavelength timing cross-correlations show a puzzlingly long MIR lag relative to the optical, though at limited  significance. We compile archival MIR and X-ray luminosities of transient black hole systems, including previously unreported detections of \gx. These trace the evolution of the MIR--to--X-ray flux ratio with accretion state, and also reveal high MIR luminosities for GX339-4 across all states.
\end{abstract}




\section{Introduction}
\label{sec:intro}

X-ray binaries (XRBs) are ideal targets for studying transient multiwavelength emission arising from accretion activity. They spend most of their lives in a state of low-level `quiescence,' as a result of weak accretion on to the central compact object (either a black hole or a neutron star) from the binary companion \citep[e.g., ][]{lasota00, Campana98, heinke03}. This is punctuated by occasional accretion `outbursts' lasting months to years and characterised by a prolific rise in radiation output and flux variability \citep[e.g., ][]{dunn10}. Several physical  components including the accretion disc, collimated and uncollimated outflows, the companion star, and more, can contribute to the observed multiwavelength emission \citep[e.g., ][]{done07}. Spectral-timing observations offer the possibility to disentangle these components based on their differing scales and characteristic variability timescales \citep[e.g., ][]{uttleycasella14}. 

The mid-infrared (MIR) spectral regime ($\lambda$\,$\sim$\,5--100\,\micron) remains ill-studied for XRBs. There are few atmospheric transmission windows and many sources of instrumental and natural background spanning this wavelength range, so observations are usually only feasible with sensitive space instrumentation. There have been efforts to detect XRBs with all mid-infrared missions, though the poor spatial resolution of historical infrared missions  made robust counterpart identification difficult \citep{smith90}. The ISOCAM camera aboard the {\em Infrared Space Observatory} ({\em ISO}) detected several bright or persistent sources \citep{mirabel96, kaper98, fuchs03}. The era of \spitzer\ and \wise\ opened up MIR studies of XRBs more widely, with many photometric detections in quiescence as well as in outburst \citep[e.g., ][]{migliari06, gallo07, wachter08, Gandhi-2011, wang14, john24}. Spectroscopic detections of bright and persistent sources also followed \citep{fuchs06, rahoui10, rahoui11}. 

Transient outbursts grow from quiescence through the `hard' state during which the hot accretion flow close to the central source strengthens \citep[e.g., ][]{fender04,done07}. Both quiescence and the hard-state are typically dominated by non-thermal emission processes, including synchrotron radiation in the MIR \citep{migliari06, gallo07, Gandhi-2011, rahoui11}. Near peak, this can approach flux levels bright enough to be detectable from the ground in the MIR 10\,\micron\ transmission window \citep{vanparadijs94, russell13b, baglio18, vincentelli21, saikia22, echiburutrujillo24}. Dissipative plasma shocks within compact jets are able to account for the broadband spectral as well as timing properties observed during the hard-state \citep{markoff01, malzac13, Russell-2013, malzac14, gandhi17}. 
We note that the presence of circumbinary dust discs have also been proposed as the primary driver of the MIR in quiescence \citep{muno6}, a point that we will return to later in this work. 

By contrast, transient XRB detections are even sparser in the MIR during the `soft' accretion state, when thermal emission from a fully active accretion disc dominates the electromagnetic output. Radio jet emission appears to be heavily suppressed in these soft states \citep{russell11, rushton16, russell20, john24}, though some transient compact radio jets have been observed during this phase \citep{russell20_transientsoftjet}. It remains unclear whether this is a result of a complete failure to launch the jet in the soft states, or a result of inefficient (re-)acceleration of collimated plasma as it travels away from the central source, due to e.g. a change in the jet magnetic field configuration. Current constraints on optical and IR flux suppression in the soft states of most sources could support either scenario \citep[e.g., ][]{russell06}.

In the canonical model of compact jets \citep{Blandford-1979}, the characteristic frequency of synchrotron emission at any given zone scales with distance along the jet. Radio frequencies probe the outer regions of the jet, so higher frequencies (including the IR and optical) need to be utilised in order to probe physical zones closer in to the jet base. Much observational evidence supports the presence of re-acceleration along the jet in order to account for the flat IR--to--radio spectral energy distributions (SEDs) of hard-state compact jets \citep[e.g., ][though see also \citealt{kaiser06}]{Blandford-1979, Meier-2001, malzac13}. 

Here, we present one of the most sensitive MIR detections of a black hole XRB in a disc-dominated accretion state. This enables deep constraints on the inner jet behaviour during this state, as well as allowing us to test for the presence of other components often invoked in the infrared such as dust, a circumbinary disc, and the companion star. 

For this purpose, we used \jwst\ to observe the XRB \gx,  a dynamically confirmed black hole system with an orbital period $P$\,=\,1.759\,d \citep{hynes03,heida17}, located at a minimum heliocentric distance $d$\,$\gtsim$\,5\,kpc \citep{hynes04,heida17}. 
\gx\ undergoes frequent, bright transient outbursts, as a result of which it has become a cornerstone of multiwavelength X-ray binary studies over many decades \citep{dunn08, makishima86, corbel02, belloni05, Gandhi-2008, Gandhi-2010, Casella-2010, corbel13, vincentelli19}. It has been detected in the mid- and even the far-IR \citep{Gandhi-2011, corbel13b} during the hard outbursting state. No MIR {\em spectroscopic} observations have been reported so far. 

The system parameters remain under debate, with a black hole mass range of $M_{\rm BH}$\,=\,4--11\,\Msun\ \citep{zdziarski19}. Some of the key parameters, including the values adopted in this work, are listed in Table\,\ref{tab:pars}. When available or computed herein, quoted uncertainties correspond to 68\,\% intervals throughout the paper, unless stated otherwise.

\begin{table}
    \centering
    \caption{Physical parameters of \gx\ adopted in this work}
    \begin{tabular}{l|r}
    \hline
    Distance $d$      &  8\,($>$\,5)~kpc\\
    Orbital period $P$      &  1.7587\,\p\,0.0005~d\\
    Black hole mass \mbh & 6\,(\,4--11)~\Msun \\ 
    Inclination $i$ & 50\,(\,40--60)~deg \\
    Mass ratio $q (=M_2/M_1)$  & 0.18\,\p\,0.05 \\
    Binary separation $a$ & 11.77\,R$_\odot$ = 27.3~lt-sec\\
    Donor Roche Lobe radius $R_{\rm L}$ & 3.88\,R$_\odot$ = 9.1~lt-sec\\
    Extinction $A_{\rm V}$ & 3.5\,\p\,0.5 mag\\
    \hline
    \end{tabular}~\\
    Approximate ranges or constraints are noted in brackets. 1-$\sigma$ uncertainties are stated for measured/adopted quantities, when available (\citealt{Gandhi-2011,heida17, zdziarski19}).
    \label{tab:pars}
\end{table}

\section{Observations}
\label{sec:obs}

\begin{figure*}
    \centering
    \hspace*{-1cm}\includegraphics[angle=90,width=1.095\textwidth]{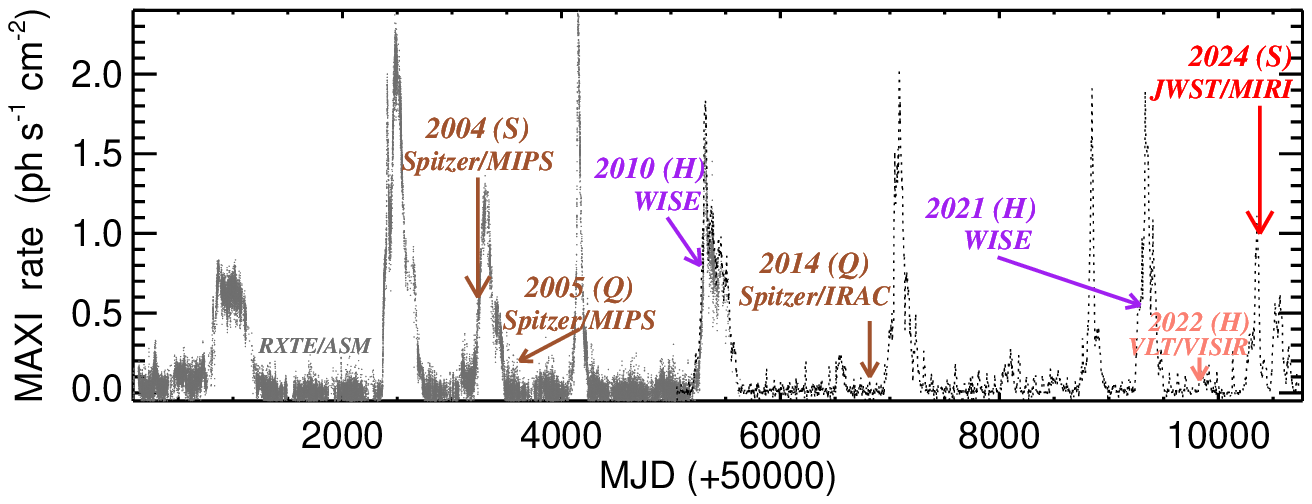}
    \includegraphics[angle=0,width=\columnwidth]{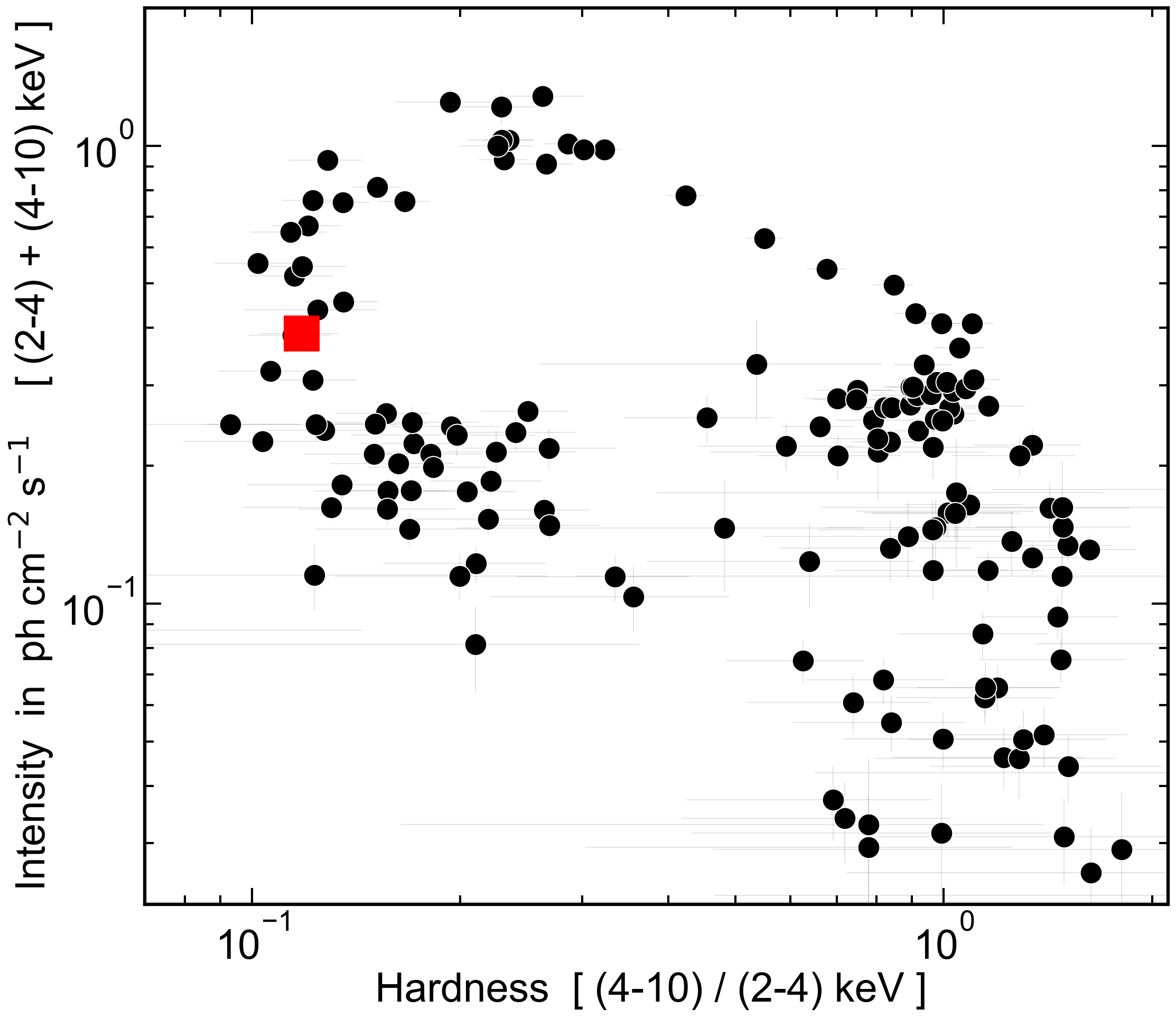}
    \hfill
    \includegraphics[angle=0,width=\columnwidth]{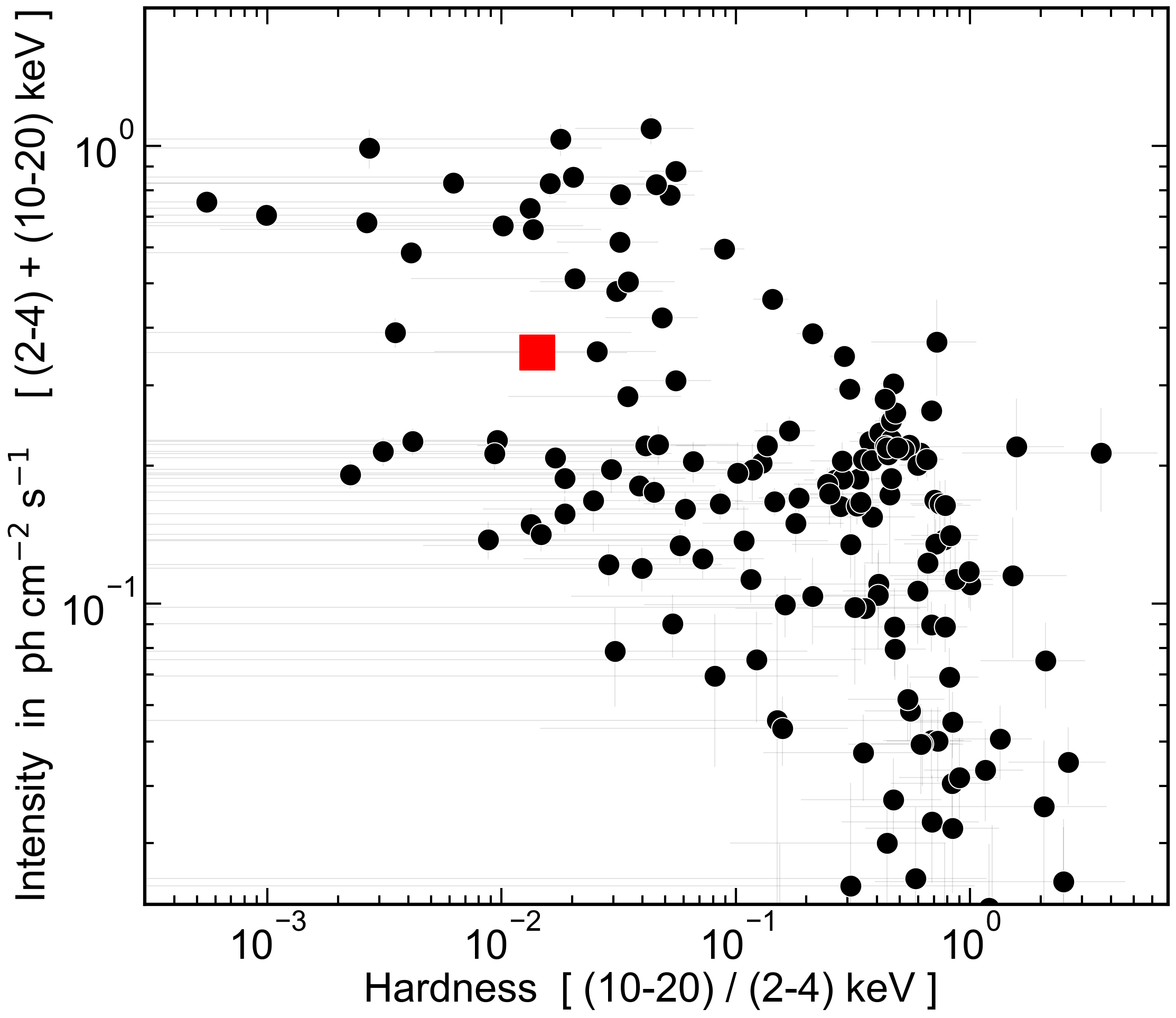}
    \caption{({\em Top}) Long-term X-ray light curve of \gx, with all known historical MIR observational epochs annotated and denoted by arrows. Recent X-ray data are from the MAXI camera, covering the 2--20\,keV band. Pre-MAXI data from \rxte/ASM are plotted as gray dots before $\sim$\,MJD\,55,000, normalised to the MAXI photon flux by multiplying by a factor of 0.032, which converts between the respective full energy bands of the two cameras assuming a 1 keV blackbody model approximately appropriate for outburst peaks. X-ray accretion state is given in parentheses: soft (S), hard (H) and quiescent (Q). {\em (Bottom)} X-ray Hardness-Intensity diagrams using MAXI data, covering the 2023/24 outburst. The panels differ in the definition of the harder band: (4--10\,keV) on the left and (10--20\,keV) on the right, highlighting differences relevant to the dominance of the high-energy power-law tail. The red point denotes the observation date coinciding with \jwst. 
    }
    \label{fig:longterm}
\end{figure*}

\subsection{{\em JWST}/MIRI}
\label{sec:miriobs}

We first triggered \jwst\ in September 2022, during the rising part of an outburst at that time as part of a Target-of-Opportunity Phase Cycle\,1 program (ID 1586). The observation was unsuccessful due to a guide star acquisition error. The cause of the failure was analysed by the STScI mission team and attributed to a faulty entry in the mission's guide star catalogue, which was later rectified by the mission team (WOPR 88542; STScI, priv. comm.). 

\gx\ was subsequently retriggered on, and observed, on 2024 March 10 (MJD~60379) using MIRI \citep{miri}, set up for Slitless Low Resolution Spectroscopy \citep[LRS; ][]{kendrew15}. The SLITLESS PRISM sub-array was utilised to obtain continuous time-series monitoring for an on-source stare lasting about 2\,hours. Source acquisition was performed using an offset pointing slew from the nearby bright star Gaia\,DR3\,5938658049055243136 located 13\farcs 5 to the south-east of \gx. 

The first on-source science frame began at UT\,06:36:16.31. Observations were performed in the standard non-destructive {\tt FASTR1} mode (cf. \citealt{ressler15, dyrek24}), for a total of {\tt nINT}\,=\,1,470 integrations (each 4.77\,s long, with a tagged-on reset dead-time of approximately 0.16\,s). The time of the final science frame was UT\,08:37:03.78.

Data were processed using the latest \jwst\ pipeline, version 1.17, available at the time of writing, with calibration database context 1322. The pipeline conducts several steps of calibration and artefact correction to ensure a linear signal response and to flag outliers. There are still a number of shortcomings in the pipeline (e.g. background uncertainties are not included within the final reported errors, there appears to be excess r.m.s. variability towards the red end of the bandpass), which have been discussed in our prior work \citep{gandhi25}. These were manually implemented where possible and relevant for the analysis below. Other relevant pipeline settings were adopted from our prior work. We experimented with modifying parameters (e.g. the jump step detection threshold which can identify sudden ramp changes due to cosmic rays), but these did not impact the final extraction significantly. 

The source was found to be faint, but clearly detected in all integrations. The signal weakens towards longer wavelengths where calibrations remain uncertain, and background artefacts become more difficult to account for. We thus restricted our analysis to the wavelength regime of 5--10\,\micron. 

\jwst\ caught \gx\ during a state characterised by strong disc emission, as the source was declining from the peak of its 2023--24 outburst (Fig.\,\ref{fig:longterm}). The overall outburst turned out to be weaker compared to prior prominent outbursts, though the source still completed a full traversal of the X-ray hardness-intensity diagram (HID) common to this system \citep[e.g. ][]{belloni05}. The source was classified as being in the X-ray soft state by \citet{mastroserio25}, who used MAXI X-ray sky monitor hardness ratios to place it in a region of the HID characteristic of soft sources (Fig.\,\ref{fig:longterm} bottom panels). Their detailed spectral analysis confirmed the presence of a strong accretion disc, together a Comptonised tail at energies above 10\,keV carrying $\sim$\,1\% of the X-ray broadband flux. Furthermore, X-ray variability was found to be weak and no significant X-ray polarisation was detected, self-consistent with a soft state. Our analysis presented later in this paper confirms these general properties. The panel on the bottom-right of Fig.\,\ref{fig:longterm} shows hardness values based on the highest-energy MAXI band (10--20\,keV, as opposed to 4--10\,keV on the bottom-left). The scatter in this panel is more sensitive to the strength of the high-energy power-law tail than HIDs based on lower-energy bands, demonstrating some scatter within this accretion state which should be kept in mind for any detailed comparisons of the accretion state with other observations.

\subsection{Multiwavelength Campaign} 
\label{sec:mwlcampaign}

We arranged for several multiwavelength facilities to coordinate with the \jwst\ observation. Due to a failure in unrelated observations scheduled just prior to ours, the \jwst\ schedule changed dynamically and the \gx\ observing window slid forward by approximately 2\,hours without our knowledge, resulting in imperfect overlap with some facilities. Nevertheless, we were able to build in sufficient redundancy within our campaign to ensure strict simultaneity with a few facilities, and quasi-simultaneity with the rest to within 1\,day (and usually just a few hours) across the electromagnetic spectrum. A figure denoting the coodination timeline may be found in Appendix\,\ref{sec:coordination}.

\subsubsection{Swift} 
The X-ray Telescope \citep{xrt} onboard the {\em Neil Gehrels Swift Observatory} \citep{swift} observed the source for a duration of 1029.6 s, starting at MJD 60379.45545 (ObsID 00014052219) in windowed timing (WT) mode. This observation began 2.3 hours after the end of the \jwst\ observation. 

Data products were extracted using UK Swift tools \citep{evans09}. The spectrum was grouped to a minimum of 20\,counts for fitting. The source was  bright, with a  net count rate of 224 ct\,s$^{-1}$ over the 0.3--10 keV band, suggesting pile-up even in WT mode (the automated extraction corrects for this using an annulus). Spectral analysis was restricted to 1--10\,keV, in order to avoid residuals related to bright absorbed sources at soft energies.\footnote{\url{https://www.swift.ac.uk/analysis/xrt/digest\_cal.php\#abs}}

\subsubsection{NuSTAR}
The {\em Nuclear Spectroscopic Telescope Array} \citep[\nustar;][]{harrison13} observed \gx\ for approximately 20\,ks starting on 2024 March 10 02:30 UTC (ObsID: 80902342002, PI: Shaw). Data were reprocessed with the FTOOL {\tt nupipeline} included in HEASoft v6.35.1 (NuSTARDAS v2.1.5). Event files from each focal plane module (FPMA and FPMB) were corrected to the solar system barycentre using {\tt barycen}. Source spectra and light curves were extracted from FPMA and FPMB using a circular region of radius $70\arcsec$ centred on the source and background spectra and light curves were extracted from a circular region of the same size, centred on a source-free region on the same chip as \gx. Spectral responses were generated using the FTOOL {\tt nusproducts}. Light curves were binned into intervals of 4.93\,s duration (closely matching the MIRI integration spacing), and spectra were grouped such that each bin contained a minimum of 20 counts.

\subsubsection{LCO}
Optical monitoring of GX 339--4 was performed with telescopes at the Las Cumbres Observatory (LCO) before, during, and after the \jwst\ observation. Data were taken as part of a monitoring campaign of $\sim$50 low-mass X-ray binaries coordinated with the Faulkes Telescope Project \citep{Lewis2008,Lewis2018}. GX 339--4 has been regularly monitored by the 2-m Faulkes Telescope South (at Siding Spring Observatory, Australia), which is part of LCO, since 2007 \citep[e.g.][]{Russell2008ATel,CadolleBel2011}. To increase the cadence close in time to the \jwst\ observation, we observed with LCO on 12 epochs within $\pm 2$ d of \jwst. The 1-m LCO network telescopes at Cerro Tololo Inter-American Observatory (Chile), the South African Astronomical Observatory (SAAO; South Africa), and Siding Spring Observatory were used, as well as the 2-m at Siding Spring Observatory \citep[see also][for LCO magnitudes reported in the weeks--months prior to the \jwst\ observation]{Alabarta2024ATel}. Images were taken using the Bessel $B$, $V$ and $R$ filters (Vega magnitudes), the SDSS $g^{\prime}$, $r^{\prime}$ and $i^{\prime}$ filters (AB magnitudes), and Pan-STARRS $z_{\rm s}$ filter (AB magnitudes).

To reduce and analyse the images, the real-time data analysis pipeline, \textsc{XB-NEWS} \citep[the X-ray Binary New Early Warning System; see][]{Russell2019XBNEWS,Goodwin2020} was used. The \textsc{XB-NEWS} pipeline downloads images of targets from the LCO archive soon after they are taken by the telescopes, and their associated calibration data. The pipeline then carries out quality control steps so that only good quality images are analysed, and produces astrometric solutions for each image using {\em Gaia} DR2 positions\footnote{\url{https://www.cosmos.esa.int/web/gaia/dr2}}. Aperture photometry was performed on all detected stars in each image, solving for zero-point calibrations between epochs \citep{Bramich2012}. The ATLAS All-Sky Stellar Reference Catalog \citep[ATLAS-REFCAT2;][]{Tonry2018} was adopted for flux calibration. Multi-aperture photometry \citep[azimuthally-averaged PSF profile fitting photometry,][]{Stetson1990} was also performed for point sources. We detect GX 339--4 with high significance. Modified Julian Dates and Barycentric Julian Dates at mid-exposure were output by XB-NEWS. For the broadband spectrum, all measurements within $\pm 1.0$ d of 
\jwst\ were retained and averaged in each filter.

\subsubsection{REM}
We observed GX 339–4 in the $H$ band with the 60-cm Robotic Eye Mount (REM) telescope at La Silla (Chile) on March 10, 2024 (MJD 60379) under good sky conditions, strictly simultaneously with \jwst, as part of a monitoring campaign during the ongoing outburst. Observations were conducted between 06:51:07 UT and 06:53:37 UT and consisted of one set of exposures composed of five dithered 30-second integrations, which were combined for optimal background subtraction.

We performed aperture photometry using {\tt daophot} \citep{daophot}, with an aperture radius of 1.5 times the average full width at half maximum (FWHM) of the flux profiles of the field stars. Flux calibration was performed using the 2MASS catalog\footnote{\url{https://irsa.ipac.caltech.edu/Missions/2mass.html}} \citep{skrutskie06}.

\subsubsection{HAWK-I}
Near-IR photometry in the $K_s$ band was obtained with HAWK-I, an IR imager with 2$\times$2 Hawaii 2RG detectors mounted at the Very Large Telescope UT-4 on Cerro Paranal, Chile \citep{hawki}. The observation ended up being strictly simultaneous with 
\jwst. 

Data were collected in Fast-phot mode, i.e. reducing the field of view to one stripe (2048 x 64 pixels) per detector. This mode allows to obtain data cubes of 250 frames, each with 0.125 s time resolution. The instrument rotation and pointing was set to place GX\,339-4 and a bright reference star in the lower quadrant (Q1). 

The average source flux was measured through aperture photometry with an annular background region, and absolute photometry was calibrated against the bright reference star. The average flux density was measured to be $F_{\rm 2.2\,\mu{\rm m}}$\,=\,0.55\,\p\,0.05\,mJy. HAWK-I's fast-phot allows very rapid sub-second photometry, which requires dedicated calibration against changing seeing, and will be presented in future work.

\subsubsection{AstroSat}

\astrosat\ \citep{Singh2014SPIE.9144E..1SS} observed \gx\ between 2024-03-10 10:51:45.5 and 2024-03-12 14:06:46.3 UTC (observation ID: A13\_028T01\_9000006122). To capture the broadband evolution of the source, the Soft X-ray Telescope \citep[SXT;][]{Singh2017JApA...38...29S} and Large Area X-ray Proportional Counter \citep[LAXPC;][]{Yadav2017CSci..113..591Y, Agrawal2017JApA...38...30A} observed the source in fast window (FW) and event analysis (EA) mode respectively. To verify whether we detect any UV emission from the source, we also configured observations with the Ultra Violet Imaging Telescope \citep[UVIT;][]{Tandon2017AJ....154..128T, Tandon2020AJ....159..158T} using various filters in the far-ultraviolet.


SXT level 2 data for individual orbits were downloaded from AstroBrowse\footnote{\url{https://astrobrowse.issdc.gov.in/astro_archive/archive/Home.jsp}}  and merged using the \textsc{julia} tool \texttt{SXTMerger}. To mitigate core pile-up, spectra were extracted using an annular region confined to 1--4\,arcmin. Additionally, lightcurve analysis showed intervals of 0 counts during some `good time' intervals, suggesting data drops during these segments, which were excluded. The final spectral analysis was conducted with a recommended 2\% systematic error included \citep[e.g.][]{Bhargava2019MNRAS.488..720B, Bhargava2023ApJ...955..102B}.
Noticing strong residuals close to 2~keV, we applied additional gain corrections, keeping the gain slope fixed at 1.0 and varying the gain offset parameter freely. 
For bright sources, the standard ARF shows additional residuals, which can be mitigated through dedicated modelling, and we utilized the ARF developed for the SXT observation of the bright transient MAXI\,J1820+070  \citep{Banerjee2024ApJ...964..189B}. Since the extraction regions differ between the observations, this results in strong change in the cross-normalization constant for SXT (Table~\ref{tab:fit}). 

\textsc{LAXPCsoftware22Aug15}\footnote{\url{http://astrosat-ssc.iucaa.in/uploads/laxpc/LAXPCsoftware22Aug15.zip}} \citep{Antia2021JApA...42...32A, Misra2021JApA...42...55M} was used to process level 1 LAXPC data, including uniform merging of individual orbits and gain correction, and extraction of spectra and responses. Due to the steep spectral slope of the source in the operational energy range of LAXPC, we extracted the data only from layer 1 of the instrument to minimise background contribution. As a result, the LAXPC spectra are usable only in the 3--20~keV energy range. We included a recommended 3\% additional systematic error \citep[e.g.][]{Bhargava2019MNRAS.488..720B, Bhargava2023ApJ...955..102B}. 


Jointly with X-ray observations, UVIT observed the source with four FUV filters consecutively, in order: F148W~(CaF2-1), F154W~(BaF2), F169M~(Sapphire) and F172M~(Silica); see \citet{Tandon2020AJ....159..158T} for filter details. The level 2 processed data were downloaded from AstroBrowse and images from each filter were inspected to identify if the source was detected. Due to a minor pointing offset between UVIT and SXT (which was the primary instrument for the observation as required by the FW mode constraint), we verified the astrometric correction and determined the expected source location using {\em GALEX} imaging \citep{galex}. The source was not detected in any of the filters and therefore we placed upper limits in the UV bands, which are fully consistent with our broadband modelling discussed below. 

\subsubsection{ATCA}

The Australia Telescope Compact Array (ATCA) observed \gx\ between 2024-03-10 12:40:40 UT and 16:16:50 UT (starting $\sim$2\,hours after the final \jwst\ science frame). Data were recorded simultaneously at 5.5\,GHz and 9\,GHz, with 2\,GHz of bandwidth at each central frequency. We used PKS~B1934$-$638 for bandpass and flux density calibration, and the nearby source J1664$-$50 for gain calibration. Data were first flagged for radio frequency interference before being calibrated and imaged following standard procedures within the Common Astronomy Software Application (CASA v. 5.3; \citealt{CASA2022}). We used a Briggs robust parameter of 0 and fit for a point source in the image plane. 

We detected a radio source associated with a jet ejection event from \gx\ with measured flux densities of 1.36$\pm$0.03\,mJy at 5.5\,GHz and 0.89$\pm$0.03\,mJy at 9\,GHz. This gives a radio spectral index ($F_\nu$\,$\propto$\,$\nu^{\alpha_{\rm radio}}$) of $\nu^{\alpha_{\rm radio}}$\,=\,--0.9\,$\pm$\,0.2, consistent with optically-thin synchrotron emission from ejecta. Imaging the radio data on shorter time-intervals to explore intra-observational variability, we find the radio flux density to be steadily rising throughout the radio observation. See Section~\ref{sec:radiommresults} for further discussion, and full details will be presented by Russell et al. (in prep.).

\subsubsection{ALMA}
\label{sec:obs_alma}
GX 339--4 was observed with the Atacama Large (Sub-)Millimeter Array (ALMA; Project Code: 2023.A.00018.T) on 2024 March 10 between 10:55--11:49 UTC, for a total on-source observation time of $\sim$17 min. Data were taken in Bands 3, 4, and 6 at central frequencies of 97.5, 145.0, and 233.0 GHz, respectively. The ALMA correlator was set up to yield $4\times2$ GHz wide base-bands, with a 2.0-s correlator dump time. During our observations, the array was in its Cycle 10 C1 configuration with 43 antennas. The median precipitable water vapour ranged from 3.8--4.2 mm during the observations. We reduced the data within the Common Astronomy Software Application package (\textsc{casa} v. 6.2; \citealt{2007ASPC..376..127M}), using the ALMA pipeline to calibrate the data. We used J1617--5848 as a band-pass/flux calibrator and J1650--5044  as a phase calibrator. Similar to the ATCA results, we detected a source offset from the known source position of \gx, indicating a jet ejection (Section~\ref{sec:radiommresults}; Russell et al. in prep.). 

To extract flux densities of this source, we performed multi-frequency synthesis imaging with the \texttt{tclean} task within \textsc{casa}, using a natural weighting scheme to maximise sensitivity. Flux densities were measured from these images by fitting a point source in the image plane (with the \texttt{imfit} task within \textsc{casa}). We significantly detect the a source at 97.5 GHz and 145.0 GHz, with flux densities of $0.35\pm0.03$ mJy and $0.26\pm0.04$ mJy, respectively. However, we only achieve a non-detection at 233.0 GHz, with a 3--$\sigma$ upper limit of 0.18 mJy.

\subsection{Reddening}
\label{sec:reddening}

Reddening to the source is somewhat uncertain, with values spanning $E$($B$--$V$)\,$\approx$\,1.0--1.2 and corresponding extinctions \av\,$\approx$\,3.0--4.2\,mag quoted in the literature \citep[e.g., ][]{zdziarski98, Gandhi-2011, zdziarski19}. Estimates based on reddening of field stars \citep[e.g., ][]{grindlay79}, on intervening absorption components \citep{hynes04}, and from modelling the spectrum of \gx\ itself \citep{CadolleBel2011, kosenkov20}, are all broadly consistent within uncertainties. 

Here, we utilise a nominal extinction value \av\,=\,3.5\,mag, but with an uncertainty of $\Delta$\av\,=\,0.5\,mag propagated through all corrections necessary for model fitting, assuming a normal distribution. The extinction curve wavelength dependence follows the widely adopted \citet{cardelli89} law over the ultraviolet to optical range, and this is extended to the mid-infrared using the recommended interstellar extinction from \citet{chiartielens06}. Dereddening is most important for the optical--UV regime. In the MIRI range, the flux correction factors for \gx\ are much milder, with an average value of 1.2, and peaking around 1.4 near 10\,\micron. The adopted extinction curve is also tabulated in the Appendix\,\ref{sec:extinction} for reference. 

One caveat to note is that the inferred gas column density that we derive later from the X-ray analysis ($N_{\rm H}$\,$\approx$\,7\,$\times$\,10$^{21}$\,cm$^{-2}$; see \S\,\ref{sec:sedresults}) suggests somewhat smaller optical extinction of \av\,$\approx$\,2.4\,mag using recent gas-to-dust absorption-to-extinction conversion relations \citep[e.g.,][]{bahramian15, foight16}. This would not reduce the accretion disc reprocessing inferred in the optical, but would not impact the MIR corrections significantly. 

\section{Results}
\label{sec:results}

\subsection{MIRI spectral-timing}
\label{sec:miriresults}

\begin{figure}
    \centering
    \includegraphics[angle=90,width=\columnwidth]{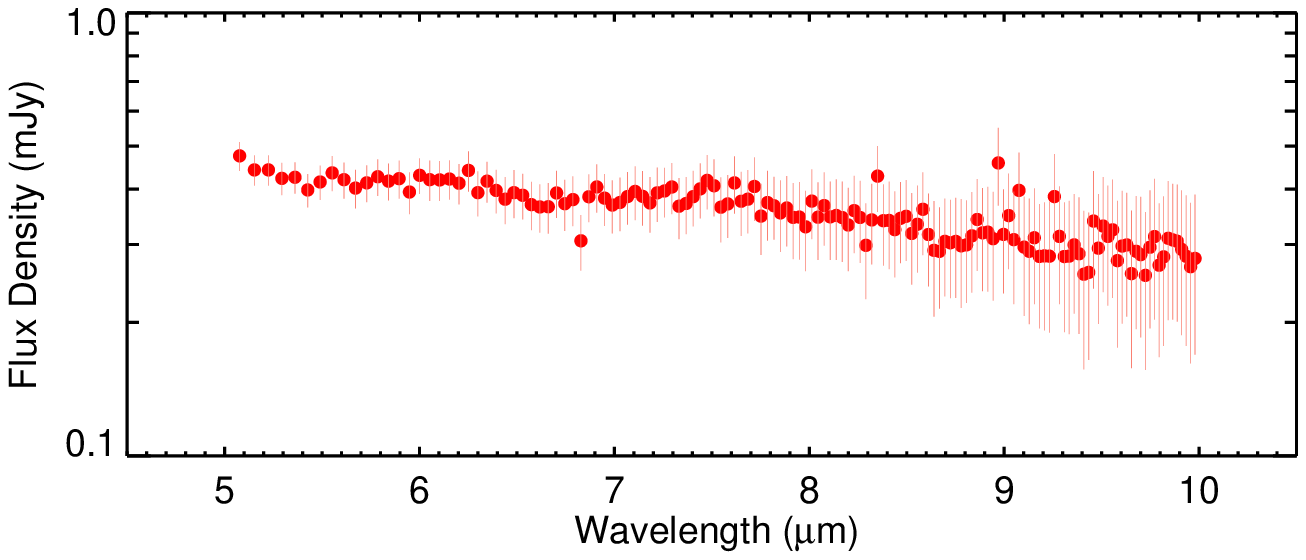}
    \vspace*{-0.cm}   
    \caption{Observed MIRI low-resolution spectrum of \gx\ during the 2024 March soft state. The points denote the median across all integrations, and error bars denote the standard deviation incorporating background and instrumental errors together with intrinsic source variability. 
    }
    \label{fig:medspec}
\end{figure}

\gx\ is detected with MIRI at flux densities spanning 0.5--0.3\,mJy across the wavelength range of 5--10\,\micron. The median spectrum, shown in Fig.\,\ref{fig:medspec}, can be characterised by a simple power-law with exponent $\alpha_{\rm MIRI}^{\rm obs}$\,=\,+0.39\,\p\,0.07 (with flux density $F_{\rm \nu} \propto \nu^{\alpha_{\rm MIRI}^{\rm obs}}$). The spectrum is featureless to within uncertainties, with no significant emission or absorption features. By artificially injecting emission lines, we estimate a line flux upper-limit of $\ltsim$\,4\,$\times$\,10$^{-16}$\,erg\,s$^{-1}$\,cm$^{-2}$ near $\lambda$\,=\,7.5\,\micron, the wavelength of the hydrogen Pfund(6--5) emission line -- one of the strongest recombination emission lines expected in the MIRI range. 

\begin{figure}
    \centering
    \includegraphics[width=1.01\columnwidth]{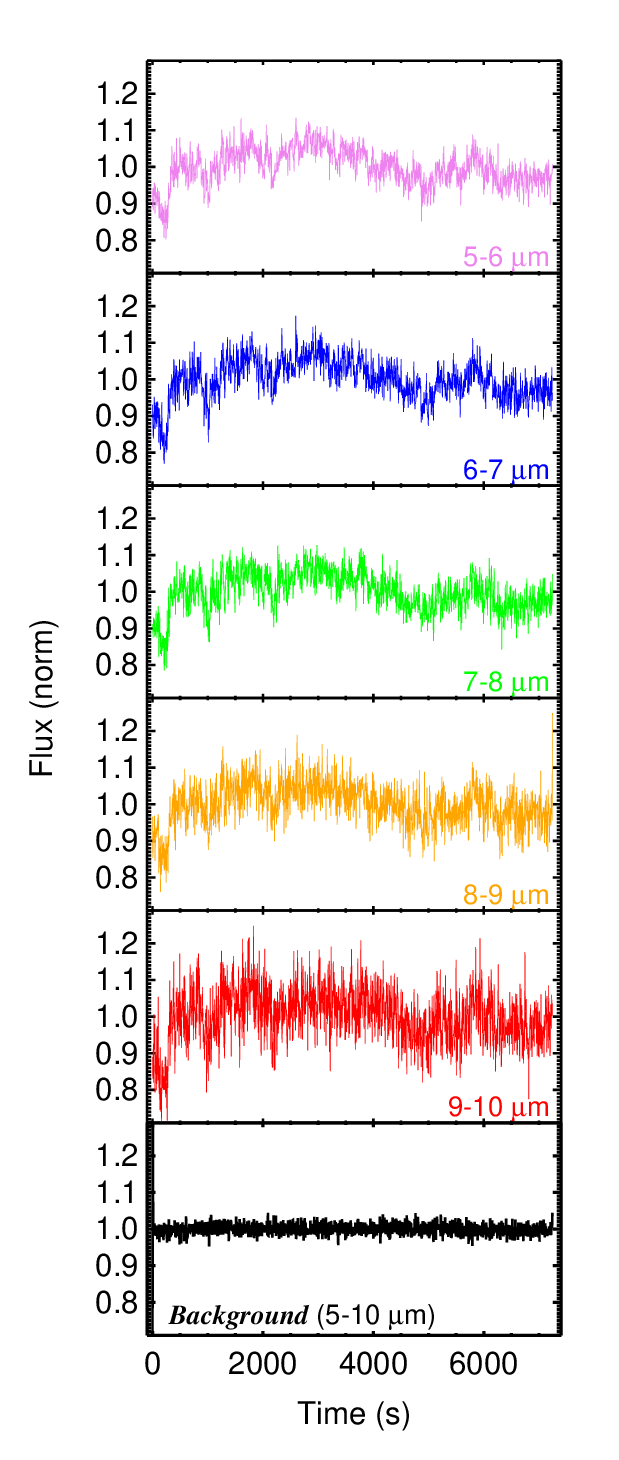}
    \vspace*{-0.5cm}
    \caption{Multi-wavelength time-series across all 1,400 integrations, split into annotated wavelength bins and normalised to the mean in each case. The bottom panel shows the background level with the same normalisation for the full (5--10\,\micron) band. 
    }
    \label{fig:lc}
\end{figure}

\begin{figure}
    \centering
        \includegraphics[angle=0,width=1.01\columnwidth]{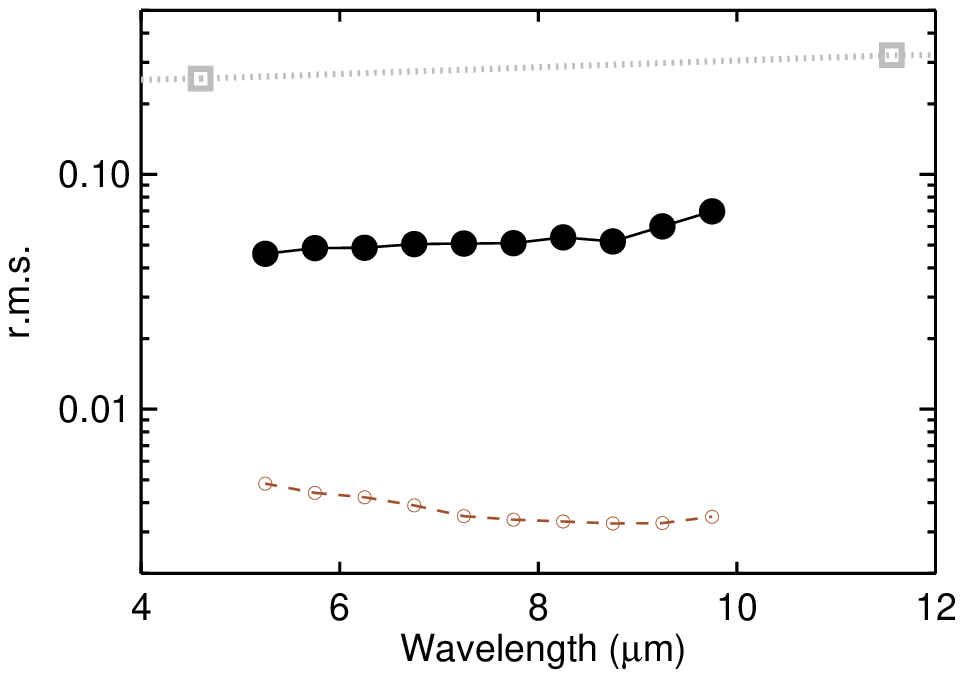}
        \hfill
        \includegraphics[angle=0,width=1.01\columnwidth]{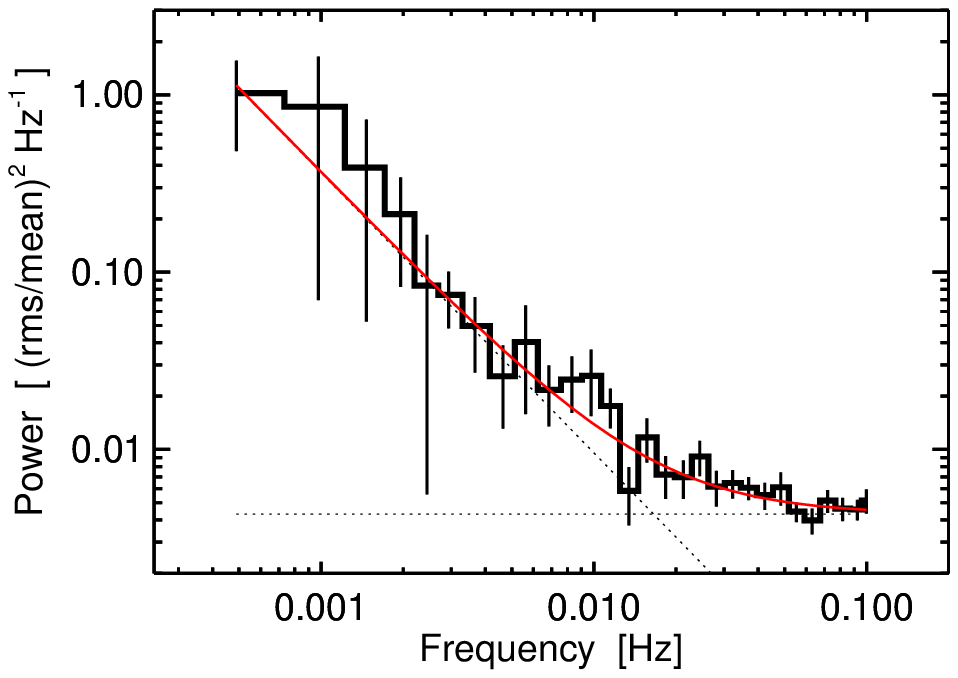}
        \hfill
        \includegraphics[angle=0,width=1.01\columnwidth]{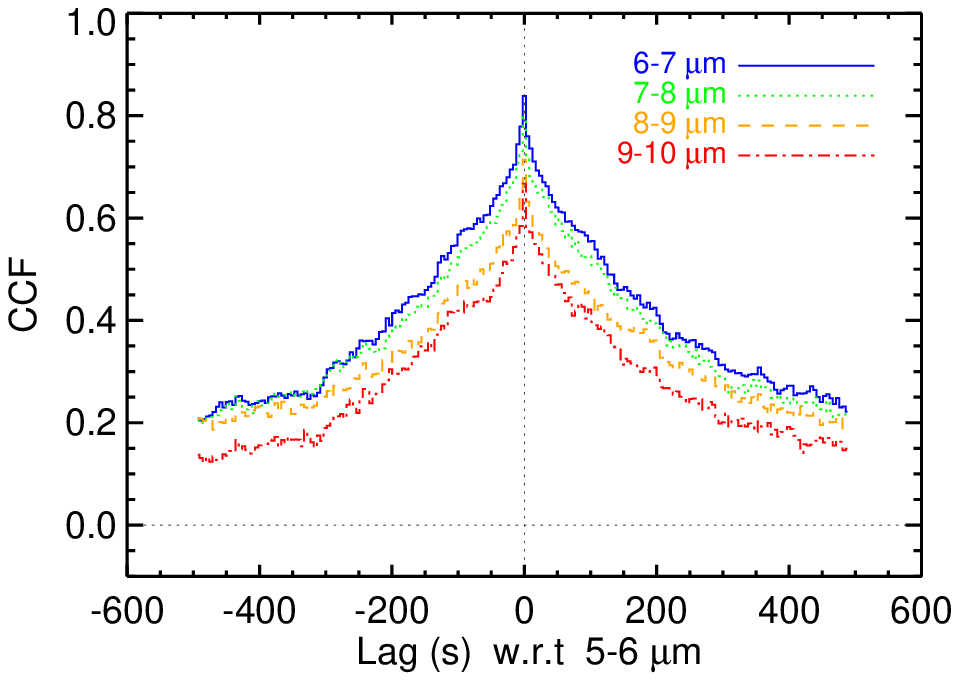}
    \caption{MIRI spectral-timing properties. {\em (a)} The excess fractional variability r.m.s. of the MIRI data as a function of wavelength (filled black circles), compared to that of the background (empty brown circles with dashed curve) and that of \wise\ hard state observations from 2010 (gray empty squares with dotted lines; \citealt{Gandhi-2011}). {\em (b)} The power spectral density (PSD) of the white-light time series, with a power-law (slope\,=\,--1.58\,\p\,0.14) and constant fit overplotted. {\em (c)} Cross-correlation functions (CCFs) between the respective annotated light curves and the shortest wavelength 5--6\,\micron\ light curve (used as a baseline reference). The excellent match between the wavelengths is apparent.}
    \label{fig:miritiming}
\end{figure}

Extracting light curves, the source is found to be significantly variable. Fig.\,\ref{fig:lc} shows this, with the time-series photometry split by wavelength. The observed stochastic variations are qualitatively different to those seen during the hard accretion states of the source at other wavelengths -- the overall variability is damped w.r.t. the hard state, and there is an absence of obvious strong flares and characteristic quasi-periodicities \citep[cf., ][ for comparison]{Gandhi-2010}. The strength of time-series variability can be characterised by the excess r.m.s. of variability amplitude above statistical noise (cf. \citealt{vaughan03}). 
Utilising this formalism for different portions of the MIRI spectrum, we find a mean excess r.m.s. value of 0.04--0.05 (4--5\,\%) across most of the spectral range, as shown in Fig.\,\ref{fig:miritiming}. Bootstrapping the full light curve yields a mean and uncertainty of r.m.s.\,=\,4.7\,(\p\,0.1)\,$\times$\,10$^{-2}$. Though there appears to be a slight rise towards the red end, we caution that systematic uncertainties in the MIRI pipeline may be underestimating the long-wavelength noise, so this should be treated with caution (see more extensive discussion and Appendix in \citealt{gandhi25}). Irrespective, the figure shows that the r.m.s. is well above that measured in the background. Also overplotted are the r.m.s. measurements from the bright 2010 hard state as observed by \wise\ in the $W2$\,(4.6\,\micron) and $W3$\,(12\,\micron) filters \citep{Gandhi-2011}.\footnote{Note that \wise\ sampling was far more sparse than MIRI, comprising 13 epochs over a period of $\approx$\,24\,h.}

Stochastic variability can be characterised by its Fourier power spectral density (PSD), shown in panel (b) of the same figure for the white-light data as being approximately representative across wavelengths. At high frequencies, the PSD flattens to a constant value, characteristic of white-noise. We fitted the PSD with a combination of a simple power-law--plus--constant as a function of Fourier frequency $f$, finding an acceptable fit with slope $\beta$\,=\,--1.58\,\p\,0.14 (Power $P_f$\,$\propto$\,$f^\beta$). To our knowledge, this is the first reported PSD of a transient XRB in the MIR, so comparisons with previous measurements do not exist. In the NIR and red optical filters, however, the hard state PSDs are generally flatter, indicative of fast flickering activity \citep{Gandhi-2010, Casella-2010, vincentelli19, paice19} which is absent here. 

The light curves in Fig.\,\ref{fig:lc} show excellent correspondence across all wavelengths. This is confirmed through a cross-correlation function (CCF) analysis, which correlates two light curves as a function of time lag. Panel (c) in Fig.\,\ref{fig:miritiming} shows high CCF peak values at zero lag, demonstrating the close match between bands with no significant delay. The longer wavelength CCFs are marginally asymmetric, showing a skew to long delays of $\approx$\,500\,s, but this is too subtle to be robustly interpreted without more dedicated analysis, especially given the aforementioned systematic uncertainties and rising statistical noise issues towards the red.  

\subsection{Broadband Spectral Energy Distribution}
\label{sec:sedresults}

\begin{figure*}
    \centering
    \includegraphics[angle=90,width=0.8\textwidth]{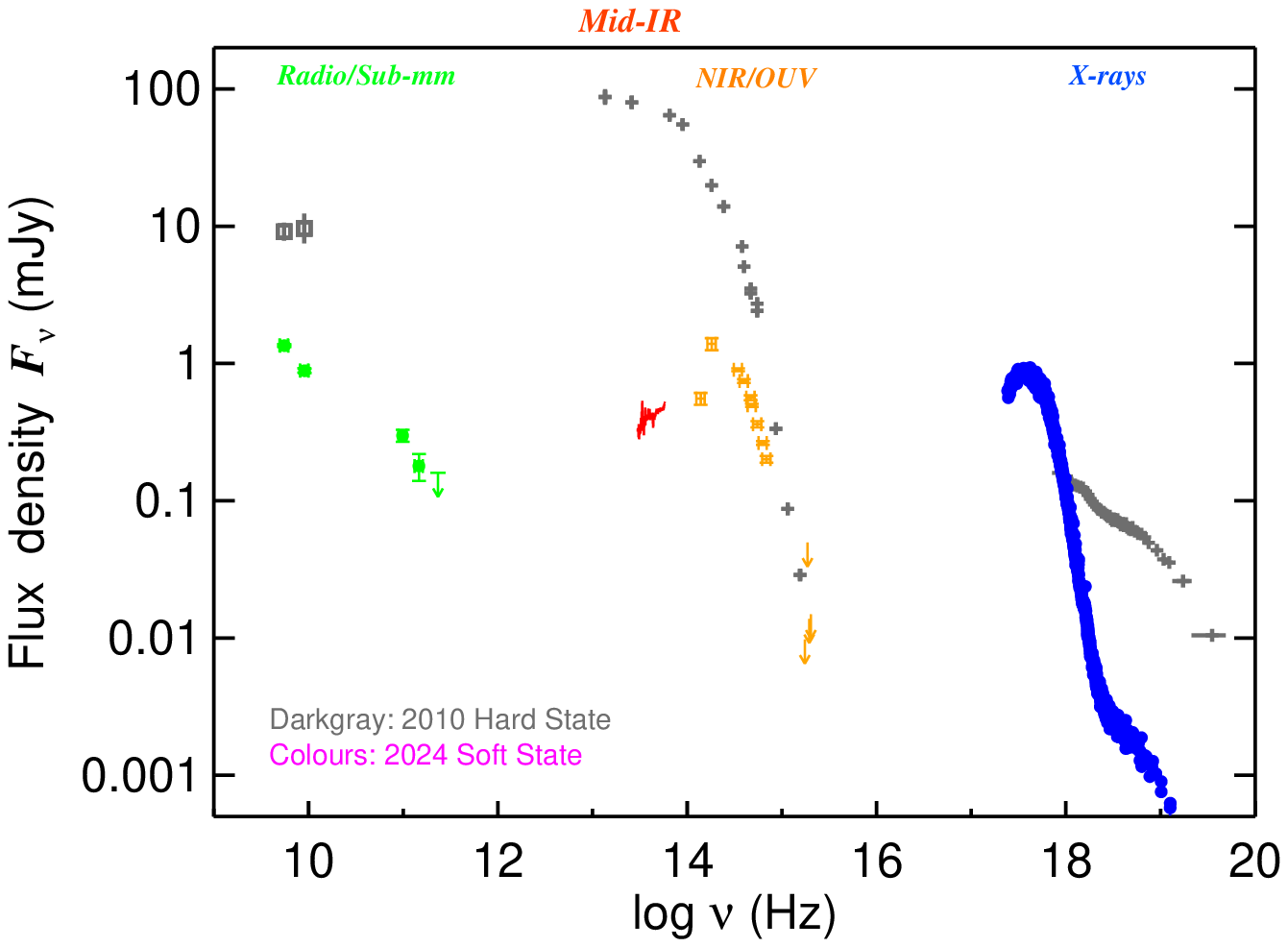}
    \vspace*{-0.cm}    
    \caption{Observed broadband SED during the 2024 soft state in flux density units, compared to the hard state 2010 SED in gray. These are the observed values, not corrected for absorption. X-ray detector spectral responses have been folded out and instrumental cross-calibration removed for display. The instruments used are described in \S\,\ref{sec:obs} and in Appendix\,\ref{sec:coordination}. 
    }
    \label{fig:wisecomparison}
\end{figure*}

\begin{figure*}
    \centering
    \includegraphics[angle=90,width=0.8\textwidth]{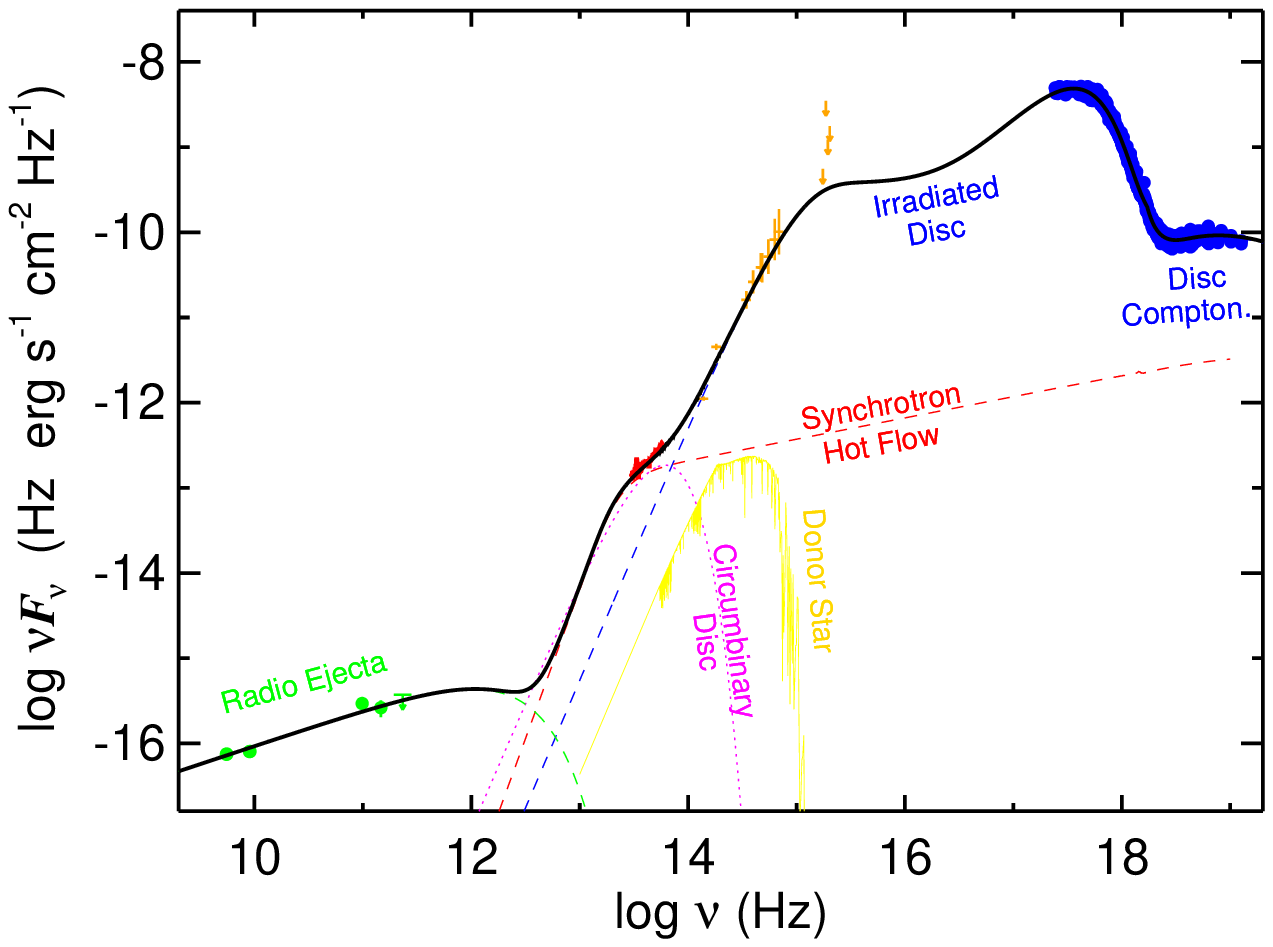}
    \vspace*{-0.cm}
    \caption{Broad-band SED of \gx\ in $\nu F_\nu$ units. The individual spectral components are a viscous inner accretion disc and Comptonisation in X-rays, the irradiated outer disc in the optical-to-near-IR, a synchrotron-emitting hot flow dominating the MIR, and a cutoff power-law to approximate the radio-to-sub-mm. In order to show the intrinsic spectral shape, absorption corrections have been applied to the data and model components where relevant. The thick black curve denotes the sum of individual spectral components. A blackbody simulating a circumbinary disc is shown as the magenta dotted curve, depicting an alternative to the synchrotron MIR component. Finally, a K2III spectral template ($T$\,=\,4500\,K, log\,$g$\,=\,2.0) is shown in yellow, normalised to the (dereddened) $H$-band flux density quoted in \citet{heida17}. 
    }
    \label{fig:sedfit}
\end{figure*}

The observed broadband spectral energy distribution (SED) is shown in Fig.\,\ref{fig:wisecomparison}. Our collated data spans approximately nine orders of magnitude of the electromagnetic spectrum from radio to X-rays, comprising a rich multiwavelength soft-state coverage. The intrinsic continuum is dimmed by reddening and obscuration from the MIR to X-rays. 
When dereddening the optical to near-IR (ONIR) range using the prescription from \S\,\ref{sec:reddening}, the ONIR slope approximates a simple power-law with $\alpha_{\rm ONIR}$\,=\,+1.80\,\p\,0.15, consistent with the long-wavelength tail of optically-thick emission (Figure~\ref{fig:sedfit}). Dereddening corrections have only a small impact on the MIRI band, resulting in a corrected slope $\alpha_{\rm MIRI}$\,=\,+0.36\,\p\,0.07. By contrast, the radio--to--sub-mm (RS) power-law is optically-thin, with exponent $\alpha_{\rm RS}$\,=\,--0.57\,\p\,0.03.\\

These observations occurred during an X-ray bright soft state, when disc emission is expected to be prominent. \citet{mastroserio25} have presented such a fit incorporating a Kerr disc, thermal Comptonisation and other narrow features to model the X-ray continuum. Our modelling is qualitatively similar to theirs, except that we are here interested in covering the full SED including energies much lower than X-rays alone. We thus attempted a {\tt diskir} model fit, which combines a standard multi-colour disc \citep{mitsuda84} with a Comptonised high-energy tail that irradiates both the inner and outer disc regions. The irradiated outer disc reprocesses this emission to the OIR, in a manner consistent with its local temperature and the reprocessing fraction \citep{diskir}. 
This combination allows joint modelling of the X-rays and the ONIR regime. The primary fits were conducted in the {\tt xspec} package \citep{xspec}, using Levenberg-Marquardt minimisation, and quoted confidence ranges for the SED fit are 90\,\% intervals corresponding to $\Delta$\,$\chi^2$\,=\,2.71 for 1 parameter of interest. The fraction $f_{\rm out}$ of the inner disc radiation which is reprocessed into the OIR is $f_{\rm out}$\,=\,7.9$_{-5.1}^{+15.8}$\,$\times$\,10$^{-3}$. The outer disc radius ranges over sizes of 10$^{4.4-4.6}$ times the inner disc radius. Comptonisation of disc photons to higher energies accounts for the weak high-energy X-ray tail seen in \nustar. 

The MIRI spectral slope is not consistent with such an irradiated disc, and we will discuss its origin in the following section, along with full model parameters for the combined multiwavelength fit. Here, we simply note that the secondary star is proposed to be of spectral type K2III \citep{heida17}, with a peak in the ONIR regime, and with only a minor flux contribution to the MIRI band of at most a few percent. 

\subsection{Multiwavelength Variability}
\label{sec:simultaneous}

\begin{figure}
    \centering
    \hspace*{-1cm}
    \includegraphics[width=1.15\columnwidth]{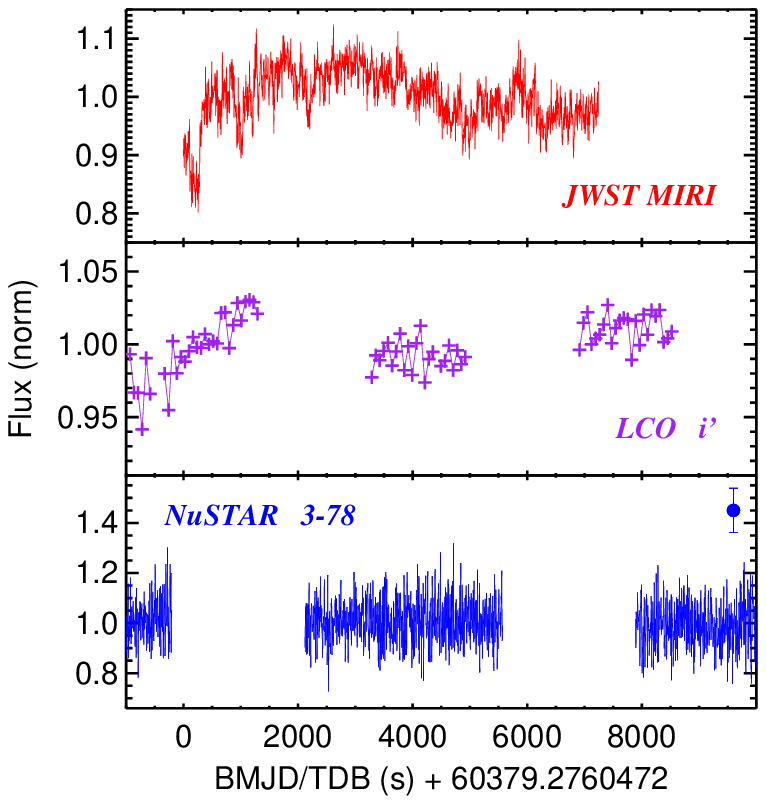}
    \vspace*{-0.5cm}
    \caption{Strictly simultaneous overlap between the time series datasets gathered with \jwst\ MIRI (5--10\,\micron), LCO ($i'$ band), and \nustar\ FPMA+B (3--78\,keV). Note the different y-axis ranges used across the panels, in order to highlight the variability. Statistical uncertainties are important only for the \nustar\ dataset in the bottom panel, where a representative error bar is overplotted in the top corner.}
    \label{fig:simultaneous_mwl_lcs}
\end{figure}

At least two of the datasets with time series measurements in our campaign were able to be coordinated strictly simultaneously with the MIRI observation: \nustar\ and LCO in the $i'$ band. The respective light curve segments spanning the \jwst\ window are shown in Fig.\,\ref{fig:simultaneous_mwl_lcs}. X-ray variability is typically suppressed during the soft states of XRB outbursts, and we found an upper-limit of only 0.02 to the fractional r.m.s. of the \nustar\ lightcurve. 

By contrast, the $i'$ band shows significant fluctuations, with a qualitatively similar trend to that seen in MIRI: there is a rise near the beginning of the observing window, followed by a fall in the mean level by a few per cent and then a slight rise towards the end again. The $i'$ fractional r.m.s. is measured to be 2.0\,(\p\,0.5)\,$\times$\,10$^{-2}$.  

In order to quantify the relationship between the LCO and MIRI bands, we cross-correlate the two lightcurves. Specifically, we utilise discrete correlation function (DCF) analysis, which measures the degree of linear correlation while allowing for non-uniform time sampling \citep{dcf}. Both sets of light curves were transformed to the common Solar System Barycentric Dynamical Time (TDB) frame before cross-correlating. We note that LCO timing is GPS-controlled, and the absolute timing accuracy of \jwst\ has been validated to be within the demands of lag measurement relevant here \citep{shaw25}. 

\begin{figure}
    \centering
    \includegraphics[angle=90,width=0.5\textwidth]{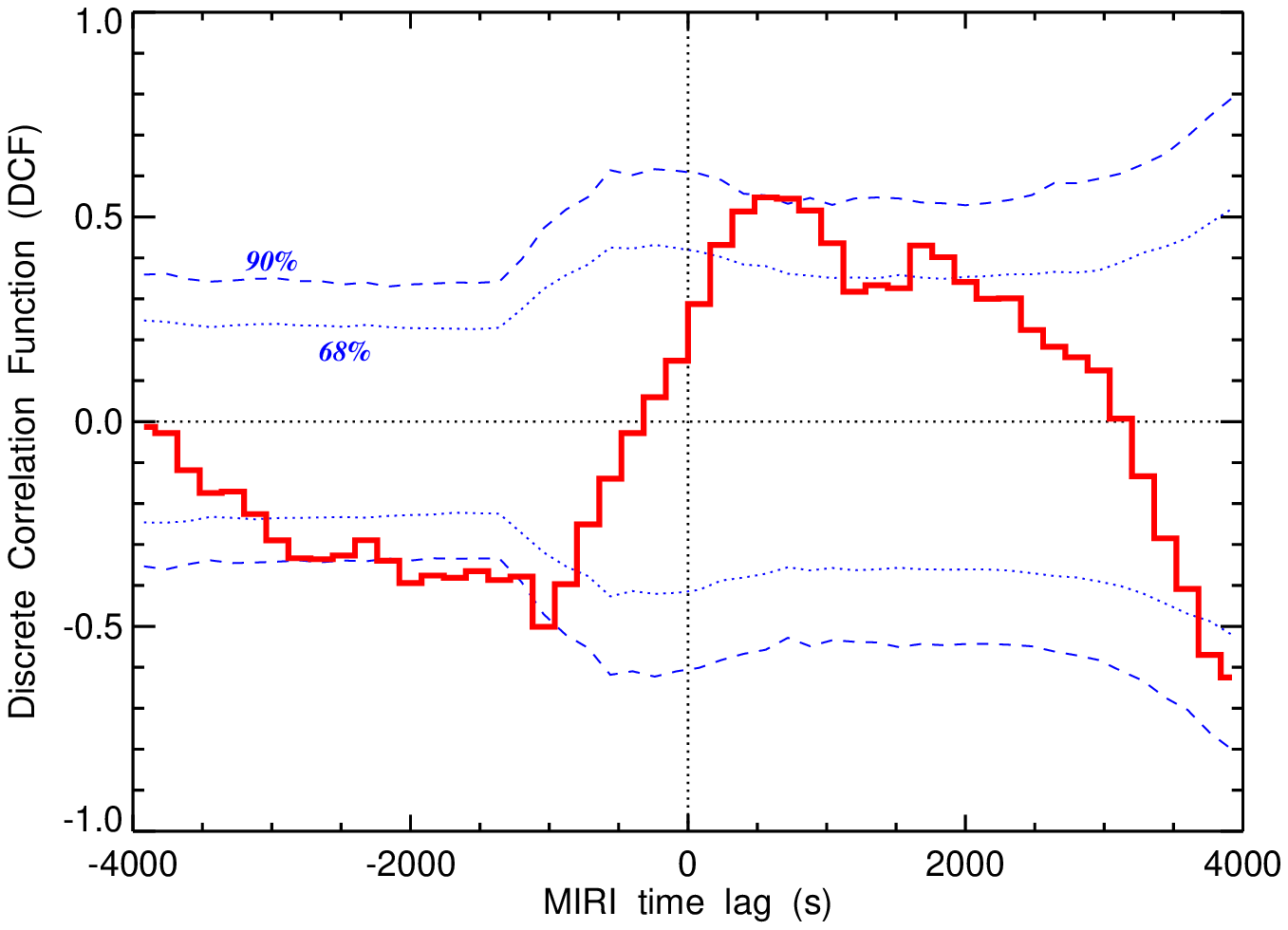}
    \caption{Discrete Correlation Function (DCF) between MIRI and LCO $i'$. Positive lags denote a time delay of MIRI w.r.t the optical. The blue dotted and dashed contours denote 68\% and 90\% confidence intervals, respectively, based on 5,000 simulated DCFs and power-law noise lightcurves.}
    \label{fig:dcf}
\end{figure}

Fig.\,\ref{fig:dcf} shows the DCF result. The peak lag at a few hundred seconds indicates the  presence of a time delay with the MIR lagging behind $i'$ by $\approx$\,750$_{-350}^{+1100}$\,s, and uncertainties based on bootstrap resampling. As will be discussed in the following sections, this lag is too long to be easily explained by any of the physical scenarios we believe to be relevant in this system. Furthermore, testing for lags using different segments of the data does not yield consistent results on the lag estimate. 

The sparse time sampling, particularly in the optical, raises the possibility of a spurious signal arising from chance coincidence amidst stochastic variability. In order to test this hypothesis, we conducted simulation tests of the DCF significance by using randomised light curves. An ensemble of 5,000 light curves were created with the red-noise generation algorithm of \citet{timmerkoenig} and the MIRI periodogram from \S\,\ref{sec:results} as a baseline. These light curves match the {\em statistical} properties of the MIRI data while being completely uncorrelated with our observed data. Thus, the mean DCFs of these randomised MIRI light curves with the LCO data should result in zero mean cross-correlation, and their scatter will be informative of the significance of any particular DCF value. Fig.\,\ref{fig:dcf} shows the result of this test as the dotted and dashed contours, representing the 68\% and 90\% confidence scatter amongst our 5,000 random DCFs. This test demonstrates that DCF values as large as the peak seen in our data {\em can} arise by coincidence, with our confidence limited to no more than $\approx$\,90\%. Thus, any inference of a lag here should be treated with caution.

Finally, we note that there are two additional simultaneous data sets which require some dedicated analysis and will be presented in follow-up works. The {\em NICER} camera aboard the International Space Station \citep{nicer14} observed the source at the same time as MIRI (Obs ID 7702010121), but the observation was conducted during orbit {\em daytime}, and thus requires custom processing to ensure appropriate screening and analysis. In addition, there are HAWK-I NIR {\em timing} data which also require dedicated analysis. Both of these will be presented in a follow-up work focused on the multiwavelength timing properties of the source. 

\section{Discussion}
\label{sec:discussion}

The unabsorbed bolometric luminosity of \gx\ during the observed state is dominated by the {\tt diskir} model, \lbol\,=\,1.2\,$\times$\,10$^{38}$\,$(d/{\rm 8\,kpc})^2$\,erg\,s$^{-1}$.  The corresponding Eddington fraction is \lbol/\ledd\,=\,0.15\,$(d/{\rm 8\,kpc})^2$\,$(M_{\rm BH}/{6\,{\rm M}_{\odot}})^{-1}$. These estimates are nearly identical to the source power during the \wise\ observations of the 2010 hard state for the same assumed $M_{\rm BH}$, though the spectrum peaked in higher-energy X-rays at that time (\citealt{Gandhi-2011}).  
The 2010 broadband SED is overplotted in Fig.\,\ref{fig:wisecomparison} in gray. The MIR band fluxes are separated by $\approx$\,2.5\,dex between the two campaigns, a difference larger than in any other portion of the spectrum. Radio fluxes are now lower by at least 1 dex, with luminosity $L_{\rm Radio-Sub\,mm}^{\rm 10^{9}-10^{12}\,Hz }$\,=\,1.1\,$\times$\,10$^{31}$\,erg\,s$^{-1}$, or $\nu L_{\rm \nu}$\,(5\,GHz)\,=\,5.7\,$\times$\,10$^{29}$\,erg\,s$^{-1}$. 
Strong quenching of the jet in the soft state can easily account for these changes. Pronounced soft state jet decrements have previously been reported in the radio for 4U\,1957--11 \citep[][up to 1500--3300$\times$]{maccarone20}, MAXI\,J1535--571 \citep[][]{russell20}, and Swift\,J1753.5--0127 \citep{rushton16}, amongst others.

What, then, is the origin of the mid-IR continuum observed by MIRI? There was no detectable compact jet activity during our campaign, with the radio and sub-mm detections likely reflecting residual ejecta emission from earlier in the outburst (Russell et al. in prep). 
The MIR slope $\alpha_{\rm MIRI}$ matches neither $\alpha_{\rm ONIR}$ nor $\alpha_{\rm RS}$, so the MIRI continuum cannot be connected seamlessly at either end without invoking the presence of a spectral break. So the origin of the MIR emission cannot be the compact jet, as in the hard state \cite[e.g., ][]{Gandhi-2011}. 
We consider several possible physical scenarios below.

\subsection{Circumbinary Disc?}
\label{sec:cbd}

Thermal emission from dust with characteristic temperatures of a few hundred Kelvin contributes copiously to the infrared emission of active galaxies (AGN). 
The possibility of dust emission peaking in MIR XRB observations has previously been raised, where it has been attributed to a putative circumbinary disc \citep[CBD; ][]{muno6}. Such discs may be the result of fallback of ejecta from the progenitor supernova, or owe their origin to outflows as part of the accretion--feedback cycle of the compact object. Sensitive observations such as ours, which are not strongly impacted by bright non-thermal jet emission, provide an excellent opportunity to test this scenario. 

We included a single-temperature blackbody model in our broadband SED fit to test for any putative CBD component. Both its temperature and normalisation were left free to vary. This was fitted to the broadband data in conjunction with the absorbed {\em diskir} component, using \xspec\ as before. A simple cutoff power-law was additionally included to account for the radio--sub-mm regime, though this does not significantly impact on the infrared fit. The result is shown in Fig.\,\ref{fig:sedfit} and the complete set of fitted model parameters are listed in Table\,\ref{tab:fit}, and the model components are shown on the broadband SED in Fig.\,\ref{fig:sedfit}. 

\begin{table}
  \begin{center}
  \caption{Results of broadband spectral fitting with an absorbed, irradiated and Comptonised disc, a thermally emitting circumbinary disc and radio-emitting jet ejecta power-law\label{tab:fit}}
  \begin{tabular}{lrlr}
\hline
Component          &  \multicolumn{2}{c}{Parameter}         & Units \\
\hline
\hline
\multicolumn{4}{c}{}\\
\multicolumn{4}{c}{\em \underline{Nuclear Absorption}}\\
{\tt tbabs}  & $N_{\rm H}$             & $6.8\,(\pm\,{0.2})$\,$\times$\,10$^{21}$   &  cm$^{-2}$\\ 
\multicolumn{4}{c}{}\\
\multicolumn{4}{c}{\em \underline{Irradiated and Comptonised Disc}}\\
{\tt diskir} & $kT_{\rm disk}$        & 0.63\,$\pm$\,0.01 & keV\\
             & $\Gamma$           & 2.03$_{-0.04}^{+0.05}$  & \\
             & $kT_{\rm e}$           & 300$^f$  & keV \\
             & $L_{\rm covr}/L_{\rm d}$         & 4.1$_{-0.2}^{+0.1}$\,$\times$\,10$^{-2}$ & \\
             & $f_{\rm in}$             & 0.28$_{-0.03}^{+0.02}$ & \\
             & $r_{\rm irr}$            & 1.000$_{-0.001}^{+0.002}$ & \\
             & $f_{\rm out}$            & 2$_{-1}^{+3}$\,$\times$\,10$^{-2}$ & \\
             & log ${r_{\rm out}}$         & 4.49$_{-0.07}^{+0.06}$ & \\
             & norm            & 2.7$_{-0.2}^{+0.1}$\,$\times$\,10$^{3}$ & \\
{\tt smedge} & $E_{\rm edge}$  & $12.5_{-0.1}^{+0.2}$ &  keV \\
             & Width $W$       & $7.9_{-1.2}^{+2.0}$ & keV\\
{\tt gauss}  & $E$  & 6.56$_{-0.08}^{+0.07}$ & keV \\
             & $\sigma$        & 0.18$_{-0.09}^{+0.11}$ & keV\\
             & norm            & $1.9_{-0.6}^{+1.2}$\,$\times$\,10$^{-4}$ & ph\,cm$^{-2}$\,s$^{-1}$\\
\multicolumn{4}{c}{}\\
\multicolumn{4}{c}{\em \underline{Circumbinary Disc}}\\
{\tt bbody}  & $kT_{\rm dust}$  & 6.5\,(\p\,0.4)\,$\times$\,10$^{-5}$ &  keV \\
             & norm       & 3.0\,($\pm$\,0.3)\,$\times$\,10$^{-6}$ & 10$^{36}$\,erg\,s$^{-1}$\,kpc$^{-2}$\\
\multicolumn{4}{c}{}\\
\multicolumn{4}{c}{\em \underline{Jet Power-law}}\\
{\tt cutoffpl}& $\Gamma$  & $1.58_{-0.04}^{+0.03}$ &  \\
             & norm       & $8_{-3}^{+4}\times 10^{-5}$ & ph\,cm$^{-2}$\,s$^{-1}$\,keV$^{-1}$\\
\multicolumn{4}{c}{}\\
\multicolumn{4}{c}{\em \underline{X-ray Cross-normalisation Constants}}\\
$C_{\rm FPMB}^{\rm FPMA}$  & {\sc const}   & 0.97\,\p\,0.01     & \\
$C_{\rm SXT}^{\rm FPMA}$  & {\sc const}   & $3.97_{-0.05}^{+0.04}$ & \\
{\sc Gain (SXT)}$^g$ & {slope} & 1.0$^f$ & \\
 & {offset} & +66.2\,\p\,0.5 & eV\\
$C_{\rm LAXPC}^{\rm FPMA}$  & {\sc const}   & 0.83\,\p\,0.02 & \\
$C_{\rm XRT}^{\rm FPMA}$  & {\sc const}   & $1.28_{-0.02}^{+0.01}$ & \\
                   &                  & & \\
$\chi^2$/dof       &                  & 1650.9/1659            & \\
\hline
\end{tabular}~\par
The full model in \xspec\ notation is\,\,\,  {\sc const}\, ({\tt tbabs}\, ({\tt smedge}\, ({\tt diskir} + {\tt gauss}) + {\tt bbody} + {\tt cutoffpl} ) ).\\
$^u$unconstrained. $^f$fixed.\\
$^g$Additional gain offset allowance for {\em AstroSat}-SXT as a free variable.
\end{center}
\end{table}

An acceptable fit with $T_{\rm CBD}$\,=\,748$_{-46}^{+50}$\,K was found. The blackbody flux is 2.5\,$\times$\,10$^{-13}$\,erg\,s$^{-1}$\,cm$^{-2}$, or $L_{\rm CBD}$\,=\,1.9\,$\times$\,10$^{33} (d/{\rm 8 kpc})^2$\,erg\,s$^{-1}$, which implies an emitting radius $r_{\rm CBD}$\,=\,3\,$\times$\,10$^{12}$\,$\sqrt{(4\pi/\Omega)}$\,cm, with $\Omega$ being the effective solid angle of the emitting CBD surface. For a binary separation $a$\,$\approx$\,12\,\Rsun\, this radius corresponds to 3.7\,$\times$\,$a$ at minimum. CBDs are subject to tidal disruption which, for a circularised binary with separation $a$, occurs at a inner radius $r_{\rm CBD}$\,$\approx$\,1.7\,$a$ \citep{artymowicz94}, so the inferred radius is not in conflict with the tidal disruption radius. Stated systematic uncertainties on the binary separation are approximately 10\,\% \citep{zdziarski19}.

At face value, a CBD is an attractive interpretation of the MIRI spectrum. CBDs may encourage accretion on to the central engine \citep{muno6}\footnote{though the devil lies in the details of the source configuration and interactions \citep[cf., ][]{valli24}.}, so the presence of a CBD could naturally explain the frequent outbursting behaviour of \gx. 
This cannot, however, easily explain the observed multiwavelength variability. The dynamical timescale of a CBD can be written as 

\begin{eqnarray}
t_{\mathrm{dyn}} & = & 2\pi\, \sqrt{\frac{r_{\mathrm{CBD}}^3}{G M}}\\
                 & \ge & 12.3\, \left(\frac{r_{\rm CBD}}{3\times10^{12}\,{\rm cm}}\right)^{3/2}\, \left(\frac{7.1\,{\rm M}_\odot}{M_{\rm tot}}\right)^{1/2}\,\,{\rm days.}\nonumber
\end{eqnarray}

\noindent
Any viscous or other timescales are expected to be longer still, by factors of $\alpha_{\rm eff}^{-1}$, the effective viscosity parameter. We also note that small solid angles $\Omega$, as expected for CBDs, could easily push $r_{\rm CBD}$ to be much larger still. Thus, we do not expect substantial CBD variability on timespans of a few thousand seconds, as observed in our MIRI light curves. Finally, a combination of an irradiated outer disc and a CBD (say) would be expected to produce a larger $i'$ variability r.m.s than in MIRI -- this is opposite to what we observe.

\subsection{Wind reprocessing?}
\label{sec:wind}

The presence of mid-infrared winds has been inferred in two other black hole XRBs observed by \jwst. \grs\ shows evidence of a prolific, albeit transient, failed wind in its `X-ray obscured' state \citep{gandhi25}. In A\,0620--00, the presence of a wind has been inferred from the detection of likely bremsstrahlung emission in its quiescent state at an accretion rate which is orders-of-magnitude lower \citep{zuo25}. 

Equatorial {\rm X-ray} winds are typical of soft-state outbursts in XRBs \citep{ponti12}, and \gx\ itself is also known to host a wind in the soft state as inferred from X-ray \citep{miller04} and  optical spectroscopy of the source 
\citep[][building on prior works including \citealt{soria99, wu01, rahoui14}]{ambrifi25}. The characteristics of the MIRI spectrum are, however, not easily ascribable to a wind. Optically-thin bremsstrahlung should produce a nearly flat spectral continuum, as seen in \grs. For an optically-thick constant mass outflow rate, instead, the frequency dependence scales as $v^{2/3}$ \citep{wrightbarlow75}. Our inferred MIRI slope of $\alpha_{\rm MIRI}$\,=\,+0.36\,$\pm$\,0.07 fits neither of these. Accounting for contaminating emission from the donor star or accretion disc would not help to alleviate this discrepancy. Furthermore, bremsstrahlung would over-predict the observed radio power, unless there is an additional spectral break between the ALMA and \jwst\ band. 

A wind could also provide an explanation for the long infrared lag relative to the optical, {\em if} this lag is real (cf. discussion and simulation tests in \S\,\ref{sec:simultaneous}). This is because the recombination timescale ($t_{\rm rec}$) within a gaseous windy medium can be estimated as 

\begin{equation}
    t_{\rm rec} \sim 800\,\,\alpha_{\rm B}^{-1}\, \left(\frac{n_{\rm H}}{5\times10^{9}\,{\rm cm^{-3}}}\right)^{-1}\,\,\,{\rm s.}
\end{equation}

\noindent
where $\alpha_{\rm B}$ is the recombination coefficient with a value  $\approx$\,2.5\,$\times$\,10$^{-13}$\,cm$^3$\,s$^{-1}$ for temperatures typical of recombining winds \citep{osterbrock}, and $n_{\rm H}$ is the gas density in the wind medium.

However, we do not see evidence of recombination emission lines that may be expected from such a wind. In \S\,\ref{sec:results}, a flux limit of $F_{\rm Pf\alpha}^{\rm lim}$\,$\ltsim$\,4\,$\times$\,10$^{-16}$\,erg\,s$^{-1}$\,cm$^{-2}$ was placed on the presence of H(6--5) Pf\,$\alpha$ -- amongst the brightest emission lines in the MIRI regime. Using the \ha\ equivalent width $W$\,=\,10.5\,\AA\ reported by \citeauthor{ambrifi25} during the 2021 soft state as a template together with our measured optical SED continuum, we predict a dereddened \ha\ emission line flux of $F_{\rm H\alpha}$\,$\approx$\,10$^{-13}$\,erg\,s$^{-1}$\,cm$^{-2}$ at the time of the MIRI campaign. This, in turn, predicts an H(6--5) flux at $\lambda$\,=\,7.5\,\micron\ of $F_{\rm Pf\alpha}^{\rm pred}$\,$\approx$\,2\,$\times$\,10$^{-15}$\,erg\,s$^{-1}$\,cm$^{-2}$ under standard Case B photoionisation conditions \citep{hummerstorey87}. Continuum uncertainties, including reddening corrections, are approximately a factor of 2. This predicted flux is about a factor of 5 larger than the actual detection limit, arguing against recombination. 

One way to circumvent these limitations and allow for a MIR wind is if the \ha\ line were primarily driven by another process (e.g. disc irradiation), with only a small proportion originating in the wind. Another way to `hide' lines from the wind would be to boost the gas density $n_{\rm H}$ by several orders-of-magnitude to make the wind optically thick, but this would push its mass-loss rates up unrealistically.\footnote{Related discussions on high mass-loss rates inferred from the MIRI observations of \grs\ can be found in \citet{gandhi25}.}

In summary, even if a wind does exist in the soft state of \gx, we do not see any smoking-gun signatures of it in the MIRI band. Future observations should target simultaneous high resolution ONIR spectroscopy in order to conduct direct searches for wind signatures. There are also two hydrogen recombination edges around 5.8\,\micron\ ($n$\,=\,8) and 7.4\,\micron\ ($n$\,=\,9) which could be searched for with higher S/N spectroscopy.

\subsection{Synchrotron from a hot flow overlying the disc?}
\label{sec:hotflow}

An alternative means to power the infrared emission is via synchrotron emission from the hot electron population in the inner accretion flow also responsible for producing the hard X-rays \citep[e.g., ][]{Veledina-2013}. The same hot flow has previously been considered to subsume the role of a jet base \citep{markoff05}, though we do not observe extended/ continuous jet emission typically associated with the hard state here. 

Such hot flow models are not generally available for fitting within \xspec. However, the essential predicted spectral shape of these models is a featureless and relatively flat continuum over a moderate range of wavelengths, breaking at low frequencies when the overall self-absorption optical depth becomes larger than unity. Such a shape can thus mimic spectral breaks inferred from other processes, such as the CBD component from the previous section, which can now be replaced by synchrotron with a self-absorption cutoff at low frequencies. At higher frequencies, this model shows a flat flux density up to some undetectable cooling break swamped underneath the accretion disc emission. Detailed predictions would rely on a far better understanding of the geometry of this emitting medium in the soft state, but toy models can be used to make a feasibility test. 

We start by assuming that the hard tail in the X-rays is from disc Comptonisation in the hot plasma, and the infrared emission arises from synchrotron generated by the same electrons. We can then equate the synchrotron power and that emerging from Comptonisation to the respective magnetic and seed photon energy densities $\frac{L_{\rm Sync}}{L_{\rm Compton}} = \frac{U_B}{U_{\rm ph}}$.  If we model this coronal medium as a disc with height $h$ and radius $r_{\rm C}$, then $U_{\rm ph}$ will be $\frac{L_{\rm disc}}{\pi r_{\rm C}^2\ c}$, where $L_{\rm disc}$ is the total luminosity of the seed disc photons, and $c$ is the speed of light. The value of $h$ cancels, because the volume is linearly proportional to $h$, as is the time photons spent in the corona. We can then solve this for the magnetic field and find:

\begin{equation}
    B=\sqrt{\frac{8\, (L_{\rm Sync}/L_{\rm Compton})\,L_{\rm disc}}{r_{\rm C}^2\ c}}.
    \label{eq:b_sync_compton}
\end{equation}

\noindent
A characteristic size scale of the corona may be estimated from Eq.\,1 of \citet{Veledina-2013}, for photons emitting optically-thin synchrotron at $\nu$\,$\sim$\,10$^{14}$\,Hz and an electron temperature of 300\,keV, as fixed for our broadband fit (Table\,\ref{tab:fit} for the high-energy power-law tail):

\begin{equation}
    r_{\rm C}^{\rm min} = 1.5\times10^8 \sqrt{\left(\frac{L_{\rm Sync}}{10^{33}\,{\rm erg\,s^{-1}}}\right)\,\left(\frac{\nu}{10^{14}\,{\rm Hz}}\right)^{-3}\,\left(\frac{kT_{\rm e}}{300\,{\rm keV}}\right)^{-1}}\,\,{\rm cm.}
\end{equation}

\noindent
Setting $r_{\rm C}$\,=\,$r_{\rm C}^{\rm min}$ and taking $L_{\rm Sync}/L_{\rm Compton}=10^{-3}$ (an estimate based on the ratio of the hard X-ray flux to the MIR flux), we find that $B\approx3.4\times10^4$ G. 

Here, we have assumed that the MIRI band probes synchrotron emission near the self-absorption (SSA) break, as Fig.\,\ref{fig:sedfit} indicates. An independent estimate of the $B$ field is possible from the position of this break, for which we take $\nu_{\rm SSA}$\,=\,10$^{14}$ Hz in the formulation presented by \citet{marscher83}:

{\footnotesize 
\begin{equation}
B_{\mathrm{SSA}} = 3.1 \times 10^4 \; \left(\frac{b}{3.5}\right) \;
\left( \frac{\nu_{\rm SSA}}{10^{14}\,\mathrm{Hz}} \right)^5
\left( \frac{F_{\nu,\,\rm SSA}}{0.7\,\mathrm{mJy}} \right)^{-2}
\left( \frac{R}{1.5\times10^8 \, \mathrm{cm}} \right)^4
\quad \mathrm{G}
\end{equation}
}
\noindent
where $b$ is a constant, taken to be 3.5 based upon a typical synchrotron optically-thin spectral slope $\alpha$\,=\,--0.7 (cf. Table\,1 of \citealt{marscher83}). This estimated $B$ field is encouragingly similar to that obtained from energy density arguments (Eq.\,\ref{eq:b_sync_compton}).

With our inferred size, we can also consider the variability of the system. The viscous timescale of a thin accretion disc, which is likely powering the corona, is given, from \citet{frankkingraine02}, by 
{\footnotesize 
\begin{equation}
    t_{\rm visc}=1.6\times10^3\,\left(\frac{\alpha}{0.3}\right)^{-4/5}\left(\frac{\dot{M}}{10^{18} {\rm g/s}}\right)^{-3/10}\left(\frac{M_{\rm BH}}{6\,{\rm M_{\rm \odot}}}\right)^{1/4}\left(\frac{r_{\rm C}}{10^8\,{\rm cm}}\right)^{5/4}\,\,{\rm s}.
\end{equation}}
\hspace*{-0.1cm}This timescale is of the same order as the characteristic timescale of MIRI variability. A model SED qualtitatively satisfying these criteria is presented in Fig.\,\ref{fig:sedfit}, with radii extending over $r$\,=\,(0.1--1.5)\,$\times$\,10$^8$\,cm ($\approx$\,10--170\,gravitational radii $R_{\rm G}$), $B$\,=\,3\,$\times$\,10$^4$\,G, with a uniform $B$ field, optical depth $\tau$\,$\propto$\,$r^{-2}$, and a luminosity matching that in the MIRI band. The plasma Lorentz factor ($\gamma$) distribution follows a power-law distribution expected from shock acceleration with $N(\gamma)$\,$\propto$\,$\gamma^{p}$ with $p$\,=\,--2.5, and a space density at the outer radius of the hot flow of $n_e$\,$\approx$\,10$^{15}$\,cm$^{-3}$. This model can thus self-consistently account for both the MIRI timing properties as well as the MIR spectrum. 

There are a few caveats to note. Firstly, the model requires the presence of an extended synchrotron emitting medium in the soft state. The same medium is presumably responsible for the Comptonised emission we see in the X-rays and can constitute an extended corona or disc `atmosphere' sandwiching the accretion disc, as supported by recent polarimetric observations with the {\em IXPE} mission \citep[e.g.][]{ixpe_cygx1,ixpe_swiftj1727, ingram24,ixpe_j1727decay,ixpe_igr17091,mastroserio25}. In this picture, our results are the first to probe the physical conditions of this medium directly in the infrared. Another apparent coincidence for this scenario to work is the requirement for the SSA break to lie in the MIRI band, straddling frequencies very similar to those inferred for the optically-thick--to--thin break associated with the hard-state jet base \citep{Gandhi-2011, russell13}, despite dramatic changes in flux and timing properties between the two states (Fig.\,\ref{fig:wisecomparison}; \citealt{Gandhi-2008,Casella-2010}). Finally, we note that the compact size scales and high ionisation fractions inherent to such a coronal medium cannot accommodate the long MIR lag tentatively identified in \S\,\ref{sec:simultaneous} (though the veracity of said lag remains questionable). 

Given that the MIR band is the only portion of the spectrum where the synchrotron component dominates the broadband SED (Fig.\,\ref{fig:sedfit}), we have explored the limits of what the present data allow. Further tests could come from longer time sampling, or probing far-infrared wavelengths (a gap in our current coverage between MIRI and ALMA), as we discuss later. 

\subsection{On the origin of the radio and sub-mm emission: a discrete jet ejection}
\label{sec:radiommresults}

The radio and sub-mm detection is suggestive of jet activity at the time of our observational campaign. However, detailed imaging analysis of these data indicate that the radio and sub-mm counterparts do not arise from the core position of \gx; instead the detections at these wavelengths are shifted from the optical {\em Gaia} DR3 position of \gx\ \citep{gaiamission, gaiadr3summarycontents} to the West by $\approx$\,1$\farcs$2, along its known jet axis \citep[e.g.,][]{2004MNRAS.347L..52G,mastroserio25}. The radio/sub-mm spectrum is consistent with the measured optically-thin long-wavelength slope $\alpha_{\rm RS}$\,=\,--0.57\,\p\,0.03. In addition, imaging the radio data on shorter timescales reveals that this source was brightening over time, from $\approx$\,0.65\,mJy at the start of the observation to $\sim$\,1\,mJy at the end (over $\sim$\,3.5\,hours, with errors of order 0.06\,mJy). 

These results indicate that the radio and sub-mm emission originated from a downstream jet knot that was ejected from the system, likely near the hard-to-soft state transition, and was propagating outwards from the compact object. The angular offset corresponds to a large physical projected separation from the core of $\approx$\,0.05\,pc at the distance of \gx. As such, the observed radio/sub-mm activity can not be causally linked to  that seen at other wavelengths (e.g., in X-rays, which must arise from the core).  Our imaging requires that the compact jet emission at the core has been quenched \citep[e.g.,][]{fender04,corbel13b,russell20}, however, we are unable to place deep constraints on the quenching factor due to the presence of the jet knot. Full details of the radio and sub-mm will be presented in an upcoming publication (Russell et al. in prep).

\subsection{Archival Mid-IR detections of \gx}
\label{sec:archivalgx}

Fig.\,\ref{fig:longterm} shows that \gx\ has been observed at several historical epochs with MIR-sensitive facilities. Some of these datasets, especially the \spitzer\ observations, have not been reported in the literature thus far. While attempting to pin down the origin of the MIRI emission, we returned to these historical datasets to see if they could shed further light on our investigations. 

Table\,\ref{tab:archivalmir} lists detections of \gx\ using all observations from the \spitzer\ Heritage archive. Where fluxes were unavailable, we measured these ourselves from the Level 2 post-BCD mosaic images using aperture photometry and corrections as described in the \spitzer\ analysis handbooks. Systematic uncertainties on the photometry were estimated by varying the sizes and locations of the source and background apertures. The table shows quiescent detections in 2005 and 2014, together with one early detection in the soft state from 2004 (\citealt{tomsick04} further classified this as a `Steep Power-law' state). 

The table shows that the MIR flux density of \gx\ always bottoms out around $\approx$\,0.15--0.2\,mJy, to within uncertainties. Quiescent flux measurements are within a factor of $\approx$\,2 of our soft state detection, though the source undergoes brightening by orders of magnitude during the intervening hard states. Unless this is a 
coincidence, it requires a source that is stable and  unrelated to the standard outer disc, to a hard-state compact jet, as well as to the companion star (cf.\,Fig.\,\ref{fig:sedfit}). 

It is worth checking the immediately surrounding field of \gx\ for any contaminants. The nearest (and brightest within $\approx$4\,arcsec) star is Obj\_ID 3333279450787716846 from the Dark Energy Camera Plane Survey (DECaPS2; \citealt{decaps1, decaps2}), which lies 1\farcs 03 to the north-west of the {\em Gaia} DR3 optical centroid of \gx, well within the nominal \spitzer\ PSF of $\approx$\,2\,arcsec. The DECaPS2-reported reddest filter ($Y$-band) flux density is 3.0\,(\p\,0.2)\,$\times$\,10$^{-8}$ Mgy.\footnote{Maggies (\url{https://www.sdss3.org/dr8/algorithms/magnitudes.php}).} Assuming a late spectral class (K, say) and similar extinction correction to \gx, this flux density could account for $\approx$\,0.6\,$\times$ the \spitzer\ IRAC 3.6 and 4.5\,\micron\ quiescent fluxes listed in Table\,\ref{tab:archivalmir}. Contaminations fractions reduce steeply to only $\approx$\,0.1 of the \spitzer/MIPS 24\,\micron\ band quiescent flux density. These estimates imply that contamination by close neighbours may account for a substantial fraction of the \spitzer\ quiescent fluxes at shorter MIR wavelength but not at the long end. \jwst's excellent spatial resolution rules out any contamination to our MIRI extraction. 

\subsection{The MIR vs. X-ray luminosity parameter space}
\label{sec:mirx}

\begin{figure*}
    \centering
   \includegraphics[angle=0,width=0.8\textwidth]{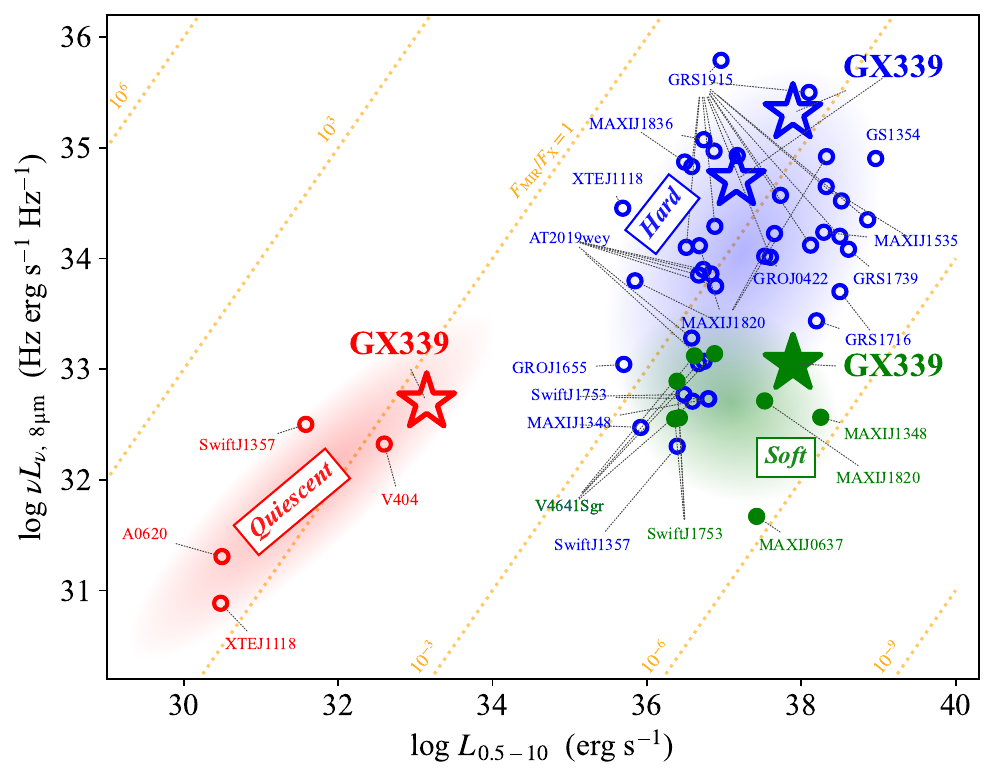}
    \caption{X-ray (0.5--10\,keV) vs. mid-IR (8\,\micron) monochromatic luminosity for a sample of MIR-detected black hole XRBs. The red, blue and green colours are for sources in the quiescent, hard and soft states, respectively, with \gx\ highlighted as a larger star. 
    Lines of constant IR/X-ray flux ratio are denoted in dotted orange. Source names are abbreviated for clarity. The shaded ellipses denote the 2-$\sigma$ covariance regions for the respective states. 
    \label{fig:mirx}}
\end{figure*}

In Fig.\,\ref{fig:mirx}, we collate X-ray and MIR detections of transient black hole XRBs in the quiescent, hard, and soft accretion states, in an attempt to 
identify 
any underlying trends between these bands. In addition to the analyses presented herein, the data come from archival studies described in Appendix\,\ref{sec:appendixmirx} and Table\,\ref{tab:mirx}. Other salient details can also be found therein. 

Quiescent detections in Fig.\,\ref{fig:mirx} are clustered around $F_{\rm MIR}/F_{\rm X-ray}$\,$\sim$\,$\mathcal{O}(1)$.\footnote{Correcting for partial near-neighbour contamination as detailed in \S\,\ref{sec:archivalgx} does not substantially affect the inferences herein.} Hard-state sources show the highest $L_{\rm X-ray}$ and $L_{\rm MIR}$ values. There is a well-known fundamental plane in XRBs coupling the {\em radio} and X-ray fluxes, and extending from the hard outburst down to quiescence seamlessly for a group of objects on the so-called `radio-loud' branch which includes \gx\ \citep[e.g., ][]{corbel03, plotkin12, corbel13, gallo14, tremou20}. A simple linear regression fit to our quiescent sources with the form: 

\begin{equation}
    {\rm log}\, \left(\frac{\nu L_{\rm \nu,\,8\mu m}}{10^{32}\,{\rm erg\,s^{-1}}}\right)\,=\alpha+\,\beta\,{\rm log}\,\left(\frac{L_{0.5-10\,{\rm keV}}}{\rm 10^{32}\,erg\,s^{-1}}\right)
\end{equation}

\noindent
yields best-fit parameters of $\alpha$\,=\,0.07\,$\pm$\,0.19 and $\beta$\,=\,0.53\,$\pm$\,0.04 for the three quiescent and hard state detections of \gx\ alone. Interestingly, the slope $\beta$ matches that obtained when fitting only the quiescent-state sources (for which we find $\alpha$\,=\,0.14\,\p\,0.20 and $\beta$\,=\,0.59\,\p\,0.18). With the small number of sources included, these fits should be used with some caution. Nevertheless, it is also interesting that these slopes are close to that of the universal radio–X-ray correlation -- e.g., the 5\,GHz versus 1--10\,keV correlation for \gx\ presented in \citealt{gallo14} has a slope of 0.62\,\p\,0.04. 

At face value, these similarities would argue for a close connection between the physical processes manifesting the radio and the MIR bands across these two states. However, it is also immediately apparent from the figure that there is a cloud of hard-state detections with enhanced scatter, with many lying at systematically lower MIR luminosities for any given X-ray power. There could be several possible factors at play here, as follows. Our collated sample includes sources that are known to display weak radio jet emission during outburst (e.g., Swift\,J1753.5--0127, MAXI\,J1348--630; \citealt{plotkin17, carotenuto22}), which could result in a corresponding weakness in Fig.\,\ref{fig:mirx} if the MIR were also dominated by (inner) jet emission. In some instances, other systematic issues such as distance uncertainties could move their location in the plotted plane (cf. discussion on AT\,2019wey, GRS\,1716--249 and Swift\,J1357.2--0933 in Appendix\,\ref{sec:appendixmirx}).  Moreover, the cloud of hard-state detections at the luminous end are dominated by \grs, which has been in a unique `X-ray obscured' state for the past few years, and cannot be neatly described as fitting into any of the canonical accretion states \citep{miller20, gandhi25}. Finally, unlike the universal jet origin for radio emission, the origin of the MIR is likely to be inherently diverse: e.g., we have inferred the presence of hot flow synchrotron herein for \gx\ in the soft state; the quiescent state emission in A\,0620--00 is free-free--wind dominated \citep{zuo25}, and a jet is thought to drive the flux and variability in V404\,Cyg \citep{borowski25}. 

The plotted soft state detections are, unsurprisingly, strongly dominated by X-rays, with $F_{\rm MIR}/F_{\rm X-ray}$\,$\sim$\,$\mathcal{O}(10^{-5})$, and the distribution of our plotted quiescent and soft state sources in MIR--X-ray space is qualitatively similar to that previously seen in fundamental plane correlations in radio--X-ray or NIR--X-ray parameter space \citep{coriat09, corbel13}. Intriguingly, \gx\ lies near the upper end of the distribution of XRBs in all accretion states (soft, hard and quiescent), approaching or rivalling systems with much larger accretion reservoirs such as V404\,Cyg and \grs. There is no obvious {\em a priori} physical reason for this, and we can only speculate that the inner synchrotron radiation zone (either the hot flow in the soft state, or the compact jet in the hard state) is particularly prominent in \gx. In order to understand the underlying physical conditions driving this, we suggest that detailed studies of the SED and timing properties in quiescence could be enlightening.

There remains much to be unravelled in understanding the location of sources in this MIR--X-ray plane. At minimum, it serves to demonstrate reasonably clear empirical parameter space segregation between the various accretion states, and highlights a progression of high $\rightarrow$ medium $\rightarrow$ low $F_{\rm MIR}$/$F_{\rm X-ray}$ flux ratios, from quiescence $\rightarrow$ hard $\rightarrow$ soft state, successively. 

Finally, it should be borne in mind that persistent systems are not retained here, and that the `snapshot' instantaneous measurements included may provide a biased overview of the population. It will thus be important to trace the evolution of individual sources across this diagram through various states, in order to circumvent such biases and to better understand how the strengths of components such as the accretion disc, winds, jet, hot flow synchrotron and donor star vary across the MIR/X-ray parameter space. 

\begin{table*}
    \caption{Newly Measured mid-infrared flux densities of \gx\ in archival \spitzer\ observations}
    \centering
    \begin{tabular}{lccr}
\hline
         Date & Mission/Band & Flux Density & State\\
         UTC &  & mJy& \\
         \hline
         2004 Aug 20 & Spitzer MIPS 24 & 0.14\,\p\,0.04$^*$ & Soft/Steep Power-law State\\
         2005 Sep 04 & Spitzer MIPS 24 & 0.19\,\p\,0.03 & Quiescence\\
         2014 Jun 08& Spitzer IRAC 3.6 & 0.22\,\p\,0.02 & Quiescence\\
         2014 Jun 08 & Spitzer IRAC 4.5 & 0.18\,\p\,0.03 & Quiescence\\
         \hline
    \end{tabular}
    ~\\
    $^*$see also initial report by \citet{tomsick04}.
    \label{tab:archivalmir}
\end{table*}

\section{Summary}

Our core motivation here has been to uncover the physical processes driving the radiative power in accreting black hole systems, and to investigate how these may change with accretion state. \gx\ is a prolific multiwavelength transient XRB, and so was a natural target of choice for \jwst\ observations. 

Our observation window fell on MJD\,60379, during (what turned out to be) the declining phase of the 2023/24 outburst. We caught the source in a disc-dominated X-ray state with a clearly detectable high-energy power-law tail (\S\,\ref{sec:obs}). The overall source characteristics, including its position in the X-ray hardness-intensity diagrams (Fig.\,\ref{fig:longterm}), a very low X-ray variability r.m.s (\S\,\ref{sec:simultaneous}) and the disc dominance are consistent with a fairly typical soft state, despite some uncertainty on the prevalence of the high-energy tail. 

We utilised MIRI low-resolution spectroscopy to facilitate infrared spectral-timing investigations over the wavelength range of $\approx$\,5--10\,\micron. A comprehensive multiwavelength campaign was coordinated close in-time with the MIRI observation window. The observations revealed a flat MIR continuum with a flux $\gtsim$\,300\,$\times$ fainter than previous hard-state photometric MIR detections (\S\,\ref{sec:sedresults}). This can be accounted for by suppression of compact jet emission in the soft state, and is supported by the absence of contemporaneous core radio and sub-mm detections (\S\,\ref{sec:radiommresults}).  The MIRI continuum shows significant red-noise variations on timescales extending to several hundreds of seconds (\S\,\ref{sec:miriresults}), which are too short to originate in an extended circumbinary disc (\S\,\ref{sec:cbd}). A warm mid-infrared--emitting wind (\S\,\ref{sec:wind}), on the other hand, cannot easily produce the broadband spectrum without corresponding recombination emission lines.

The most plausible scenario for powering the observed MIR emission is synchrotron radiation from a coronal hot flow (\S\,\ref{sec:hotflow}). With a magnetic field $B$\,$\approx$\,3\,$\times$\,10$^4$\,G and plasma extended out to $r_{\rm C}$\,$\sim$\,170\,$R_{\rm G}$, the same medium could also Comptonise disc photons to account for the high-energy power-law tail, while self-consistently accounting for the MIRI timing properties. In this scenario, the MIR provides a new probe of physical conditions in the coronal hot medium overlaying the disc in the soft state, which presumably subsumes the role of an accretion disc `atmosphere' and the base of the hard-state extended jet. 

There is weak evidence of a curiously long time lag between optical and mid-infrared, with the latter lagging the former by several hundred seconds (\S\,\ref{sec:simultaneous}). A lag of this magnitude would be impossible to reproduce by a hot flow, but simulated significance tests suggest that the inferred lag is instead likely to be a false positive, caused by sparseness in time sampling of the data. The hot flow scenario also requires the synchrotron self-absorption break frequency to remain approximately unchanged between the soft state and the peak of the (far-more MIR-luminous) hard state. This remains puzzling, but simply be a coincidence. Given these caveats, such a scenario remains to be validated, though encouragingly, the presence of similarly extended coronae has recently been inferred in other XRB X-ray polarisation studies. Observations at the redder end (e.g. with MIRI/MRS extending to $\approx$\,30\,\micron) will help to better quantify the underlying spectral slope, any long-wavelength time variability, and also aid in firmly ruling out competing physical models such as a wind or circumbinary disc. 

To our knowledge, these are the first reported spectroscopic detections of \gx\ in the mid-infrared. Its significant detection -- despite the MIR-weak flux level -- demonstrates \jwst's incredible sensitivity to probe new spectral-timing parameter space. In the process of unravelling the origin of the MIR emission, we analysed and presented several unpublished archival observations of \gx\ (\S\,\ref{sec:archivalgx}) showing that (i) the source's quiescent MIR power lies within a factor of just a few of that seen in the soft state; (ii) in the context of the wider population of black hole transients, \gx\ can, at times, be amongst the most luminous of MIR--emitting XRBs across all accretion states (\S\,\ref{sec:mirx}). With synchrotron (either coronal hot flow or compact jet emission) dominating the soft as well as the hard states, a better understanding of processes driving the quiescent MIR emission will next be important. The MIR--X-ray luminosity plane of black hole transients (Fig.\,\ref{fig:mirx}) allows isolation of characteristic \fmirfx\ flux ratios as a function of accretion state, and enables tracing evolution of the relative fluxes in these two bands as outbursts evolve.

\section*{Data Availability}

The data analysed herein are all publicly available in online telescope archives or can be made available upon reasonable request to the authors. The \jwst\ data can be identified using the Program identifier 1586.

\section*{Acknowledgements}

We are grateful to the mission and instrument teams at STScI for their patience and expert help with our numerous queries, especially MIRI scientist S. Kendrew and staff astronomers I. Wong and G. Sloan. 

PG acknowledges funding from The Royal Society (SRF$\backslash$R1$\backslash$241074). PG, CDB and CK thank UKRI Science \& Technology Facilities Council for support. AJT \& EC, GRS, and COH acknowledge the support of the Natural Sciences and Engineering Research Council of Canada (NSERC Discovery Grants: RGPIN--2024--04458, RGPIN--2021--0400, and RGPIN--2023--04264). AJT additionally acknowledges that this research was undertaken thanks to funding from the Canada Research Chairs Program. TJM acknowledges support from JWST-GO-01586.002. AWS acknowledges support from 80NSSC24K0345 and 80NSSC24K1257. RIH and ESB acknowledge support from JWST-GO-01586.007. JAT acknowledges support from JWST-GO-01586.010-A.
TS and FMV acknowledge financial support from the Spanish Ministry of Science, Innovation and Universities (MICIU) under grants PID2020-114822GB-I00
and PID2023-151588NB-I00. JAP acknowledges support from STFC consolidated grant ST/X001075/1.
DP acknowledges the support from ISRO (India), under the ISRO RESPOND program. The work of MER was carried out at the Jet Propulsion Laboratory, California Institute of Technology, under a contract with the National Aeronautics and Space Administration. MCB and TDR acknowledge support from the INAF-Astrofit fellowship. DMR, PS, KA and SKR are supported by Tamkeen under the NYU Abu Dhabi Research Institute grant CASS. SM is supported by a European Research Council (ERC) Synergy Grant "BlackHolistic" grant No. 10107164. Partial support for KL's effort on the project was provided by NASA through grant numbers HST-GO-16489 and HST-GO-16659 and from the Space Telescope Science Institute, which is operated by AURA, Inc., under NASA contract NAS 5-26555. FMV is supported by the European Union’s Horizon Europe research and innovation programme with the Marie Sk\l{}odowska-Curie grant agreement No. 101149685. 
We thank Dan Bramich for his contributions to the development of the XB-NEWS pipeline that processed the LCO data. PG thanks J.\,Malzac for discussion regarding the possibility of dark jets in the system, and is grateful to A.\,Veledina for discussions on the hot flow model. 

This work is based on observations made with the NASA/ESA/CSA {\em James Webb Space Telescope}. The data were obtained from the Mikulski Archive for Space Telescopes at the Space Telescope Science Institute, which is operated by the Association of Universities for Research in Astronomy, Inc., under NASA contract NAS 5-03127 for JWST. These observations are associated with program \#1586 \citep{jwst1586}.  
Support for program \#1586 was provided by NASA through a grant from the Space Telescope Science Institute, which is operated by the Association of Universities for Research in Astronomy, Inc., under NASA contract NAS 5-03127. This paper makes use of the following ALMA data: ADS/JAO.ALMA\#2023.A.00018.T. ALMA is a partnership of ESO (representing its member states), NSF (USA) and NINS (Japan), together with NRC (Canada), NSTC and ASIAA (Taiwan), and KASI (Republic of Korea), in cooperation with the Republic of Chile. The Joint ALMA Observatory is operated by ESO, AUI/NRAO and NAOJ. The National Radio Astronomy Observatory is a facility of the National Science Foundation operated under cooperative agreement by Associated Universities, Inc. We thank the ALMA director for granting our DDT request, used to obtain the ALMA mm data analysed in this paper.

This research has made use of the NASA/IPAC Infrared Science Archive, which is funded by the National Aeronautics and Space Administration and operated by the California Institute of Technology. The DOI of the \spitzer\ Enhanced IRS Products is 10.26131/IRSA399. 

This research has made use of MAXI data provided by RIKEN, JAXA and the MAXI team. 

This publication uses the data from the \astrosat\ mission of the Indian Space Research Organisation (ISRO), archived at the Indian Space Science Data Centre (ISSDC). This publication uses UVIT data processed by the payload operations centre at the Indian Institute of Astrophysics. 

This research has made use of data and/or software provided by the High Energy Astrophysics Science Archive Research Center (HEASARC), which is a service of the Astrophysics Science Division at NASA/GSFC.

Based on observations with ISO, an ESA project with instruments funded by ESA Member States (especially the PI countries: France, Germany, the Netherlands and the United Kingdom) and with the participation of ISAS and NASA.

SMARTNet \citep{middleton17} helped to coordinate observations.
We have made use of software and web tools from the High
Energy Astrophysics Science Archive Research Center (HEASARC)
and made use of data and the ‘Build XRT Products’ tool supplied by the UK Swift Science Data Centre at the University of
Leicester.

\bibliographystyle{mnras}
\bibliography{GX339_JWST}

\begin{thebibliography}{}
\makeatletter
\relax
\def\mn@urlcharsother{\let\do\@makeother \do\$\do\&\do\#\do\^\do\_\do\%\do\~}
\def\mn@doi{\begingroup\mn@urlcharsother \@ifnextchar [ {\mn@doi@} {\mn@doi@[]}}
\def\mn@doi@[#1]#2{\def\@tempa{#1}\ifx\@tempa\@empty \href {http://dx.doi.org/#2} {doi:#2}\else \href {http://dx.doi.org/#2} {#1}\fi \endgroup}
\def\mn@eprint#1#2{\mn@eprint@#1:#2::\@nil}
\def\mn@eprint@arXiv#1{\href {http://arxiv.org/abs/#1} {{\tt arXiv:#1}}}
\def\mn@eprint@dblp#1{\href {http://dblp.uni-trier.de/rec/bibtex/#1.xml} {dblp:#1}}
\def\mn@eprint@#1:#2:#3:#4\@nil{\def\@tempa {#1}\def\@tempb {#2}\def\@tempc {#3}\ifx \@tempc \@empty \let \@tempc \@tempb \let \@tempb \@tempa \fi \ifx \@tempb \@empty \def\@tempb {arXiv}\fi \@ifundefined {mn@eprint@\@tempb}{\@tempb:\@tempc}{\expandafter \expandafter \csname mn@eprint@\@tempb\endcsname \expandafter{\@tempc}}}

\bibitem[\protect\citeauthoryear{{Agrawal} et~al.,}{{Agrawal} et~al.}{2017}]{Agrawal2017JApA...38...30A}
{Agrawal} P.~C.,  et~al., 2017, \mn@doi [Journal of Astrophysics and Astronomy] {10.1007/s12036-017-9451-z}, \href {https://ui.adsabs.harvard.edu/abs/2017JApA...38...30A} {38, 30}

\bibitem[\protect\citeauthoryear{{Alabarta} et~al.,}{{Alabarta} et~al.}{2024}]{Alabarta2024ATel}
{Alabarta} K.,  et~al., 2024, The Astronomer's Telegram, \href {https://ui.adsabs.harvard.edu/abs/2024ATel16460....1A} {16460, 1}

\bibitem[\protect\citeauthoryear{{Ambrifi} et~al.,}{{Ambrifi} et~al.}{2025}]{ambrifi25}
{Ambrifi} A.,  et~al., 2025, \mn@doi [\aap] {10.1051/0004-6361/202451024}, \href {https://ui.adsabs.harvard.edu/abs/2025A&A...694A.109A} {694, A109}

\bibitem[\protect\citeauthoryear{{Antia} et~al.,}{{Antia} et~al.}{2021}]{Antia2021JApA...42...32A}
{Antia} H.~M.,  et~al., 2021, \mn@doi [Journal of Astrophysics and Astronomy] {10.1007/s12036-021-09712-8}, \href {https://ui.adsabs.harvard.edu/abs/2021JApA...42...32A} {42, 32}

\bibitem[\protect\citeauthoryear{{Arnaud}}{{Arnaud}}{1996}]{xspec}
{Arnaud} K.~A.,  1996, in {Jacoby} G.~H.,  {Barnes} J.,  eds,  Astronomical Society of the Pacific Conference Series Vol. 101, Astronomical Data Analysis Software and Systems V. p.~17

\bibitem[\protect\citeauthoryear{{Artymowicz} \& {Lubow}}{{Artymowicz} \& {Lubow}}{1994}]{artymowicz94}
{Artymowicz} P.,  {Lubow} S.~H.,  1994, \mn@doi [\apj] {10.1086/173679}, \href {https://ui.adsabs.harvard.edu/abs/1994ApJ...421..651A} {421, 651}

\bibitem[\protect\citeauthoryear{{Arzoumanian} et~al.,}{{Arzoumanian} et~al.}{2014}]{nicer14}
{Arzoumanian} Z.,  et~al., 2014, in {Takahashi} T.,  {den Herder} J.-W.~A.,   {Bautz} M.,  eds,  Society of Photo-Optical Instrumentation Engineers (SPIE) Conference Series Vol. 9144, Space Telescopes and Instrumentation 2014: Ultraviolet to Gamma Ray. p. 914420, \mn@doi{10.1117/12.2056811}

\bibitem[\protect\citeauthoryear{{Baglio} et~al.,}{{Baglio} et~al.}{2018}]{baglio18}
{Baglio} M.~C.,  et~al., 2018, \mn@doi [\apj] {10.3847/1538-4357/aae532}, \href {https://ui.adsabs.harvard.edu/abs/2018ApJ...867..114B} {867, 114}

\bibitem[\protect\citeauthoryear{{Bahramian}, {Heinke}, {Degenaar}, {Chomiuk}, {Wijnands}, {Strader}, {Ho}  \& {Pooley}}{{Bahramian} et~al.}{2015}]{bahramian15}
{Bahramian} A.,  {Heinke} C.~O.,  {Degenaar} N.,  {Chomiuk} L.,  {Wijnands} R.,  {Strader} J.,  {Ho} W. C.~G.,   {Pooley} D.,  2015, \mn@doi [\mnras] {10.1093/mnras/stv1585}, \href {https://ui.adsabs.harvard.edu/abs/2015MNRAS.452.3475B} {452, 3475}

\bibitem[\protect\citeauthoryear{{Banerjee} et~al.,}{{Banerjee} et~al.}{2024}]{Banerjee2024ApJ...964..189B}
{Banerjee} S.,  et~al., 2024, \mn@doi [\apj] {10.3847/1538-4357/ad24ef}, \href {https://ui.adsabs.harvard.edu/abs/2024ApJ...964..189B} {964, 189}

\bibitem[\protect\citeauthoryear{{Belloni}, {Homan}, {Casella}, {van der Klis}, {Nespoli}, {Lewin}, {Miller}  \& {M{\'e}ndez}}{{Belloni} et~al.}{2005}]{belloni05}
{Belloni} T.,  {Homan} J.,  {Casella} P.,  {van der Klis} M.,  {Nespoli} E.,  {Lewin} W.~H.~G.,  {Miller} J.~M.,   {M{\'e}ndez} M.,  2005, \mn@doi [\aap] {10.1051/0004-6361:20042457}, \href {https://ui.adsabs.harvard.edu/abs/2005A&A...440..207B} {440, 207}

\bibitem[\protect\citeauthoryear{{Bhargava}, {Belloni}, {Bhattacharya}  \& {Misra}}{{Bhargava} et~al.}{2019}]{Bhargava2019MNRAS.488..720B}
{Bhargava} Y.,  {Belloni} T.,  {Bhattacharya} D.,   {Misra} R.,  2019, \mn@doi [\mnras] {10.1093/mnras/stz1774}, \href {https://ui.adsabs.harvard.edu/abs/2019MNRAS.488..720B} {488, 720}

\bibitem[\protect\citeauthoryear{{Bhargava}, {Bhattacharyya}, {Homan}  \& {Pahari}}{{Bhargava} et~al.}{2023}]{Bhargava2023ApJ...955..102B}
{Bhargava} Y.,  {Bhattacharyya} S.,  {Homan} J.,   {Pahari} M.,  2023, \mn@doi [\apj] {10.3847/1538-4357/acee7a}, \href {https://ui.adsabs.harvard.edu/abs/2023ApJ...955..102B} {955, 102}

\bibitem[\protect\citeauthoryear{{Blandford} \& {K{\"o}nigl}}{{Blandford} \& {K{\"o}nigl}}{1979}]{Blandford-1979}
{Blandford} R.~D.,  {K{\"o}nigl} A.,  1979, \mn@doi [\apj] {10.1086/157262}, \href {http://adsabs.harvard.edu/abs/1979ApJ...232...34B} {232, 34}

\bibitem[\protect\citeauthoryear{{Borowski} et~al.,}{{Borowski} et~al.}{2025}]{borowski25}
{Borowski} E.~S.,  et~al., 2025, \mn@doi [arXiv e-prints] {10.48550/arXiv.2506.20536}, \href {https://ui.adsabs.harvard.edu/abs/2025arXiv250620536B} {p. arXiv:2506.20536}

\bibitem[\protect\citeauthoryear{{Bramich} \& {Freudling}}{{Bramich} \& {Freudling}}{2012}]{Bramich2012}
{Bramich} D.~M.,  {Freudling} W.,  2012, \mn@doi [\mnras] {10.1111/j.1365-2966.2012.21385.x}, \href {https://ui.adsabs.harvard.edu/abs/2012MNRAS.424.1584B} {424, 1584}

\bibitem[\protect\citeauthoryear{{Bright} et~al.,}{{Bright} et~al.}{2025}]{bright25}
{Bright} J.~S.,  et~al., 2025, \mn@doi [\mnras] {10.1093/mnras/staf1098}, \href {https://ui.adsabs.harvard.edu/abs/2025MNRAS.541.1851B} {541, 1851}

\bibitem[\protect\citeauthoryear{{Brocksopp}, {Bandyopadhyay}  \& {Fender}}{{Brocksopp} et~al.}{2004}]{brocksopp04}
{Brocksopp} C.,  {Bandyopadhyay} R.~M.,   {Fender} R.~P.,  2004, \mn@doi [\na] {10.1016/j.newast.2003.11.002}, \href {https://ui.adsabs.harvard.edu/abs/2004NewA....9..249B} {9, 249}

\bibitem[\protect\citeauthoryear{{Burrows} et~al.,}{{Burrows} et~al.}{2005}]{xrt}
{Burrows} D.~N.,  et~al., 2005, \mn@doi [\ssr] {10.1007/s11214-005-5097-2}, \href {https://ui.adsabs.harvard.edu/abs/2005SSRv..120..165B} {120, 165}

\bibitem[\protect\citeauthoryear{{CASA Team} et~al.,}{{CASA Team} et~al.}{2022}]{CASA2022}
{CASA Team} et~al., 2022, \mn@doi [\pasp] {10.1088/1538-3873/ac9642}, \href {https://ui.adsabs.harvard.edu/abs/2022PASP..134k4501C} {134, 114501}

\bibitem[\protect\citeauthoryear{{Cadolle Bel} et~al.,}{{Cadolle Bel} et~al.}{2011}]{CadolleBel2011}
{Cadolle Bel} M.,  et~al., 2011, \mn@doi [\aap] {10.1051/0004-6361/201117684}, \href {https://ui.adsabs.harvard.edu/abs/2011A&A...534A.119C} {534, A119}

\bibitem[\protect\citeauthoryear{{Campana}, {Colpi}, {Mereghetti}, {Stella}  \& {Tavani}}{{Campana} et~al.}{1998}]{Campana98}
{Campana} S.,  {Colpi} M.,  {Mereghetti} S.,  {Stella} L.,   {Tavani} M.,  1998, \mn@doi [\aapr] {10.1007/s001590050012}, \href {https://ui.adsabs.harvard.edu/abs/1998A&ARv...8..279C} {8, 279}

\bibitem[\protect\citeauthoryear{{Cao} et~al.,}{{Cao} et~al.}{2022}]{cao22}
{Cao} H.-M.,  et~al., 2022, \mn@doi [\aap] {10.1051/0004-6361/202142241}, \href {https://ui.adsabs.harvard.edu/abs/2022A&A...657A.104C} {657, A104}

\bibitem[\protect\citeauthoryear{{Cardelli}, {Clayton}  \& {Mathis}}{{Cardelli} et~al.}{1989}]{cardelli89}
{Cardelli} J.~A.,  {Clayton} G.~C.,   {Mathis} J.~S.,  1989, \mn@doi [\apj] {10.1086/167900}, \href {http://adsabs.harvard.edu/abs/1989ApJ...345..245C} {345, 245}

\bibitem[\protect\citeauthoryear{{Carotenuto}, {Corbel}  \& {Tzioumis}}{{Carotenuto} et~al.}{2022}]{carotenuto22}
{Carotenuto} F.,  {Corbel} S.,   {Tzioumis} A.,  2022, \mn@doi [\mnras] {10.1093/mnrasl/slac087}, \href {https://ui.adsabs.harvard.edu/abs/2022MNRAS.517L..21C} {517, L21}

\bibitem[\protect\citeauthoryear{{Casares} et~al.,}{{Casares} et~al.}{2023}]{casares23}
{Casares} J.,  et~al., 2023, \mn@doi [\mnras] {10.1093/mnras/stad3068}, \href {https://ui.adsabs.harvard.edu/abs/2023MNRAS.526.5209C} {526, 5209}

\bibitem[\protect\citeauthoryear{{Casella} et~al.,}{{Casella} et~al.}{2010}]{Casella-2010}
{Casella} P.,  et~al., 2010, \mn@doi [\mnras] {10.1111/j.1745-3933.2010.00826.x}, \href {http://adsabs.harvard.edu/abs/2010MNRAS.404L..21C} {404, L21}

\bibitem[\protect\citeauthoryear{{Charles}, {Matthews}, {Buckley}, {Gandhi}, {Kotze}  \& {Paice}}{{Charles} et~al.}{2019}]{charles19}
{Charles} P.,  {Matthews} J.~H.,  {Buckley} D. A.~H.,  {Gandhi} P.,  {Kotze} E.,   {Paice} J.,  2019, \mn@doi [\mnras] {10.1093/mnrasl/slz120}, \href {https://ui.adsabs.harvard.edu/abs/2019MNRAS.489L..47C} {489, L47}

\bibitem[\protect\citeauthoryear{{Chiar} \& {Tielens}}{{Chiar} \& {Tielens}}{2006}]{chiartielens06}
{Chiar} J.~E.,  {Tielens} A.~G.~G.~M.,  2006, \mn@doi [\apj] {10.1086/498406}, \href {https://ui.adsabs.harvard.edu/abs/2006ApJ...637..774C} {637, 774}

\bibitem[\protect\citeauthoryear{{Corbel} \& {Fender}}{{Corbel} \& {Fender}}{2002}]{corbel02}
{Corbel} S.,  {Fender} R.~P.,  2002, \mn@doi [\apjl] {10.1086/341870}, \href {http://adsabs.harvard.edu/abs/2002ApJ...573L..35C} {573, L35}

\bibitem[\protect\citeauthoryear{{Corbel}, {Nowak}, {Fender}, {Tzioumis}  \& {Markoff}}{{Corbel} et~al.}{2003}]{corbel03}
{Corbel} S.,  {Nowak} M.~A.,  {Fender} R.~P.,  {Tzioumis} A.~K.,   {Markoff} S.,  2003, \mn@doi [\aap] {10.1051/0004-6361:20030090}, \href {https://ui.adsabs.harvard.edu/abs/2003A&A...400.1007C} {400, 1007}

\bibitem[\protect\citeauthoryear{{Corbel}, {Coriat}, {Brocksopp}, {Tzioumis}, {Fender}, {Tomsick}, {Buxton}  \& {Bailyn}}{{Corbel} et~al.}{2013a}]{corbel13}
{Corbel} S.,  {Coriat} M.,  {Brocksopp} C.,  {Tzioumis} A.~K.,  {Fender} R.~P.,  {Tomsick} J.~A.,  {Buxton} M.~M.,   {Bailyn} C.~D.,  2013a, \mn@doi [\mnras] {10.1093/mnras/sts215}, \href {https://ui.adsabs.harvard.edu/abs/2013MNRAS.428.2500C} {428, 2500}

\bibitem[\protect\citeauthoryear{{Corbel} et~al.,}{{Corbel} et~al.}{2013b}]{corbel13b}
{Corbel} S.,  et~al., 2013b, \mn@doi [\mnras] {10.1093/mnrasl/slt018}, \href {https://ui.adsabs.harvard.edu/abs/2013MNRAS.431L.107C} {431, L107}

\bibitem[\protect\citeauthoryear{{Coriat}, {Corbel}, {Buxton}, {Bailyn}, {Tomsick}, {K{\"o}rding}  \& {Kalemci}}{{Coriat} et~al.}{2009}]{coriat09}
{Coriat} M.,  {Corbel} S.,  {Buxton} M.~M.,  {Bailyn} C.~D.,  {Tomsick} J.~A.,  {K{\"o}rding} E.,   {Kalemci} E.,  2009, \mn@doi [\mnras] {10.1111/j.1365-2966.2009.15461.x}, \href {https://ui.adsabs.harvard.edu/abs/2009MNRAS.400..123C} {400, 123}

\bibitem[\protect\citeauthoryear{{Corral-Santana}, {Casares}, {Mu{\~n}oz-Darias}, {Bauer}, {Mart{\'\i}nez-Pais}  \& {Russell}}{{Corral-Santana} et~al.}{2016}]{blackcat}
{Corral-Santana} J.~M.,  {Casares} J.,  {Mu{\~n}oz-Darias} T.,  {Bauer} F.~E.,  {Mart{\'\i}nez-Pais} I.~G.,   {Russell} D.~M.,  2016, \mn@doi [\aap] {10.1051/0004-6361/201527130}, \href {https://ui.adsabs.harvard.edu/abs/2016A&A...587A..61C} {587, A61}

\bibitem[\protect\citeauthoryear{{Done}, {Gierli{\'n}ski}  \& {Kubota}}{{Done} et~al.}{2007}]{done07}
{Done} C.,  {Gierli{\'n}ski} M.,   {Kubota} A.,  2007, \mn@doi [\aapr] {10.1007/s00159-007-0006-1}, \href {https://ui.adsabs.harvard.edu/abs/2007A&ARv..15....1D} {15, 1}

\bibitem[\protect\citeauthoryear{{Dunn}, {Fender}, {K{\"o}rding}, {Cabanac}  \& {Belloni}}{{Dunn} et~al.}{2008}]{dunn08}
{Dunn} R.~J.~H.,  {Fender} R.~P.,  {K{\"o}rding} E.~G.,  {Cabanac} C.,   {Belloni} T.,  2008, \mn@doi [\mnras] {10.1111/j.1365-2966.2008.13258.x}, \href {https://ui.adsabs.harvard.edu/abs/2008MNRAS.387..545D} {387, 545}

\bibitem[\protect\citeauthoryear{{Dunn}, {Fender}, {K{\"o}rding}, {Belloni}  \& {Cabanac}}{{Dunn} et~al.}{2010}]{dunn10}
{Dunn} R.~J.~H.,  {Fender} R.~P.,  {K{\"o}rding} E.~G.,  {Belloni} T.,   {Cabanac} C.,  2010, \mn@doi [\mnras] {10.1111/j.1365-2966.2010.16114.x}, \href {https://ui.adsabs.harvard.edu/abs/2010MNRAS.403...61D} {403, 61}

\bibitem[\protect\citeauthoryear{{Dyrek}, {Ducrot}, {Lagage}, {Tremblin}, {Kendrew}, {Bouwman}  \& {Bouffet}}{{Dyrek} et~al.}{2024}]{dyrek24}
{Dyrek} A.,  {Ducrot} E.,  {Lagage} P.~O.,  {Tremblin} P.,  {Kendrew} S.,  {Bouwman} J.,   {Bouffet} R.,  2024, \mn@doi [\aap] {10.1051/0004-6361/202347127}, \href {https://ui.adsabs.harvard.edu/abs/2024A&A...683A.212D} {683, A212}

\bibitem[\protect\citeauthoryear{{Echibur{\'u}-Trujillo} et~al.,}{{Echibur{\'u}-Trujillo} et~al.}{2024}]{echiburutrujillo24}
{Echibur{\'u}-Trujillo} C.,  et~al., 2024, \mn@doi [\apj] {10.3847/1538-4357/ad1a10}, \href {https://ui.adsabs.harvard.edu/abs/2024ApJ...962..116E} {962, 116}

\bibitem[\protect\citeauthoryear{{Edelson} \& {Krolik}}{{Edelson} \& {Krolik}}{1988}]{dcf}
{Edelson} R.~A.,  {Krolik} J.~H.,  1988, \mn@doi [\apj] {10.1086/166773}, \href {https://ui.adsabs.harvard.edu/abs/1988ApJ...333..646E} {333, 646}

\bibitem[\protect\citeauthoryear{{Evans} et~al.,}{{Evans} et~al.}{2009}]{evans09}
{Evans} P.~A.,  et~al., 2009, \mn@doi [\mnras] {10.1111/j.1365-2966.2009.14913.x}, \href {https://ui.adsabs.harvard.edu/abs/2009MNRAS.397.1177E} {397, 1177}

\bibitem[\protect\citeauthoryear{{Ewing} et~al.,}{{Ewing} et~al.}{2025}]{ixpe_igr17091}
{Ewing} M.,  et~al., 2025, \mn@doi [\mnras] {10.1093/mnras/staf859}, \href {https://ui.adsabs.harvard.edu/abs/2025MNRAS.541.1774E} {541, 1774}

\bibitem[\protect\citeauthoryear{{Fender}, {Belloni}  \& {Gallo}}{{Fender} et~al.}{2004}]{fender04}
{Fender} R.~P.,  {Belloni} T.~M.,   {Gallo} E.,  2004, \mn@doi [\mnras] {10.1111/j.1365-2966.2004.08384.x}, \href {http://adsabs.harvard.edu/abs/2004MNRAS.355.1105F} {355, 1105}

\bibitem[\protect\citeauthoryear{{Foight}, {G{\"u}ver}, {{\"O}zel}  \& {Slane}}{{Foight} et~al.}{2016}]{foight16}
{Foight} D.~R.,  {G{\"u}ver} T.,  {{\"O}zel} F.,   {Slane} P.~O.,  2016, \mn@doi [\apj] {10.3847/0004-637X/826/1/66}, \href {https://ui.adsabs.harvard.edu/abs/2016ApJ...826...66F} {826, 66}

\bibitem[\protect\citeauthoryear{{Frank}, {King}  \& {Raine}}{{Frank} et~al.}{2002}]{frankkingraine02}
{Frank} J.,  {King} A.,   {Raine} D.~J.,  2002, {Accretion Power in Astrophysics: Third Edition}.
{Cambridge University Press}

\bibitem[\protect\citeauthoryear{{Fuchs}, {Mirabel}  \& {Claret}}{{Fuchs} et~al.}{2003}]{fuchs03}
{Fuchs} Y.,  {Mirabel} I.~F.,   {Claret} A.,  2003, \mn@doi [\aap] {10.1051/0004-6361:20030562}, \href {https://ui.adsabs.harvard.edu/abs/2003A&A...404.1011F} {404, 1011}

\bibitem[\protect\citeauthoryear{{Fuchs}, {Koch Miramond}  \& {{\'A}brah{\'a}m}}{{Fuchs} et~al.}{2006}]{fuchs06}
{Fuchs} Y.,  {Koch Miramond} L.,   {{\'A}brah{\'a}m} P.,  2006, \mn@doi [\aap] {10.1051/0004-6361:20042160}, \href {https://ui.adsabs.harvard.edu/abs/2006A&A...445.1041F} {445, 1041}

\bibitem[\protect\citeauthoryear{{Gaia Collaboration} et~al.,}{{Gaia Collaboration} et~al.}{2016}]{gaiamission}
{Gaia Collaboration} et~al., 2016, \mn@doi [\aap] {10.1051/0004-6361/201629272}, \href {https://ui.adsabs.harvard.edu/abs/2016A&A...595A...1G} {595, A1}

\bibitem[\protect\citeauthoryear{{Gaia Collaboration} et~al.,}{{Gaia Collaboration} et~al.}{2023}]{gaiadr3summarycontents}
{Gaia Collaboration} et~al., 2023, \mn@doi [\aap] {10.1051/0004-6361/202243940}, \href {https://ui.adsabs.harvard.edu/abs/2023A&A...674A...1G} {674, A1}

\bibitem[\protect\citeauthoryear{{Gallo}, {Corbel}, {Fender}, {Maccarone}  \& {Tzioumis}}{{Gallo} et~al.}{2004}]{2004MNRAS.347L..52G}
{Gallo} E.,  {Corbel} S.,  {Fender} R.~P.,  {Maccarone} T.~J.,   {Tzioumis} A.~K.,  2004, \mn@doi [\mnras] {10.1111/j.1365-2966.2004.07435.x}, \href {https://ui.adsabs.harvard.edu/abs/2004MNRAS.347L..52G} {347, L52}

\bibitem[\protect\citeauthoryear{{Gallo}, {Migliari}, {Markoff}, {Tomsick}, {Bailyn}, {Berta}, {Fender}  \& {Miller-Jones}}{{Gallo} et~al.}{2007}]{gallo07}
{Gallo} E.,  {Migliari} S.,  {Markoff} S.,  {Tomsick} J.~A.,  {Bailyn} C.~D.,  {Berta} S.,  {Fender} R.,   {Miller-Jones} J. C.~A.,  2007, \mn@doi [\apj] {10.1086/521524}, \href {https://ui.adsabs.harvard.edu/abs/2007ApJ...670..600G} {670, 600}

\bibitem[\protect\citeauthoryear{{Gallo} et~al.,}{{Gallo} et~al.}{2014}]{gallo14}
{Gallo} E.,  et~al., 2014, \mn@doi [\mnras] {10.1093/mnras/stu1599}, \href {https://ui.adsabs.harvard.edu/abs/2014MNRAS.445..290G} {445, 290}

\bibitem[\protect\citeauthoryear{{Gandhi} et~al.,}{{Gandhi} et~al.}{2008}]{Gandhi-2008}
{Gandhi} P.,  et~al., 2008, \mn@doi [\mnras] {10.1111/j.1745-3933.2008.00529.x}, \href {http://adsabs.harvard.edu/abs/2008MNRAS.390L..29G} {390, L29}

\bibitem[\protect\citeauthoryear{{Gandhi} et~al.,}{{Gandhi} et~al.}{2010}]{Gandhi-2010}
{Gandhi} P.,  et~al., 2010, \mn@doi [\mnras] {10.1111/j.1365-2966.2010.17083.x}, \href {http://adsabs.harvard.edu/abs/2010MNRAS.407.2166G} {407, 2166}

\bibitem[\protect\citeauthoryear{{Gandhi} et~al.,}{{Gandhi} et~al.}{2011}]{Gandhi-2011}
{Gandhi} P.,  et~al., 2011, \mn@doi [\apjl] {10.1088/2041-8205/740/1/L13}, \href {http://adsabs.harvard.edu/abs/2011ApJ...740L..13G} {740, L13}

\bibitem[\protect\citeauthoryear{{Gandhi} et~al.,}{{Gandhi} et~al.}{2017}]{gandhi17}
{Gandhi} P.,  et~al., 2017, \mn@doi [Nature Astronomy] {10.1038/s41550-017-0273-3}, \href {https://ui.adsabs.harvard.edu/abs/2017NatAs...1..859G} {1, 859}

\bibitem[\protect\citeauthoryear{{Gandhi} et~al.,}{{Gandhi} et~al.}{2021}]{jwst1586}
{Gandhi} P.,  et~al., 2021, {Black Hole Jet Launching Physics with MIRI}, JWST Proposal. Cycle 1, ID. \#1586

\bibitem[\protect\citeauthoryear{{Gandhi} et~al.,}{{Gandhi} et~al.}{2025}]{gandhi25}
{Gandhi} P.,  et~al., 2025, \mn@doi [\mnras] {10.1093/mnras/staf036}, \href {https://ui.adsabs.harvard.edu/abs/2025MNRAS.537.1385G} {537, 1385}

\bibitem[\protect\citeauthoryear{{Gehrels} et~al.,}{{Gehrels} et~al.}{2004}]{swift}
{Gehrels} N.,  et~al., 2004, \mn@doi [\apj] {10.1086/422091}, \href {https://ui.adsabs.harvard.edu/abs/2004ApJ...611.1005G} {611, 1005}

\bibitem[\protect\citeauthoryear{{Gierli{\'n}ski}, {Done}  \& {Page}}{{Gierli{\'n}ski} et~al.}{2009}]{diskir}
{Gierli{\'n}ski} M.,  {Done} C.,   {Page} K.,  2009, \mn@doi [\mnras] {10.1111/j.1365-2966.2008.14166.x}, \href {https://ui.adsabs.harvard.edu/abs/2009MNRAS.392.1106G} {392, 1106}

\bibitem[\protect\citeauthoryear{{Goodwin} et~al.,}{{Goodwin} et~al.}{2020}]{Goodwin2020}
{Goodwin} A.~J.,  et~al., 2020, \mn@doi [\mnras] {10.1093/mnras/staa2588}, \href {https://ui.adsabs.harvard.edu/abs/2020MNRAS.498.3429G} {498, 3429}

\bibitem[\protect\citeauthoryear{{Grindlay}}{{Grindlay}}{1979}]{grindlay79}
{Grindlay} J.~E.,  1979, \mn@doi [\apjl] {10.1086/183031}, \href {https://ui.adsabs.harvard.edu/abs/1979ApJ...232L..33G} {232, L33}

\bibitem[\protect\citeauthoryear{{Harrison} et~al.,}{{Harrison} et~al.}{2013}]{harrison13}
{Harrison} F.~A.,  et~al., 2013, \mn@doi [\apj] {10.1088/0004-637X/770/2/103}, \href {http://adsabs.harvard.edu/abs/2013ApJ...770..103H} {770, 103}

\bibitem[\protect\citeauthoryear{{Heida}, {Jonker}, {Torres}  \& {Chiavassa}}{{Heida} et~al.}{2017}]{heida17}
{Heida} M.,  {Jonker} P.~G.,  {Torres} M.~A.~P.,   {Chiavassa} A.,  2017, \mn@doi [\apj] {10.3847/1538-4357/aa85df}, \href {https://ui.adsabs.harvard.edu/abs/2017ApJ...846..132H} {846, 132}

\bibitem[\protect\citeauthoryear{{Heinke}, {Grindlay}, {Lugger}, {Cohn}, {Edmonds}, {Lloyd}  \& {Cool}}{{Heinke} et~al.}{2003}]{heinke03}
{Heinke} C.~O.,  {Grindlay} J.~E.,  {Lugger} P.~M.,  {Cohn} H.~N.,  {Edmonds} P.~D.,  {Lloyd} D.~A.,   {Cool} A.~M.,  2003, \mn@doi [\apj] {10.1086/378885}, \href {https://ui.adsabs.harvard.edu/abs/2003ApJ...598..501H} {598, 501}

\bibitem[\protect\citeauthoryear{{Hummer} \& {Storey}}{{Hummer} \& {Storey}}{1987}]{hummerstorey87}
{Hummer} D.~G.,  {Storey} P.~J.,  1987, \mn@doi [\mnras] {10.1093/mnras/224.3.801}, \href {https://ui.adsabs.harvard.edu/abs/1987MNRAS.224..801H} {224, 801}

\bibitem[\protect\citeauthoryear{{Hynes}, {Steeghs}, {Casares}, {Charles}  \& {O'Brien}}{{Hynes} et~al.}{2003}]{hynes03}
{Hynes} R.~I.,  {Steeghs} D.,  {Casares} J.,  {Charles} P.~A.,   {O'Brien} K.,  2003, \mn@doi [\apjl] {10.1086/368108}, \href {https://ui.adsabs.harvard.edu/abs/2003ApJ...583L..95H} {583, L95}

\bibitem[\protect\citeauthoryear{{Hynes}, {Steeghs}, {Casares}, {Charles}  \& {O'Brien}}{{Hynes} et~al.}{2004}]{hynes04}
{Hynes} R.~I.,  {Steeghs} D.,  {Casares} J.,  {Charles} P.~A.,   {O'Brien} K.,  2004, \mn@doi [\apj] {10.1086/421014}, \href {https://ui.adsabs.harvard.edu/abs/2004ApJ...609..317H} {609, 317}

\bibitem[\protect\citeauthoryear{{Ingram} et~al.,}{{Ingram} et~al.}{2024}]{ingram24}
{Ingram} A.,  et~al., 2024, \mn@doi [\apj] {10.3847/1538-4357/ad3faf}, \href {https://ui.adsabs.harvard.edu/abs/2024ApJ...968...76I} {968, 76}

\bibitem[\protect\citeauthoryear{{John}, {De}, {Lucchini}, {Behar}, {Kara}, {MacLeod}, {Panagiotou}  \& {Wang}}{{John} et~al.}{2024}]{john24}
{John} C.,  {De} K.,  {Lucchini} M.,  {Behar} E.,  {Kara} E.,  {MacLeod} M.,  {Panagiotou} C.,   {Wang} J.,  2024, \mn@doi [\mnras] {10.1093/mnras/stae2432}, \href {https://ui.adsabs.harvard.edu/abs/2024MNRAS.535.2633J} {535, 2633}

\bibitem[\protect\citeauthoryear{{Kaiser}}{{Kaiser}}{2006}]{kaiser06}
{Kaiser} C.~R.,  2006, \mn@doi [\mnras] {10.1111/j.1365-2966.2006.10030.x}, \href {https://ui.adsabs.harvard.edu/abs/2006MNRAS.367.1083K} {367, 1083}

\bibitem[\protect\citeauthoryear{{Kaper}, {Trams}, {Barr}, {van Loon}  \& {Waters}}{{Kaper} et~al.}{1998}]{kaper98}
{Kaper} L.,  {Trams} N.~R.,  {Barr} P.,  {van Loon} J.~T.,   {Waters} L.~B.~F.~M.,  1998, \mn@doi [\apss] {10.1023/A:1001190413686}, \href {https://ui.adsabs.harvard.edu/abs/1998Ap&SS.255..199K} {255, 199}

\bibitem[\protect\citeauthoryear{{Kendrew} et~al.,}{{Kendrew} et~al.}{2015}]{kendrew15}
{Kendrew} S.,  et~al., 2015, \mn@doi [\pasp] {10.1086/682255}, \href {https://ui.adsabs.harvard.edu/abs/2015PASP..127..623K} {127, 623}

\bibitem[\protect\citeauthoryear{{Kosenkov}, {Veledina}, {Suleimanov}  \& {Poutanen}}{{Kosenkov} et~al.}{2020}]{kosenkov20}
{Kosenkov} I.~A.,  {Veledina} A.,  {Suleimanov} V.~F.,   {Poutanen} J.,  2020, \mn@doi [\aap] {10.1051/0004-6361/201936143}, \href {https://ui.adsabs.harvard.edu/abs/2020A&A...638A.127K} {638, A127}

\bibitem[\protect\citeauthoryear{{Krawczynski} et~al.,}{{Krawczynski} et~al.}{2022}]{ixpe_cygx1}
{Krawczynski} H.,  et~al., 2022, \mn@doi [Science] {10.1126/science.add5399}, \href {https://ui.adsabs.harvard.edu/abs/2022Sci...378..650K} {378, 650}

\bibitem[\protect\citeauthoryear{{Lasota}}{{Lasota}}{2000}]{lasota00}
{Lasota} J.~P.,  2000, \mn@doi [\aap] {10.48550/arXiv.astro-ph/0005073}, \href {https://ui.adsabs.harvard.edu/abs/2000A&A...360..575L} {360, 575}

\bibitem[\protect\citeauthoryear{{Lewis}}{{Lewis}}{2018}]{Lewis2018}
{Lewis} F.,  2018, Robotic Telescope, Student Research and Education Proceedings, \href {https://ui.adsabs.harvard.edu/abs/2018RTSRE...1..237L} {1, 237}

\bibitem[\protect\citeauthoryear{{Lewis}, {Russell}, {Fender}, {Roche}  \& {Clark}}{{Lewis} et~al.}{2008}]{Lewis2008}
{Lewis} F.,  {Russell} D.~M.,  {Fender} R.~P.,  {Roche} P.,   {Clark} J.~S.,  2008, arXiv e-prints, \href {https://ui.adsabs.harvard.edu/abs/2008arXiv0811.2336L} {p. arXiv:0811.2336}

\bibitem[\protect\citeauthoryear{{Maccarone}, {Osler}, {Miller-Jones}, {Atri}, {Russell}, {Meier}, {McHardy}  \& {Longa-Pe{\~n}a}}{{Maccarone} et~al.}{2020}]{maccarone20}
{Maccarone} T.~J.,  {Osler} A.,  {Miller-Jones} J. C.~A.,  {Atri} P.,  {Russell} D.~M.,  {Meier} D.~L.,  {McHardy} I.~M.,   {Longa-Pe{\~n}a} P.~A.,  2020, \mn@doi [\mnras] {10.1093/mnrasl/slaa120}, \href {https://ui.adsabs.harvard.edu/abs/2020MNRAS.498L..40M} {498, L40}

\bibitem[\protect\citeauthoryear{{Mainzer} et~al.,}{{Mainzer} et~al.}{2011}]{neowise}
{Mainzer} A.,  et~al., 2011, \mn@doi [\apj] {10.1088/0004-637X/731/1/53}, \href {https://ui.adsabs.harvard.edu/abs/2011ApJ...731...53M} {731, 53}

\bibitem[\protect\citeauthoryear{{Makishima}, {Maejima}, {Mitsuda}, {Bradt}, {Remillard}, {Tuohy}, {Hoshi}  \& {Nakagawa}}{{Makishima} et~al.}{1986}]{makishima86}
{Makishima} K.,  {Maejima} Y.,  {Mitsuda} K.,  {Bradt} H.~V.,  {Remillard} R.~A.,  {Tuohy} I.~R.,  {Hoshi} R.,   {Nakagawa} M.,  1986, \mn@doi [\apj] {10.1086/164534}, \href {https://ui.adsabs.harvard.edu/abs/1986ApJ...308..635M} {308, 635}

\bibitem[\protect\citeauthoryear{{Malzac}}{{Malzac}}{2013}]{malzac13}
{Malzac} J.,  2013, \mn@doi [\mnras] {10.1093/mnrasl/sls017}, \href {https://ui.adsabs.harvard.edu/abs/2013MNRAS.429L..20M} {429, L20}

\bibitem[\protect\citeauthoryear{{Malzac}}{{Malzac}}{2014}]{malzac14}
{Malzac} J.,  2014, \mn@doi [\mnras] {10.1093/mnras/stu1144}, \href {http://adsabs.harvard.edu/abs/2014MNRAS.443..299M} {443, 299}

\bibitem[\protect\citeauthoryear{{Markoff}, {Falcke}  \& {Fender}}{{Markoff} et~al.}{2001}]{markoff01}
{Markoff} S.,  {Falcke} H.,   {Fender} R.,  2001, \mn@doi [\aap] {10.1051/0004-6361:20010420}, \href {https://ui.adsabs.harvard.edu/abs/2001A&A...372L..25M} {372, L25}

\bibitem[\protect\citeauthoryear{{Markoff}, {Nowak}  \& {Wilms}}{{Markoff} et~al.}{2005}]{markoff05}
{Markoff} S.,  {Nowak} M.~A.,   {Wilms} J.,  2005, \mn@doi [\apj] {10.1086/497628}, \href {http://adsabs.harvard.edu/abs/2005ApJ...635.1203M} {635, 1203}

\bibitem[\protect\citeauthoryear{{Marscher}}{{Marscher}}{1983}]{marscher83}
{Marscher} A.~P.,  1983, \mn@doi [\apj] {10.1086/160597}, \href {https://ui.adsabs.harvard.edu/abs/1983ApJ...264..296M} {264, 296}

\bibitem[\protect\citeauthoryear{{Martin} et~al.,}{{Martin} et~al.}{2005}]{galex}
{Martin} D.~C.,  et~al., 2005, \mn@doi [\apjl] {10.1086/426387}, \href {https://ui.adsabs.harvard.edu/abs/2005ApJ...619L...1M} {619, L1}

\bibitem[\protect\citeauthoryear{{Mastroserio} et~al.,}{{Mastroserio} et~al.}{2025}]{mastroserio25}
{Mastroserio} G.,  et~al., 2025, \mn@doi [\apjl] {10.3847/2041-8213/ad9913}, \href {https://ui.adsabs.harvard.edu/abs/2025ApJ...978L..19M} {978, L19}

\bibitem[\protect\citeauthoryear{{Matsuoka} et~al.,}{{Matsuoka} et~al.}{2009}]{maxi}
{Matsuoka} M.,  et~al., 2009, \mn@doi [\pasj] {10.1093/pasj/61.5.999}, \href {https://ui.adsabs.harvard.edu/abs/2009PASJ...61..999M} {61, 999}

\bibitem[\protect\citeauthoryear{{McMullin}, {Waters}, {Schiebel}, {Young}  \& {Golap}}{{McMullin} et~al.}{2007}]{2007ASPC..376..127M}
{McMullin} J.~P.,  {Waters} B.,  {Schiebel} D.,  {Young} W.,   {Golap} K.,  2007, in {Shaw} R.~A.,  {Hill} F.,   {Bell} D.~J.,  eds,  Astronomical Society of the Pacific Conference Series Vol. 376, Astronomical Data Analysis Software and Systems XVI. p.~127

\bibitem[\protect\citeauthoryear{{Meier}, {Koide}  \& {Uchida}}{{Meier} et~al.}{2001}]{Meier-2001}
{Meier} D.~L.,  {Koide} S.,   {Uchida} Y.,  2001, \mn@doi [Science] {10.1126/science.291.5501.84}, \href {http://adsabs.harvard.edu/abs/2001Sci...291...84M} {291, 84}

\bibitem[\protect\citeauthoryear{{Middleton} et~al.,}{{Middleton} et~al.}{2017}]{middleton17}
{Middleton} M.~J.,  et~al., 2017, \mn@doi [\nar] {10.1016/j.newar.2017.07.002}, \href {https://ui.adsabs.harvard.edu/abs/2017NewAR..79...26M} {79, 26}

\bibitem[\protect\citeauthoryear{{Migliari}, {Tomsick}, {Maccarone}, {Gallo}, {Fender}, {Nelemans}  \& {Russell}}{{Migliari} et~al.}{2006}]{migliari06}
{Migliari} S.,  {Tomsick} J.~A.,  {Maccarone} T.~J.,  {Gallo} E.,  {Fender} R.~P.,  {Nelemans} G.,   {Russell} D.~M.,  2006, \mn@doi [\apjl] {10.1086/505028}, \href {https://ui.adsabs.harvard.edu/abs/2006ApJ...643L..41M} {643, L41}

\bibitem[\protect\citeauthoryear{{Migliari} et~al.,}{{Migliari} et~al.}{2007}]{migliari07}
{Migliari} S.,  et~al., 2007, \mn@doi [\apj] {10.1086/522023}, \href {https://ui.adsabs.harvard.edu/abs/2007ApJ...670..610M} {670, 610}

\bibitem[\protect\citeauthoryear{{Miller} et~al.,}{{Miller} et~al.}{2004}]{miller04}
{Miller} J.~M.,  et~al., 2004, \mn@doi [\apj] {10.1086/380196}, \href {https://ui.adsabs.harvard.edu/abs/2004ApJ...601..450M} {601, 450}

\bibitem[\protect\citeauthoryear{{Miller} et~al.,}{{Miller} et~al.}{2020}]{miller20}
{Miller} J.~M.,  et~al., 2020, \mn@doi [\apj] {10.3847/1538-4357/abbb31}, \href {https://ui.adsabs.harvard.edu/abs/2020ApJ...904...30M} {904, 30}

\bibitem[\protect\citeauthoryear{{Mirabel}, {Claret}, {Cesarsky}, {Boulade}  \& {Cesarsky}}{{Mirabel} et~al.}{1996}]{mirabel96}
{Mirabel} I.~F.,  {Claret} A.,  {Cesarsky} C.~J.,  {Boulade} O.,   {Cesarsky} D.~A.,  1996, \aap, \href {https://ui.adsabs.harvard.edu/abs/1996A&A...315L.113M} {315, L113}

\bibitem[\protect\citeauthoryear{{Misra}, {Roy}  \& {Yadav}}{{Misra} et~al.}{2021}]{Misra2021JApA...42...55M}
{Misra} R.,  {Roy} J.,   {Yadav} J.~S.,  2021, \mn@doi [Journal of Astrophysics and Astronomy] {10.1007/s12036-021-09734-2}, \href {https://ui.adsabs.harvard.edu/abs/2021JApA...42...55M} {42, 55}

\bibitem[\protect\citeauthoryear{{Mitsuda} et~al.,}{{Mitsuda} et~al.}{1984}]{mitsuda84}
{Mitsuda} K.,  et~al., 1984, \pasj, \href {https://ui.adsabs.harvard.edu/abs/1984PASJ...36..741M} {36, 741}

\bibitem[\protect\citeauthoryear{{Muno} \& {Mauerhan}}{{Muno} \& {Mauerhan}}{2006}]{muno6}
{Muno} M.~P.,  {Mauerhan} J.,  2006, \mn@doi [\apjl] {10.1086/507990}, \href {https://ui.adsabs.harvard.edu/abs/2006ApJ...648L.135M} {648, L135}

\bibitem[\protect\citeauthoryear{{Osterbrock} \& {Ferland}}{{Osterbrock} \& {Ferland}}{2006}]{osterbrock}
{Osterbrock} D.~E.,  {Ferland} G.~J.,  2006, {Astrophysics of gaseous nebulae and active galactic nuclei}.
University Science Books

\bibitem[\protect\citeauthoryear{{Paice} et~al.,}{{Paice} et~al.}{2019}]{paice19}
{Paice} J.~A.,  et~al., 2019, \mn@doi [\mnras] {10.1093/mnrasl/slz148}, \href {https://ui.adsabs.harvard.edu/abs/2019MNRAS.490L..62P} {490, L62}

\bibitem[\protect\citeauthoryear{{Pirard} et~al.,}{{Pirard} et~al.}{2004}]{hawki}
{Pirard} J.-F.,  et~al., 2004, in {Moorwood} A. F.~M.,  {Iye} M.,  eds,  Society of Photo-Optical Instrumentation Engineers (SPIE) Conference Series Vol. 5492, Ground-based Instrumentation for Astronomy. pp 1763--1772, \mn@doi{10.1117/12.578293}

\bibitem[\protect\citeauthoryear{{Plotkin}, {Markoff}, {Kelly}, {K{\"o}rding}  \& {Anderson}}{{Plotkin} et~al.}{2012}]{plotkin12}
{Plotkin} R.~M.,  {Markoff} S.,  {Kelly} B.~C.,  {K{\"o}rding} E.,   {Anderson} S.~F.,  2012, \mn@doi [\mnras] {10.1111/j.1365-2966.2011.19689.x}, \href {https://ui.adsabs.harvard.edu/abs/2012MNRAS.419..267P} {419, 267}

\bibitem[\protect\citeauthoryear{{Plotkin} et~al.,}{{Plotkin} et~al.}{2016}]{Plotkin-2016}
{Plotkin} R.~M.,  et~al., 2016, \mn@doi [\mnras] {10.1093/mnras/stv2861}, \href {http://adsabs.harvard.edu/abs/2016MNRAS.456.2707P} {456, 2707}

\bibitem[\protect\citeauthoryear{{Plotkin} et~al.,}{{Plotkin} et~al.}{2017}]{plotkin17}
{Plotkin} R.~M.,  et~al., 2017, \mn@doi [\apj] {10.3847/1538-4357/aa8d6d}, \href {https://ui.adsabs.harvard.edu/abs/2017ApJ...848...92P} {848, 92}

\bibitem[\protect\citeauthoryear{{Podgorn{\'y}} et~al.,}{{Podgorn{\'y}} et~al.}{2024}]{ixpe_j1727decay}
{Podgorn{\'y}} J.,  et~al., 2024, \mn@doi [\aap] {10.1051/0004-6361/202450566}, \href {https://ui.adsabs.harvard.edu/abs/2024A&A...686L..12P} {686, L12}

\bibitem[\protect\citeauthoryear{{Ponti}, {Fender}, {Begelman}, {Dunn}, {Neilsen}  \& {Coriat}}{{Ponti} et~al.}{2012}]{ponti12}
{Ponti} G.,  {Fender} R.~P.,  {Begelman} M.~C.,  {Dunn} R.~J.~H.,  {Neilsen} J.,   {Coriat} M.,  2012, \mn@doi [\mnras] {10.1111/j.1745-3933.2012.01224.x}, \href {https://ui.adsabs.harvard.edu/abs/2012MNRAS.422L..11P} {422, L11}

\bibitem[\protect\citeauthoryear{{Rahoui}, {Chaty}, {Rodriguez}, {Fuchs}, {Mirabel}  \& {Pooley}}{{Rahoui} et~al.}{2010}]{rahoui10}
{Rahoui} F.,  {Chaty} S.,  {Rodriguez} J.,  {Fuchs} Y.,  {Mirabel} I.~F.,   {Pooley} G.~G.,  2010, \mn@doi [\apj] {10.1088/0004-637X/715/2/1191}, \href {https://ui.adsabs.harvard.edu/abs/2010ApJ...715.1191R} {715, 1191}

\bibitem[\protect\citeauthoryear{{Rahoui}, {Lee}, {Heinz}, {Hines}, {Pottschmidt}, {Wilms}  \& {Grinberg}}{{Rahoui} et~al.}{2011}]{rahoui11}
{Rahoui} F.,  {Lee} J.~C.,  {Heinz} S.,  {Hines} D.~C.,  {Pottschmidt} K.,  {Wilms} J.,   {Grinberg} V.,  2011, \mn@doi [\apj] {10.1088/0004-637X/736/1/63}, \href {https://ui.adsabs.harvard.edu/abs/2011ApJ...736...63R} {736, 63}

\bibitem[\protect\citeauthoryear{{Rahoui}, {Coriat}  \& {Lee}}{{Rahoui} et~al.}{2014}]{rahoui14}
{Rahoui} F.,  {Coriat} M.,   {Lee} J.~C.,  2014, \mn@doi [\mnras] {10.1093/mnras/stu977}, \href {https://ui.adsabs.harvard.edu/abs/2014MNRAS.442.1610R} {442, 1610}

\bibitem[\protect\citeauthoryear{{Ressler} et~al.,}{{Ressler} et~al.}{2015}]{ressler15}
{Ressler} M.~E.,  et~al., 2015, \mn@doi [\pasp] {10.1086/682258}, \href {https://ui.adsabs.harvard.edu/abs/2015PASP..127..675R} {127, 675}

\bibitem[\protect\citeauthoryear{{Rieke} et~al.,}{{Rieke} et~al.}{2015}]{miri}
{Rieke} G.~H.,  et~al., 2015, \mn@doi [\pasp] {10.1086/682252}, \href {https://ui.adsabs.harvard.edu/abs/2015PASP..127..584R} {127, 584}

\bibitem[\protect\citeauthoryear{{Rushton} et~al.,}{{Rushton} et~al.}{2016}]{rushton16}
{Rushton} A.~P.,  et~al., 2016, \mn@doi [\mnras] {10.1093/mnras/stw2020}, \href {https://ui.adsabs.harvard.edu/abs/2016MNRAS.463..628R} {463, 628}

\bibitem[\protect\citeauthoryear{Russell, Fender, Hynes, Brocksopp, Homan, Jonker  \& Buxton}{Russell et~al.}{2006}]{russell06}
Russell D.~M.,  Fender R.~P.,  Hynes R.~I.,  Brocksopp C.,  Homan J.,  Jonker P.~G.,   Buxton M.~M.,  2006, \mn@doi [Mon. Not. Roy. Astron. Soc.] {10.1111/j.1365-2966.2006.10756.x}, 371, 1334

\bibitem[\protect\citeauthoryear{{Russell}, {Altamirano}, {Lewis}, {Roche}, {Markwardt}  \& {Fender}}{{Russell} et~al.}{2008}]{Russell2008ATel}
{Russell} D.~M.,  {Altamirano} D.,  {Lewis} F.,  {Roche} P.,  {Markwardt} C.~B.,   {Fender} R.~P.,  2008, The Astronomer's Telegram, \href {https://ui.adsabs.harvard.edu/abs/2008ATel.1586....1R} {1586, 1}

\bibitem[\protect\citeauthoryear{{Russell}, {Miller-Jones}, {Maccarone}, {Yang}, {Fender}  \& {Lewis}}{{Russell} et~al.}{2011}]{russell11}
{Russell} D.~M.,  {Miller-Jones} J.~C.~A.,  {Maccarone} T.~J.,  {Yang} Y.~J.,  {Fender} R.~P.,   {Lewis} F.,  2011, \mn@doi [\apjl] {10.1088/2041-8205/739/1/L19}, \href {https://ui.adsabs.harvard.edu/abs/2011ApJ...739L..19R} {739, L19}

\bibitem[\protect\citeauthoryear{{Russell} et~al.,}{{Russell} et~al.}{2013a}]{Russell-2013}
{Russell} D.~M.,  et~al., 2013a, \mn@doi [\mnras] {10.1093/mnras/sts377}, \href {http://adsabs.harvard.edu/abs/2013MNRAS.429..815R} {429, 815}

\bibitem[\protect\citeauthoryear{{Russell} et~al.,}{{Russell} et~al.}{2013b}]{russell13}
{Russell} D.~M.,  et~al., 2013b, \mn@doi [\mnras] {10.1093/mnras/sts377}, \href {https://ui.adsabs.harvard.edu/abs/2013MNRAS.429..815R} {429, 815}

\bibitem[\protect\citeauthoryear{{Russell} et~al.,}{{Russell} et~al.}{2013c}]{russell13b}
{Russell} D.~M.,  et~al., 2013c, \mn@doi [\apjl] {10.1088/2041-8205/768/2/L35}, \href {https://ui.adsabs.harvard.edu/abs/2013ApJ...768L..35R} {768, L35}

\bibitem[\protect\citeauthoryear{{Russell}, {Soria}, {Miller-Jones}, {Curran}, {Markoff}, {Russell}  \& {Sivakoff}}{{Russell} et~al.}{2014}]{russell14}
{Russell} T.~D.,  {Soria} R.,  {Miller-Jones} J.~C.~A.,  {Curran} P.~A.,  {Markoff} S.,  {Russell} D.~M.,   {Sivakoff} G.~R.,  2014, \mn@doi [\mnras] {10.1093/mnras/stt2498}, \href {http://adsabs.harvard.edu/abs/2014MNRAS.439.1390R} {439, 1390}

\bibitem[\protect\citeauthoryear{{Russell}, {Qasim}, {Bernardini}, {Plotkin}, {Lewis}, {Koljonen}  \& {Yang}}{{Russell} et~al.}{2018}]{russell18}
{Russell} D.~M.,  {Qasim} A.~A.,  {Bernardini} F.,  {Plotkin} R.~M.,  {Lewis} F.,  {Koljonen} K. I.~I.,   {Yang} Y.-J.,  2018, \mn@doi [\apj] {10.3847/1538-4357/aa9d8c}, \href {https://ui.adsabs.harvard.edu/abs/2018ApJ...852...90R} {852, 90}

\bibitem[\protect\citeauthoryear{{Russell} et~al.,}{{Russell} et~al.}{2019}]{Russell2019XBNEWS}
{Russell} D.~M.,  et~al., 2019, \mn@doi [Astronomische Nachrichten] {10.1002/asna.201913610}, \href {https://ui.adsabs.harvard.edu/abs/2019AN....340..278R} {340, 278}

\bibitem[\protect\citeauthoryear{{Russell}, {Casella}, {Kalemci}, {Vahdat Motlagh}, {Saikia}, {Pirbhoy}  \& {Maitra}}{{Russell} et~al.}{2020a}]{russell20_transientsoftjet}
{Russell} D.~M.,  {Casella} P.,  {Kalemci} E.,  {Vahdat Motlagh} A.,  {Saikia} P.,  {Pirbhoy} S.~F.,   {Maitra} D.,  2020a, \mn@doi [\mnras] {10.1093/mnras/staa1182}, \href {https://ui.adsabs.harvard.edu/abs/2020MNRAS.495..182R} {495, 182}

\bibitem[\protect\citeauthoryear{{Russell} et~al.,}{{Russell} et~al.}{2020b}]{russell20}
{Russell} T.~D.,  et~al., 2020b, \mn@doi [\mnras] {10.1093/mnras/staa2650}, \href {https://ui.adsabs.harvard.edu/abs/2020MNRAS.498.5772R} {498, 5772}

\bibitem[\protect\citeauthoryear{{Russell} et~al.,}{{Russell} et~al.}{2022}]{russell22_atel}
{Russell} D.~M.,  et~al., 2022, The Astronomer's Telegram, \href {https://ui.adsabs.harvard.edu/abs/2022ATel15596....1R} {15596, 1}

\bibitem[\protect\citeauthoryear{{Saikia} et~al.,}{{Saikia} et~al.}{2022}]{saikia22}
{Saikia} P.,  et~al., 2022, \mn@doi [\apj] {10.3847/1538-4357/ac6ce1}, \href {https://ui.adsabs.harvard.edu/abs/2022ApJ...932...38S} {932, 38}

\bibitem[\protect\citeauthoryear{{Saydjari} et~al.,}{{Saydjari} et~al.}{2023}]{decaps1}
{Saydjari} A.~K.,  et~al., 2023, \mn@doi [\apjs] {10.3847/1538-4365/aca594}, \href {https://ui.adsabs.harvard.edu/abs/2023ApJS..264...28S} {264, 28}

\bibitem[\protect\citeauthoryear{{Schlafly} et~al.,}{{Schlafly} et~al.}{2018}]{decaps2}
{Schlafly} E.~F.,  et~al., 2018, \mn@doi [\apjs] {10.3847/1538-4365/aaa3e2}, \href {https://ui.adsabs.harvard.edu/abs/2018ApJS..234...39S} {234, 39}

\bibitem[\protect\citeauthoryear{{Shaw} et~al.,}{{Shaw} et~al.}{2025}]{shaw25}
{Shaw} A.~W.,  et~al., 2025, \mn@doi [\aj] {10.3847/1538-3881/ad8eb1}, \href {https://ui.adsabs.harvard.edu/abs/2025AJ....169...21S} {169, 21}

\bibitem[\protect\citeauthoryear{{Singh} et~al.,}{{Singh} et~al.}{2014}]{Singh2014SPIE.9144E..1SS}
{Singh} K.~P.,  et~al., 2014, in {Takahashi} T.,  {den Herder} J.-W.~A.,   {Bautz} M.,  eds,  Society of Photo-Optical Instrumentation Engineers (SPIE) Conference Series Vol. 9144, Space Telescopes and Instrumentation 2014: Ultraviolet to Gamma Ray. p. 91441S, \mn@doi{10.1117/12.2062667}

\bibitem[\protect\citeauthoryear{{Singh} et~al.,}{{Singh} et~al.}{2017}]{Singh2017JApA...38...29S}
{Singh} K.~P.,  et~al., 2017, \mn@doi [Journal of Astrophysics and Astronomy] {10.1007/s12036-017-9448-7}, \href {https://ui.adsabs.harvard.edu/abs/2017JApA...38...29S} {38, 29}

\bibitem[\protect\citeauthoryear{{Skrutskie} et~al.,}{{Skrutskie} et~al.}{2006}]{skrutskie06}
{Skrutskie} M.~F.,  et~al., 2006, \mn@doi [\aj] {10.1086/498708}, \href {https://ui.adsabs.harvard.edu/abs/2006AJ....131.1163S} {131, 1163}

\bibitem[\protect\citeauthoryear{{Smith}, {Beall}  \& {Swain}}{{Smith} et~al.}{1990}]{smith90}
{Smith} H.~A.,  {Beall} J.~H.,   {Swain} M.~R.,  1990, \mn@doi [\aj] {10.1086/115326}, \href {https://ui.adsabs.harvard.edu/abs/1990AJ.....99..273S} {99, 273}

\bibitem[\protect\citeauthoryear{{Soria}, {Wu}  \& {Johnston}}{{Soria} et~al.}{1999}]{soria99}
{Soria} R.,  {Wu} K.,   {Johnston} H.~M.,  1999, \mn@doi [\mnras] {10.1046/j.1365-8711.1999.02933.x}, \href {https://ui.adsabs.harvard.edu/abs/1999MNRAS.310...71S} {310, 71}

\bibitem[\protect\citeauthoryear{{Stetson}}{{Stetson}}{1987}]{daophot}
{Stetson} P.~B.,  1987, \mn@doi [\pasp] {10.1086/131977}, \href {https://ui.adsabs.harvard.edu/abs/1987PASP...99..191S} {99, 191}

\bibitem[\protect\citeauthoryear{{Stetson}}{{Stetson}}{1990}]{Stetson1990}
{Stetson} P.~B.,  1990, \mn@doi [\pasp] {10.1086/132719}, \href {https://ui.adsabs.harvard.edu/abs/1990PASP..102..932S} {102, 932}

\bibitem[\protect\citeauthoryear{{Tandon} et~al.,}{{Tandon} et~al.}{2017}]{Tandon2017AJ....154..128T}
{Tandon} S.~N.,  et~al., 2017, \mn@doi [\aj] {10.3847/1538-3881/aa8451}, \href {https://ui.adsabs.harvard.edu/abs/2017AJ....154..128T} {154, 128}

\bibitem[\protect\citeauthoryear{{Tandon} et~al.,}{{Tandon} et~al.}{2020}]{Tandon2020AJ....159..158T}
{Tandon} S.~N.,  et~al., 2020, \mn@doi [\aj] {10.3847/1538-3881/ab72a3}, \href {https://ui.adsabs.harvard.edu/abs/2020AJ....159..158T} {159, 158}

\bibitem[\protect\citeauthoryear{{Timmer} \& {K{\"o}nig}}{{Timmer} \& {K{\"o}nig}}{1995}]{timmerkoenig}
{Timmer} J.,  {K{\"o}nig} M.,  1995, \aap, \href {https://ui.adsabs.harvard.edu/abs/1995A&A...300..707T} {300, 707}

\bibitem[\protect\citeauthoryear{{Tomsick} et~al.,}{{Tomsick} et~al.}{2004}]{tomsick04}
{Tomsick} J.~A.,  et~al., 2004, in American Astronomical Society Meeting Abstracts. p. 104.04

\bibitem[\protect\citeauthoryear{{Tonry} et~al.,}{{Tonry} et~al.}{2018}]{Tonry2018}
{Tonry} J.~L.,  et~al., 2018, \mn@doi [\apj] {10.3847/1538-4357/aae386}, \href {https://ui.adsabs.harvard.edu/abs/2018ApJ...867..105T} {867, 105}

\bibitem[\protect\citeauthoryear{{Tremou} et~al.,}{{Tremou} et~al.}{2020}]{tremou20}
{Tremou} E.,  et~al., 2020, \mn@doi [\mnras] {10.1093/mnrasl/slaa019}, \href {https://ui.adsabs.harvard.edu/abs/2020MNRAS.493L.132T} {493, L132}

\bibitem[\protect\citeauthoryear{{Uttley} \& {Casella}}{{Uttley} \& {Casella}}{2014}]{uttleycasella14}
{Uttley} P.,  {Casella} P.,  2014, \mn@doi [\ssr] {10.1007/s11214-014-0072-4}, \href {https://ui.adsabs.harvard.edu/abs/2014SSRv..183..453U} {183, 453}

\bibitem[\protect\citeauthoryear{{Valli} et~al.,}{{Valli} et~al.}{2024}]{valli24}
{Valli} R.,  et~al., 2024, \mn@doi [\aap] {10.1051/0004-6361/202449421}, \href {https://ui.adsabs.harvard.edu/abs/2024A&A...688A.128V} {688, A128}

\bibitem[\protect\citeauthoryear{{Vaughan}, {Edelson}, {Warwick}  \& {Uttley}}{{Vaughan} et~al.}{2003}]{vaughan03}
{Vaughan} S.,  {Edelson} R.,  {Warwick} R.~S.,   {Uttley} P.,  2003, \mn@doi [\mnras] {10.1046/j.1365-2966.2003.07042.x}, \href {https://ui.adsabs.harvard.edu/abs/2003MNRAS.345.1271V} {345, 1271}

\bibitem[\protect\citeauthoryear{{Veledina}, {Poutanen}  \& {Vurm}}{{Veledina} et~al.}{2013}]{Veledina-2013}
{Veledina} A.,  {Poutanen} J.,   {Vurm} I.,  2013, \mn@doi [\mnras] {10.1093/mnras/stt124}, \href {https://ui.adsabs.harvard.edu/abs/2013MNRAS.430.3196V} {430, 3196}

\bibitem[\protect\citeauthoryear{{Veledina} et~al.,}{{Veledina} et~al.}{2023}]{ixpe_swiftj1727}
{Veledina} A.,  et~al., 2023, \mn@doi [\apjl] {10.3847/2041-8213/ad0781}, \href {https://ui.adsabs.harvard.edu/abs/2023ApJ...958L..16V} {958, L16}

\bibitem[\protect\citeauthoryear{{Vincentelli} et~al.,}{{Vincentelli} et~al.}{2019}]{vincentelli19}
{Vincentelli} F.~M.,  et~al., 2019, \mn@doi [\apjl] {10.3847/2041-8213/ab5860}, \href {https://ui.adsabs.harvard.edu/abs/2019ApJ...887L..19V} {887, L19}

\bibitem[\protect\citeauthoryear{{Vincentelli} et~al.,}{{Vincentelli} et~al.}{2021}]{vincentelli21}
{Vincentelli} F.~M.,  et~al., 2021, \mn@doi [\mnras] {10.1093/mnras/stab475}, \href {https://ui.adsabs.harvard.edu/abs/2021MNRAS.503..614V} {503, 614}

\bibitem[\protect\citeauthoryear{{Wachter}}{{Wachter}}{2008}]{wachter08}
{Wachter} S.,  2008, in {Bandyopadhyay} R.~M.,  {Wachter} S.,  {Gelino} D.,   {Gelino} C.~R.,  eds,  American Institute of Physics Conference Series Vol. 1010, A Population Explosion: The Nature \& Evolution of X-ray Binaries in Diverse Environments. AIP, pp 210--214 (\mn@eprint {arXiv} {0804.1574}), \mn@doi{10.1063/1.2945044}

\bibitem[\protect\citeauthoryear{{Wang} \& {Wang}}{{Wang} \& {Wang}}{2014}]{wang14}
{Wang} X.,  {Wang} Z.,  2014, \mn@doi [\apj] {10.1088/0004-637X/788/2/184}, \href {https://ui.adsabs.harvard.edu/abs/2014ApJ...788..184W} {788, 184}

\bibitem[\protect\citeauthoryear{{Wright} \& {Barlow}}{{Wright} \& {Barlow}}{1975}]{wrightbarlow75}
{Wright} A.~E.,  {Barlow} M.~J.,  1975, \mn@doi [\mnras] {10.1093/mnras/170.1.41}, \href {https://ui.adsabs.harvard.edu/abs/1975MNRAS.170...41W} {170, 41}

\bibitem[\protect\citeauthoryear{{Wu}, {Soria}, {Hunstead}  \& {Johnston}}{{Wu} et~al.}{2001}]{wu01}
{Wu} K.,  {Soria} R.,  {Hunstead} R.~W.,   {Johnston} H.~M.,  2001, \mn@doi [\mnras] {10.1046/j.1365-8711.2001.03915.x}, \href {https://ui.adsabs.harvard.edu/abs/2001MNRAS.320..177W} {320, 177}

\bibitem[\protect\citeauthoryear{{Yadav}, {Agrawal}, {Antia}, {Manchanda}, {Paul}  \& {Misra}}{{Yadav} et~al.}{2017}]{Yadav2017CSci..113..591Y}
{Yadav} J.~S.,  {Agrawal} P.~C.,  {Antia} H.~M.,  {Manchanda} R.~K.,  {Paul} B.,   {Misra} R.,  2017, \mn@doi [Current Science] {10.18520/cs/v113/i04/591-594}, \href {https://ui.adsabs.harvard.edu/abs/2017CSci..113..591Y} {113, 591}

\bibitem[\protect\citeauthoryear{{Zdziarski}, {Poutanen}, {Mikolajewska}, {Gierlinski}, {Ebisawa}  \& {Johnson}}{{Zdziarski} et~al.}{1998}]{zdziarski98}
{Zdziarski} A.~A.,  {Poutanen} J.,  {Mikolajewska} J.,  {Gierlinski} M.,  {Ebisawa} K.,   {Johnson} W.~N.,  1998, \mn@doi [\mnras] {10.1046/j.1365-8711.1998.02021.x}, \href {https://ui.adsabs.harvard.edu/abs/1998MNRAS.301..435Z} {301, 435}

\bibitem[\protect\citeauthoryear{{Zdziarski}, {Zi{\'o}{\l}kowski}  \& {Miko{\l}ajewska}}{{Zdziarski} et~al.}{2019}]{zdziarski19}
{Zdziarski} A.~A.,  {Zi{\'o}{\l}kowski} J.,   {Miko{\l}ajewska} J.,  2019, \mn@doi [\mnras] {10.1093/mnras/stz1787}, \href {https://ui.adsabs.harvard.edu/abs/2019MNRAS.488.1026Z} {488, 1026}

\bibitem[\protect\citeauthoryear{{Zhao}, {Gandhi}, {Dashwood Brown}, {Knigge}, {Charles}, {Maccarone}  \& {Nuchvanichakul}}{{Zhao} et~al.}{2023}]{zhao23}
{Zhao} Y.,  {Gandhi} P.,  {Dashwood Brown} C.,  {Knigge} C.,  {Charles} P.~A.,  {Maccarone} T.~J.,   {Nuchvanichakul} P.,  2023, \mn@doi [\mnras] {10.1093/mnras/stad2226}, \href {https://ui.adsabs.harvard.edu/abs/2023MNRAS.525.1498Z} {525, 1498}

\bibitem[\protect\citeauthoryear{{Zuo} et~al.,}{{Zuo} et~al.}{2025}]{zuo25}
{Zuo} Z.,  et~al., 2025, \mn@doi [arXiv e-prints] {10.48550/arXiv.2505.23918}, \href {https://ui.adsabs.harvard.edu/abs/2025arXiv250523918Z} {p. arXiv:2505.23918}

\bibitem[\protect\citeauthoryear{{van Paradijs}, {Telesco}, {Kouveliotou}  \& {Fishman}}{{van Paradijs} et~al.}{1994}]{vanparadijs94}
{van Paradijs} J.,  {Telesco} C.~M.,  {Kouveliotou} C.,   {Fishman} G.~J.,  1994, \mn@doi [\apjl] {10.1086/187403}, \href {https://ui.adsabs.harvard.edu/abs/1994ApJ...429L..19V} {429, L19}

\makeatother
\end{thebibliography}

\section*{Appendix}

\renewcommand\thesubsection{\Alph{subsection}}

\subsection{Coordination of Multiwavelength Campaign}
\label{sec:coordination}

Once we were informed of the \jwst\ observation date and time window, we scheduled multiwavelength observations to coordinate with the \jwst\ window. Exact overlap was only possible with other space facilities (e.g. \nustar), and from Chilean facilities where the observation occurred during local night time. This was further complicated by an unannounced \jwst\ window change by a few hours resulting from the failure of a preceding observation. 

\begin{figure*}
    \centering
        \hspace{0.4cm}\includegraphics[angle=90,width=0.77\textwidth]{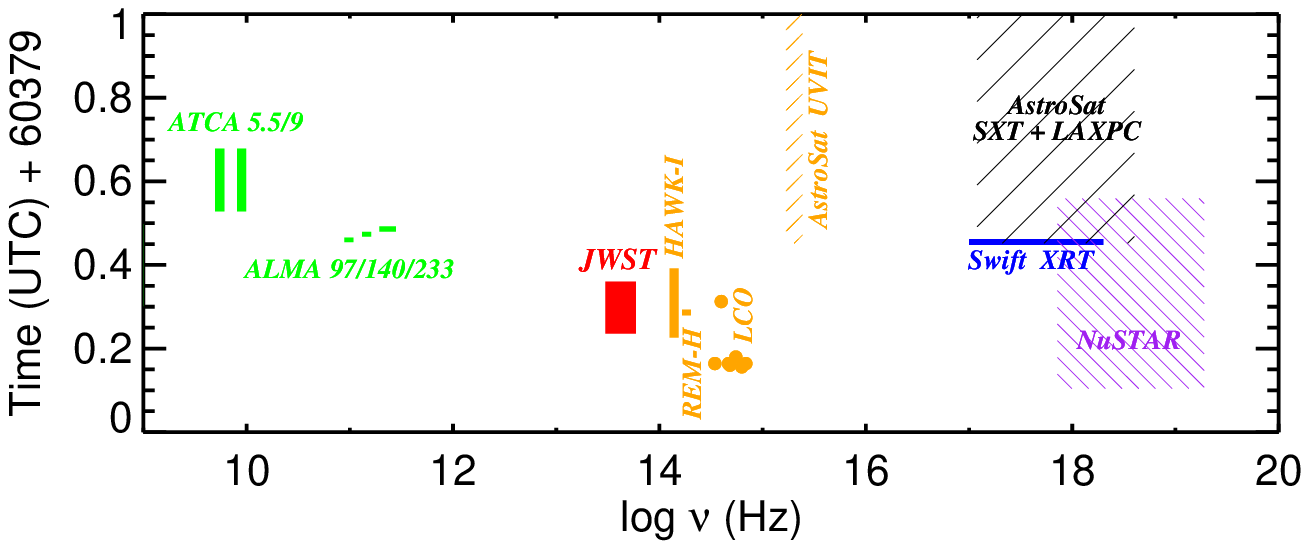}

    \vspace*{-0.45cm}\caption{Observation windows across the multiwavelength facilities from our campaign. All observations are continuous within the plotted observing window, except for LCO which is gappy and the plotted circles denote the mid-observation times; the $i'$ filter was the only LCO filter strictly simultaneous with \jwst. 
    \label{fig:coordination}}
\end{figure*}

Despite these issues, all datasets in our final campaign were coordinated to within a few hours of the \jwst\ window, all on the same calendar day. The respective observation windows for all facilities are depicted pictorially in Fig.\,\ref{fig:coordination}. Though times are not aligned to a common timeframe here, any corresponding differences across the observatories are negligible on the plot scale. 

\subsection{Extinction curve for \gx}

\label{sec:extinction}

The extinction values adopted along the sight-line to \gx\ are listed in Table\,\ref{tab:extinct} as a function of wavelength, together with the extinction curve ($A_\lambda$/$A_{\rm V}$). An \av\ value of 3.5\,mag is assumed, together with a systematic 1-$\sigma$ uncertainty of 0.5\,mag \citep{Gandhi-2011, heida17, zdziarski19}. The optical regime utilises the extinction curve of \citet{cardelli89}. This is extended to the infrared beyond 1\,\micron\ with the extinction curve of \citet{chiartielens06}, matched at the $K$ band. Extinction values at specific wavelengths of interest corresponding to some filters utilised herein are also noted for ease of reference. 

\begin{table}
    \centering
    \caption{Adopted Extinction Curve for \gx}
\begin{tabular}{lcr}
    \hline
    \hline
    $\lambda$ \dotfill\ & $A_{\lambda}$ & $A_{\lambda}$/$A_{\rm V}$ \\
    (\micron)  &  (mag) &  \\
    \hline
0.148\,(CaF2-1) \dotfill\ & 9.418 & 2.691\\
0.154\,(BaF2) \dotfill\ & 9.150 & 2.614\\
0.161\,(Sapphire) \dotfill\ & 8.941 & 2.555\\
0.172\,(Silica) \dotfill\ & 8.780 & 2.508\\
0.436\,($B$) \dotfill\ & 4.667 & 1.333\\
0.477\,($g'$) \dotfill\ & 4.236 & 1.210\\
0.55\,($V$) \dotfill\ & 3.5\,\p\,0.5 & 1.0 \\
0.622\,($r'$) \dotfill\ & 3.016 & 0.862\\
0.641\,($R$) \dotfill\ & 2.919 & 0.834\\
0.754\,($i'$) \dotfill\ & 2.367 & 0.676\\
0.870\,($z_s$) \dotfill\ & 1.787 & 0.510\\
1 \dotfill\ & 1.414 & 0.404\\
1.65\,($H$) \dotfill\ & 0.596 & 0.170\\
2 \dotfill\ & 0.430 & 0.123\\
2.159\,($K_s$) \dotfill\ & 0.382 & 0.109\\
3 \dotfill\ & 0.248 & 0.071\\
4 \dotfill\ & 0.189 & 0.054\\
5 \dotfill\ & 0.163 & 0.047\\
6 \dotfill\ & 0.151 & 0.043\\
7 \dotfill\ & 0.146 & 0.042\\
8 \dotfill\ & 0.149 & 0.043\\
9 \dotfill\ & 0.272 & 0.078\\
10 \dotfill\ & 0.378 & 0.108\\
11 \dotfill\ & 0.297 & 0.085\\
12 \dotfill\ & 0.234 & 0.067\\
13 \dotfill\ & 0.204 & 0.058\\
14 \dotfill\ & 0.206 & 0.059\\
15 \dotfill\ & 0.217 & 0.062\\
16 \dotfill\ & 0.231 & 0.066\\
17 \dotfill\ & 0.247 & 0.071\\
18 \dotfill\ & 0.263 & 0.075\\
19 \dotfill\ & 0.271 & 0.077\\
20 \dotfill\ & 0.265 & 0.076\\
21 \dotfill\ & 0.250 & 0.071\\
22 \dotfill\ & 0.234 & 0.067\\
23 \dotfill\ & 0.221 & 0.063\\
24 \dotfill\ & 0.207 & 0.059\\
25 \dotfill\ & 0.195 & 0.056\\
26 \dotfill\ & 0.184 & 0.053\\
27 \dotfill\ & 0.174 & 0.050\\
    \hline
    \hline
\end{tabular}
    \label{tab:extinct}
\end{table}

\subsection{Archival quasi-simultaneous MIR and X-ray luminosities of black hole XRBs}
\label{sec:appendixmirx}

Table\,\ref{tab:archivalmir} presents the respective MIR and X-ray luminosities of black hole XRBs. This includes all the sources from the extensive compilation of \citet{john24} featuring MAXI and NEOWISE detections \citep{maxi, neowise}, supplemented with several other works based on either ground-based MIR detections or detections in the quiescent state, including the present paper. Respective references are listed in the table. While this compilation should be fairly comprehensive in terms of MIR-detected sources, we do not claim it to be complete; e.g., there are multiple detections of specific sources such as \grs, not all of which are included herein. Multiple detections at very similar joint (overlapping) luminosities are excluded in some cases for plotting purposes. Unless otherwise referenced, the reported luminosities correspond to those from NEOWISE and MAXI from our analysis. Statistical uncertainties are typically $\approx$\,0.05--0.1 dex, and signal-to-noise values were always better than 3, as also restricted by \citet{john24}. Systematic uncertainties (e.g. those on distances discussed below, or on the background and field as discussed for \gx\ herein) often dominate for individual objects. 

Where energy-band conversions were needed, an X-ray photon-index of $\Gamma$\,=\,1.7 was used to estimate the 0.5--10\,keV power. Where NEOWISE data were used, the state assignment follows the band hardness ratios as in \citet{john24}; in some cases, there remains ambiguity in hard vs. soft states (e.g., for Swift\,J1753.5--0.127 which plateaued in a hard state for a very extended period), so state definitions should probably be taken with a grain of salt. For the MIR, we chose to report the 8\,\micron\ monochromatic luminosity as this should sample the source spectra deep in the thermal infrared within the MIRI band, while still allowing constraints from missions such as \spitzer\ and the tail-end of ground-based observatories sensitive to the so-called `N-band' atmospheric window. Any conversions from nearby wavelengths were carried out by using MIR spectral information, where available (e.g., extrapolating the spectral slope between \wise\ $W1$ and $W2$, or by assuming a flat flux density $F_{\nu}$\,$\propto$\,$\nu^0$ where not). Such band conversions are not perfect; e.g. they do not account properly for obscuration, nor for new spectral components dominating in different MIR bands. On the scale of plot, however, such differences will be minor in most cases. In selected sources such as \grs, we cannot rule out systematic differences related to its recent deep X-ray--obscured state, which could introduce scatter of 1--2 dex in its inferred X-ray powers. 

We used the custom \wise\ processing results from \citet{john24} where available, except in a few instances where we downloaded and analysed reported catalogue values from the NEOWISE survey \citep{neowise} for cross-checking or supplementing with newer data that has become available after publication of the results reported by \citeauthor{john24} Our results were broadly consistent with published values, with one caveat regarding the distances. 

Distances are mostly taken from the recent compilation of XRBs from \citet{zhao23}, or from \citet{blackcat} and references therein. In three cases, a new and significantly distance measurement results in qualitatively different inferences. These are GRS\,1716--249 for which \citet{casares23} report a best-estimate distance of 6.9\,kpc. Swift\,J1357.2--0933 with a distance of 6.3\,kpc \citep{charles19} and 6.0\,kpc for AT\,2019wey \citep{cao22}, respectively. These are all significantly larger than the values assumed in \citet{john24}, and move their location to the upper-right in the the MIR/X-ray luminosity plane, as discussed in the main text. Other relatively minor changes (including the distance of 8.0\,kpc for \gx\ used herein, as compared to 5.0\,kpc in \citeauthor{john24}), have a much smaller impact on the overall distribution. Furthermore, we stress that some of our assumed distances are likely still lower limits (e.g., Swift\,J1357.2--0933 and AT\,2019wey) which may change their location further with updated studies. 

Only observations that are simultaneous in X-rays and MIR to within 1\,day are retained for the hard as well as soft states. For the quiescent state, flux variations by orders of magnitude are unlikely, and we include non-simultaneous data in one instance for source Swift\,J1357.2--0933. Excluding this data point does not change any of our inferences qualitatively.

\begin{center}
\begin{table*}
\caption{Archival MIR and quasi-simultaneous X-ray luminosities of black hole XRBs.\label{tab:mirx}}
\begin{tabular}{lcccccr}
\hline
      Name & MJD & log\,$L_{\rm X-ray}$ &  log\,$L_{\rm MIR}$ & State & $d$ & References \\
           &  & log erg\,s$^{-1}$ &  log erg\,s$^{-1}$ &  & kpc & \\
\hline
A\,0620--00 & 53454 & 30.49 & 31.31 & Q & 1.5 & \citet{gallo07} \\
AT\,2019wey & 59100 & 35.18 & 31.52 & H & 6.0 & \citet{cao22}\\
 & 59257 & 35.03 & 31.72 & H & & \\
 & 59466 & 35.18 & 32.37 & H & & \\
 & 59830 & 35.27 & 32.30 & H & & \\
 & 59988 & 35.33 & 32.19 & H & & \\
 & 60197 & 35.12 & 32.31 & H & & \\
 & 60353 & 35.32 & 32.35 & H & & \\
GRO\,J0422+32 & 48874 & 37.53 & 34.02 & H & 2.5 &  \citet{vanparadijs94}; \citet{brocksopp04} \\
GRO\,J1655--40 & 53636 & 35.70 & 33.04 & H & 3.2 & \citet{migliari07}\\
GRS\,1716--249 & 56736 & 38.01 & 33.12 & H & 6.9 & \citet{john24}; \citet{casares23}\\
 & 57864 & 38.50 & 33.70 & H & & \citet{saikia22}\\
GRS\,1739--278 & 56736 & 38.42 & 33.77 & H & 7.2 & \citet{john24} \\
GRS\,1915+105 & 53280 & 38.50 & 34.19 & H & 9.4 & \citet{rahoui10} \\
 & 57125 & 38.82 & 34.31 & H & & \\
 & 57662 & 38.48 & 34.49 & H & & \\
 & 57854 & 38.09 & 34.08 & H & & \\
 & 58386 & 37.56 & 33.97 & O & & \\
 & 58750 & 36.83 & 34.94 & O & & \\
 & 58954 & 36.54 & 34.79 & O & & \\
 & 59114 & 37.69 & 34.53 & O & & \\
 & 59317 & 36.85 & 34.26 & O & & \\
 & 59681 & 36.47 & 34.06 & O & & \\
 & 60047 & 36.92 & 35.75 & O & & \\
 & 60101 & 38.10 & 35.50 & O & &\citetalias{gandhi25} \\
 & 60412 & 37.13 & 34.89 & O & & \\
GS\,1354--64 & 57245 & 38.78 & 34.59 & H & 25.0 & \citet{john24} \\
GX\,339--4 & 55266 & 37.89 & 35.32 & H & 8.0 & \citet{Gandhi-2011} \\
 & 56816 & 33.15 & 32.70 & Q & & This paper; \citet{tremou20}\\
 & 59828 & 37.16 & 34.71 & H & & \citet{russell22_atel}; Tremou et al. (2025, in prep.)\\
 & 60379 & 37.89 & 33.05 & S & & This paper\\
MAXI\,J0637--430 & 58923 & 37.42 & 31.75 & S & 8.7 & \citet{john24} \\
MAXI\,J1348--630 &  58527 & 37.98 & 32.94 & S & 2.2 & \\
 &  58690 & 36.35 & 32.73 & H & & \\
 &  58892 & 35.14 & 32.22 & H & & \\
MAXI\,J1535--571 & 58008 & 38.29 & 34.24 & H & 4.1 & \citet{baglio18}\\
                 & 58012 & 38.32 & 34.65 & H & & \citet{baglio18}\\
MAXI\,J1820+070 & 58258  & 37.65 & 34.22 & H & 2.9 & \citet{echiburutrujillo24}; \citet{bright25} \\
 & 58204 & 37.88 & 35.11 & H & &  \\
 & 58367 & 37.63 & 32.73 & S & &  \\
 & 58404 & 35.85 & 33.80 & H & & \citet{echiburutrujillo24}; \citet{bright25}\\
 & 58568 & 36.17 & 34.17 & H & & \\
MAXI\,J1836--194 & 55845 & 36.74 & 35.07 & H & 8.0 & \citet{russell13b}; \citet{russell14} \\
 & 55861 & 36.49 & 34.87 & H & & \citet{russell13b} \\
Swift\,J1357.2--0933 & 55581 & 35.52 & 31.43 & H & 6.3 & \citet{john24}; \citet{charles19} \\
 & 56736/55422$^\ddag$ & 31.58 & 32.50 & Q & & \citet{Plotkin-2016}; \citet{charles19} \\
Swift\,J1753.5--0127 &  55383 & 36.80 & 32.73 & H & 6.0 & \\
 &  56736 & 36.39 & 32.89 & S & & \\
 &  57101 & 36.36 & 32.55 & S & & \\
 &  57278 & 36.42 & 32.56 & S & & \\
 &  57287 & 36.48 & 32.77 & H & & \\
V404\,Cyg & 	60231 & 32.60 & 32.32 & Q & 2.2 & \citet{borowski25} \\
V4641\,Sgr &  56744 & 36.53 & 33.20 & S & 6.2 & \\
 &  58366 & 36.67 & 33.05 & H & & \\
 &  58566 & 36.59 & 32.71 & H & & \\
 &  58933 & 36.79 & 33.22 & S & & \\
XTE\,J1118+480 & 51649 & 35.69 & 34.45 & H & 1.7 & \citet{russell13}; \citet{dunn10} \\
 & 53331 & 30.48 & 30.88 & Q & & \citet{gallo07} \\
\hline
\end{tabular}
~\\
Accretion States: `Q': Quiescent, `S': Soft, `H': Hard. `O' refers to the X-ray Obscured state specific to \grs.\\
$^\ddag$X-ray/MIR mean observation dates -- non-simultaneous, as may be appropriate for quiescence, to within quiescence variability scatter (e.g. \citealt{russell18}).\\
Distances ($d$) are from \citet{zhao23} and \citet{blackcat} and references therein, unless otherwise cited.
\end{table*}
\end{center}

\section*{Affiliations}
{\small 
$^1$School of Physics \& Astronomy, University of Southampton, Southampton SO17 1BJ, UK\\
$^{2}$Center for Astrophysics and Space Science, New York University Abu Dhabi, PO Box 129188, Abu Dhabi, UAE\\
$^{3}$INAF-Osservatorio Astronomico di Brera, Via Bianchi 46, I-23807 Merate (LC), Italy\\
$^{4}$Department of Astronomy and Astrophysics, Tata Institute of Fundamental Research, 1 Homi Bhabha Road, Colaba, Mumbai 400005, India\\
$^{5}$INAF–Osservatorio Astronomico di Cagliari, Via della Scienza 5, 09047 Selargius (CA), Italy\\ 
$^{6}$Anton Pannekoek Institute for Astronomy \& Gravitation Astroparticle Physics Amsterdam (GRAPPA) Institute\\
$^{7}$University of Amsterdam, Science Park 904, 1098XH Amsterdam, The Netherlands\\
$^{8}$Department of Physics, University of Alberta, CCIS 4-181, Edmonton, AB T6G 2E1, Canada\\ 
$^{9}$Space Telescope Science Institute, 3700 San Martin Drive, Baltmore, MD, USA 21218\\
$^{10}$Eureka Scientific Inc., 2542 Delmar Avenue, Suite 100, Oakland, CA, 94602-30414, USA \\
$^{11}$Department of Physics \& Astronomy, Texas Tech University, Box 41051, Lubbock, TX 79409-1051, USA\\
$^{12}$ Dipartimento di Fisica, Universit\`{a} degli Studi di Milano, Via Celoria 16, I-20133 Milano, Italy\\
$^{13}$INAF - IASF Palermo, via Ugo La Malfa, 153, I-90146 Palermo, Italy\\
$^{14}$Department of Physics \& Astronomy, Butler University, 4600 Sunset Avenue, Indianapolis, IN 46208\\
$^{15}$Department of Physics and Astronomy, University of Lethbridge, Lethbridge, Alberta, T1K 3M4, Canada\\
$^{16}$INAF Istituto di Astrofisica e Planetologia Spaziali, Via del Fosso del Cavaliere 100, 00133 Roma, Italy\\
$^{17}$Department of Physics \& Astronomy, Louisiana State University, 202 Nicholoson Hall, Tower Drive, Baton Rouge, LA 70803, USA \\
$^{18}$South African Astronomical Observatory, P.O.Box 9, Observatory, 7935, South Africa\\
$^{19}$INAF-Osservatorio Astronomico di Roma, via Frascati 33, I-00078 Monteporzio Catone (RM), Italy\\
$^{20}$Inter-University Centre for Astronomy and Astrophysics, Post Bag 4, Ganeshkhind, Pune 411007, India\\
$^{21}$John A. Tomsick\,Space Sciences Laboratory, University of California, 7 Gauss Way, Berkeley, CA 94720-7450, USA\\
$^{22}$Faulkes Telescope Project, Cardiff, UK\\
$^{23}$The Schools' Observatory, Liverpool John Moores University, IC2, 146 Brownlow Hill, Liverpool, L3 5RF, UK\\
$^{24}$Centre for Extragalactic Astronomy, Department of Physics, Durham University, South Road, Durham DH1 3LE, UK\\
$^{25}$Department of Physics, R. J. College, Mumbai 400086, India\\
$^{26}$Jet Propulsion Laboratory, California Institute of Technology, 4800 Oak Grove Drive, Pasadena, CA 91109, USA\\
$^{27}$Instituto de Astrof\'\i{}sica de Canarias (IAC), E-38200 La Laguna, Tenerife, Spain\\
$^{28}$Departamento de  Astrof\'\i{}sica, Universidad de La Laguna (ULL), E-38206 La Laguna, Tenerife, Spain\\
}

\bsp	
\label{lastpage}
\end{document}